\documentclass[journal]{IEEEtran}

\pdfpageattr{/Group << /S /Transparency /I true /CS /DeviceRGB >>}

\usepackage[utf8]{inputenc}
\usepackage[T1]{fontenc}
\usepackage{textcomp}
\usepackage{microtype}
\usepackage{calc}
\usepackage{multirow}

\usepackage[range-phrase=--,per-mode=symbol-or-fraction,binary-units=true,range-units=single,list-units=single,detect-all]{siunitx}
\usepackage{silence}
\AtBeginDocument{
\DeclareSIUnit\byte{Byte}
\DeclareSIUnit\decibeli{dBi}
\DeclareSIUnit\decibelm{dBm}
\DeclareSIUnit\megasamplespersecond{Msps}
\DeclareSIUnit\dbm{dBm}
\DeclareSIUnit\ppm{ppm}
\DeclareSIUnit\watthour{Wh}
\sisetup{exponent-product=\ensuremath{\cdot}}
}

\usepackage{csquotes}
\usepackage[backend=biber,style=ieee,doi=false,isbn=false,mincitenames=1,maxcitenames=2]{biblatex}
\DeclareFieldFormat{sentencecase}{#1} 
\DeclareFieldFormat{titlecase}{#1} 
\usepackage{xpatch}
\xpatchbibmacro{textcite}{\addspace}{\addnbspace}{}{}
\xpatchbibmacro{Textcite}{\addspace}{\addnbspace}{}{}

\DefineBibliographyStrings{english}{
	andothers = et~al\adddot\addspace
}

\usepackage{tabularx}
\usepackage{booktabs}
\usepackage{rotating} 
\usepackage{multirow}
\newcommand\mycell[1]{%
   \begin{tabular}[t]{@{}c@{}} #1 \end{tabular}}

\usepackage[flushleft]{threeparttable}
\usepackage{xltabular}

\usepackage{amsmath}
\usepackage{amssymb}
\usepackage{amsfonts}
\usepackage{amsthm}

\usepackage{booktabs}

\usepackage[caption=false,font=footnotesize]{subfig}
\DeclareGraphicsExtensions{.pdf,.png,.jpg}
\usepackage{tikz}
\usetikzlibrary{calc}

\usetikzlibrary{mindmap}
\newsavebox\rotatenodebox
\makeatletter
\newcommand*{\rotateme}[1]{%
    \pgfmathparse{
        \tikz@grow@circle@from@start-
        (\pgfkeysvalueof{/tikz/sibling angle})*(\tikznumberofcurrentchild-1)
    }%
    \rotatebox[origin=c]{\pgfmathresult}{#1}%
}
\makeatother

\tikzset{
    conc3/.style={concept,scale=.7},
    orange/.style={
        conc3
    }
}

\usepackage[]{hyperref}
\usepackage{multicol}

\usepackage[american]{babel}
\usepackage{hyphenat}

\usepackage{algorithmic}
\usepackage{algorithm}

\usepackage{color}
\usepackage{soul} 

\usepackage{acro}
\usepackage[xindy]{glossaries}

\DeclareAcronym{ML}{short=ML, long=machine learning}
\DeclareAcronym{RL}{short=RL, long=reinforcement learning}
\DeclareAcronym{ABP}{short=ABP, long=adaptation-based programming}
\DeclareAcronym{PHY}{short=PHY, long=physical}
\DeclareAcronym{MAC}{short=MAC, long=medium access control}
\DeclareAcronym{QoS}{short=QoS, long=quality of service}
\DeclareAcronym{QoE}{short=QoE, long=quality of experience}
\DeclareAcronym{CSI}{short=CSI, long=channel state information}
\DeclareAcronym{RSS}{short=RSS, long=received signal strength}
\DeclareAcronym{GPS}{short=GPS, long=global positioning system}
\DeclareAcronym{LoS}{short=LoS, long=line of sight}
\DeclareAcronym{NLoS}{short=NLoS, long=non-line of sight}
\DeclareAcronym{AP}{short=AP, long=access point}
\DeclareAcronym{TOA}{short=TOA, long=time of arrival}
\DeclareAcronym{TDOA}{short=TDOA, long=time difference of arrival}
\DeclareAcronym{5G}{short=5G, long=fifth-generation mobile networks}
\DeclareAcronym{BS}{short=BS, long=base station}
\DeclareAcronym{WLAN}{short=WLAN, long=wireless local area network}
\DeclareAcronym{DCF}{short=DCF, long=distributed coordination function}
\DeclareAcronym{EDCA}{short=EDCA, long=enhanced distributed channel access}
\DeclareAcronym{HCF}{short=HCF, long=hybrid coordination function}
\DeclareAcronym{AIFS}{short=AIFS, long=arbitration inter-frame space}
\DeclareAcronym{AI}{short=AI, long=artificial intelligence}
\DeclareAcronym{iQRA}{short=iQRA, long=intelligent Q-learning based resource allocation}
\DeclareAcronym{CSMA}{short=CSMA, long=carrier sensing multiple access}
\DeclareAcronym{CSMA/CA}{short=CSMA/CA, long=carrier sensing multiple access with collision avoidance}
\DeclareAcronym{CSMA/ECA}{short=CSMA/ECA, long=carrier sense multiple access with enhanced collision avoidance}
\DeclareAcronym{CW}{short=CW, long=contention window}
\DeclareAcronym{COSB}{short=COSB, long=channel observation-based scaled backoff}
\DeclareAcronym{PDS}{short=PDS, long=post-decision state-based}
\DeclareAcronym{DL}{short=DL, long=deep learning}
\DeclareAcronym{DRL}{short=DRL, long=deep reinforcement learning}
\DeclareAcronym{DSL}{short=DSL, long=deep supervised learning}
\DeclareAcronym{DUSL}{short=DUSL, long=deep unsupervised learning}
\DeclareAcronym{MCS}{short=MCS, long=modulation and coding scheme}
\DeclareAcronym{V2I}{short=V2I, long=vehicle-to-infrastructure}
\DeclareAcronym{V2X}{short=V2X, long=vehicle-to-everything}
\DeclareAcronym{MLR}{short=MLR, long=machine-learned ranking}
\DeclareAcronym{LtR}{short=LTR, long=learning to rank}
\DeclareAcronym{DNN}{short=DNN, long=deep neural network}
\DeclareAcronym{CCMAC}{short=CCMAC, long=coordinated cooperative MAC}
\DeclareAcronym{HMM}{short=HMM, long=hidden Markov model}
\DeclareAcronym{NN}{short=NN, long=neural network}
\DeclareAcronym{MIMO}{short=MIMO, long=multiple-input multiple-output}
\DeclareAcronym{CNN}{short=CNN, long=convolutional neural network}
\DeclareAcronym{DCNN}{short=DCNN, long=deep convolutional neural network}
\DeclareAcronym{mmWave}{short=mmWave, long=millimeter wave}
\DeclareAcronym{LSTM}{short=LSTM, long=long short-term memory}
\DeclareAcronym{LTE-LAA}{short=LTE-LAA, long=LTE-Licensed Assisted Access}
\DeclareAcronym{NR-U}{short=NR-U, long=New Radio-Unlicensed}
\DeclareAcronym{LTE-U}{short=LTE-U, long=LTE-Unlicensed}
\DeclareAcronym{TXOP}{short=TXOP, long=transmission opportunity}
\DeclareAcronym{CSAT}{short=CSAT, long=carrier sense adaptive transmission}
\DeclareAcronym{VCFG}{short=VCFG, long=virtual coalition formation game}
\DeclareAcronym{LBT}{short=LBT, long=listen before talk}
\DeclareAcronym{SBS}{short=SBS, long=small base station}
\DeclareAcronym{MOS}{short=MOS, long=mean opinion score}
\DeclareAcronym{SDN}{short=SDN, long=software-defined networking}
\DeclareAcronym{SDR}{short=SDR, long=software-defined radio}
\DeclareAcronym{MME}{short=MME, long=mobile management entity, long-plural=mobile management entities} 
\DeclareAcronym{C-RAN}{short=C-RAN, long=LTE cloud wireless access network}
\DeclareAcronym{HARQ}{short=HARQ, long=hybrid automatic repeat request} 
\DeclareAcronym{RAT}{short=RAT, long=radio access technology,long-plural-form=radio access technologies}
\DeclareAcronym{DFS}{short=DFS, long=dynamic frame selection}
\DeclareAcronym{DCM}{short=DCM, long=duty cycle management}
\DeclareAcronym{ABS}{short=ABS, long=almost blank sub-frame}
\DeclareAcronym{LTE-A}{short=LTE-A, long=LTE-Advanced}
\DeclareAcronym{OFDMA}{short=OFDMA, long=orthogonal frequency-division multiple access}
\DeclareAcronym{MU-MIMO}{short=MU-MIMO, long=multi-user MIMO}
\DeclareAcronym{UE}{short=UE, long=user entity}
\DeclareAcronym{YOLO}{short=YOLO, long=you only look once}
\DeclareAcronym{ANN}{short=ANN, long=artificial neural network}
\DeclareAcronym{SNR}{short=SNR, long=signal to noise ratio}
\DeclareAcronym{ACK}{short=ACK, long=acknowledgment}
\DeclareAcronym{NACK}{short=NACK, long=negative acknowledgment}
\DeclareAcronym{MDP}{short=MDP, long=Markov decision process}
\DeclareAcronym{DQN}{short=DQN, long=deep Q-network}
\DeclareAcronym{DDQN}{short=DDQN, long=double deep Q-network}
\DeclareAcronym{BER}{short=BER, long=bit error rate}
\DeclareAcronym{ARF}{short=ARF, long=auto rate fallback}
\DeclareAcronym{eNB}{short=eNB, long=evolved Node B}
\DeclareAcronym{SGI}{short=SGI, long=short guard interval}
\DeclareAcronym{COTS}{short=COTS, long=commercial off-the-shelf}
\DeclareAcronym{DT}{short=DT, long=decision tree}
\DeclareAcronym{RF}{short=RF, long=random forest}
\DeclareAcronym{RFR}{short=RFR, long=random forest regressor}
\DeclareAcronym{SVM}{short=SVM, long=support vector machine}
\DeclareAcronym{LiBRA}{short=LiBRA, long=learning-based beam and rate adaptation}
\DeclareAcronym{6G}{short=6G, long=sixth-generation mobile networks}
\DeclareAcronym{IoT}{short=IoT, long=Internet of things}
\DeclareAcronym{CRN}{short=CRN, long=cognitive radio network}
\DeclareAcronym{UAV}{short=UAV, long=unmanned aerial vehicle}
\DeclareAcronym{VANET}{short=VANET, long=vehicular ad doc network}
\DeclareAcronym{HWMP}{short=HWMP, long=hybrid wireless mesh protocol}
\DeclareAcronym{SARSA}{short=SARSA, long=state action reward state action}
\DeclareAcronym{CRQ}{short=CRQ, long=collision resolution request}
\DeclareAcronym{MAB}{short=MAB, long=multi-armed bandit}
\DeclareAcronym{SS}{short=SS, long=spatial stream}
\DeclareAcronym{GBRT}{short=GBRT, long=gradient boosted regression tree}
\DeclareAcronym{SVR}{short=SVR, long=support vector regressor}
\DeclareAcronym{DPP}{short=DPP, long=determinantal point process}
\DeclareAcronym{CDF}{short=CDF, long=cumulative density function}
\DeclareAcronym{kNN}{short=kNN, long=k-nearest neighbor}
\DeclareAcronym{LTE}{short=LTE, long=Long Term Evolution}
\DeclareAcronym{CR}{short=CR, long=cognitive radio}
\DeclareAcronym{WMN}{short=WMN, long=wireless mesh network}
\DeclareAcronym{RNN}{short=RNN, long=recurrent neural network}
\DeclareAcronym{A-MSDU}{short=A-MSDU, long=aggregated MAC service data unit}
\DeclareAcronym{A-MPDU}{short=A-MPDU, long=aggregated MAC protocol data unit}
\DeclareAcronym{FCS}{short=FCS, long=frame check sequence}
\DeclareAcronym{DSRC}{short=DSRC, long=dedicated short-range communications}
\DeclareAcronym{TCP}{short=TCP, long=transmission control protocol}
\DeclareAcronym{FER}{short=FER, long=frame error rate}
\DeclareAcronym{CRAHN}{short=CRAHN, long=cognitive radio ad hoc network}
\DeclareAcronym{MANET}{short=MANET, long=mobile ad hoc network}
\DeclareAcronym{OLSR}{short=OLSR, long=optimized link state routing}
\DeclareAcronym{DTN}{short=DTN, long=delay tolerant network}
\DeclareAcronym{MEC}{short=MEC, long=multi-access edge computing}
\DeclareAcronym{NFV}{short=NFV, long=network functions virtualization}
\DeclareAcronym{LPWAN}{short=LPWAN, long=low-power wide area network}
\DeclareAcronym{WAN}{short=WAN, long=wide area network}
\DeclareAcronym{WSN}{short=WSN, long=wireless sensor network}
\DeclareAcronym{BSS}{short=BSS, long=basic service set}
\DeclareAcronym{GCN}{short=GCN, long=graph convolutional network}
\DeclareAcronym{SAP}{short=SAP, long=spatial adaptive play}
\DeclareAcronym{LASSO}{short=LASSO, long=least absolute shrinkage and selection operator}
\DeclareAcronym{OLS}{short=OLS, long=ordinary least squares}
\DeclareAcronym{LPA}{short=LPA, long=label propagation algorithm}
\DeclareAcronym{GNA}{short=GNA, long=Girvan-Newman algorithm}
\DeclareAcronym{STA}{short=STA, long=station}
\DeclareAcronym{MFNN}{short=MFNN, long=multi-layer feed-forward neural network}
\DeclareAcronym{RSSI}{short=RSSI, long=received signal strength indicator}
\DeclareAcronym{CCA}{short=CCA, long=clear channel assessment}
\DeclareAcronym{SD-WLAN}{short=SD-WLAN, long= software-defined WLAN}
\DeclareAcronym{NCA}{short=NCA, long=normalized channel access}
\DeclareAcronym{FDR}{short=FDR, long=frame delivery ratio}
\DeclareAcronym{NB}{short=NB, long=naive Bayes}
\DeclareAcronym{NBT}{short=NBT, long=Naive Bayes Tree}
\DeclareAcronym{J48DT}{short=J48DT, long=J48 decision three}
\DeclareAcronym{MLP}{short=MLP, long=multilayer perceptrons}
\DeclareAcronym{MSE}{short=MSE, long=mean square error}
\DeclareAcronym{FFNN}{short=FFNN, long=feed forward neural network}
\DeclareAcronym{ARMA}{short=ARMA, long=autoregressive moving average}
\DeclareAcronym{HetNet}{short=HetNet, long=heterogeneous network}
\DeclareAcronym{MTL}{short=MTL, long=multi-task learning}
\DeclareAcronym{MRL}{short=MRL, long=multi-resolution learning}
\DeclareAcronym{M2M}{short=M2M, long=machine to machine}
\DeclareAcronym{RU}{short=RU, long=resource unit}
\DeclareAcronym{CFO}{short=CFO, long=carrier frequency offset}
\DeclareAcronym{SR}{short=SR, long=spatial reuse}
\DeclareAcronym{UCB}{short=UCB, long=upper confidence bound}
\DeclareAcronym{TS}{short=TS, long=Thompson sampling}
\DeclareAcronym{WARP}{short=WARP, long=wireless open access research platform}
\DeclareAcronym{RAW}{short=RAW, long=restricted access window}
\DeclareAcronym{DQL}{short=DQL, long=deep Q-learning}
\DeclareAcronym{QL}{short=QL, long=Q-learning}
\DeclareAcronym{LR-WPAN}{short=LR-WPAN, long=low-rate wireless personal area network}
\DeclareAcronym{NB-IoT}{short=NB-IoT, long=narrow-band IoT}
\DeclareAcronym{SL}{short=SL, long=supervised learning}
\DeclareAcronym{USL}{short=USL, long=unsupervised learning}
\DeclareAcronym{AC}{short=AC, long=access category}
\DeclareAcronym{DBCA}{short=DBCA, long=dynamic bandwidth channel access}
\DeclareAcronym{SBCA}{short=SBCA, long=static bandwidth channel access}
\DeclareAcronym{Li-Fi}{short=Li-Fi, long=light fidelity}
\DeclareAcronym{VLC}{short=VLC, long=visible light communication}
\DeclareAcronym{ITE}{short=ITE, long=iterative trial and error}
\DeclareAcronym{HA-DBCA}{short=HA-DBCA, long=hybrid adaptive DBCA}
\DeclareAcronym{DCB}{short=DCB, long=on-demand channel bonding}
\DeclareAcronym{MADDPG}{short=MADDPG, long=multi-agent deep deterministic policy gradient}
\DeclareAcronym{GNN}{short=GNN, long=graph neural network}
\DeclareAcronym{MH-GAN}{short=MH-GAN, long=Metropolis-Hastings generative adversarial network}
\DeclareAcronym{PNN}{short=PNN, long=probabilistic neural network}
\DeclareAcronym{DDPG}{short=DDPG, long=deep deterministic policy gradient}
\DeclareAcronym{EIED}{short=EIED, long=exponential-increase exponential-decrease}
\DeclareAcronym{CARA}{short=CARA, long=collision-aware rate adaptation}
\DeclareAcronym{MUSE}{short=MUSE, long=MU-MIMO user selection}
\DeclareAcronym{IDS}{short=IDS, long=intrusion detection system}
\DeclareAcronym{MHCP}{short=MHCP, long=Matérn hard-core processes}
\DeclareAcronym{POMDP}{short=POMDP, long=partially observable Markov decision process}
\DeclareAcronym{SARA}{short=SARA, long=stochastic automata rate adaptation}
\DeclareAcronym{NNMR}{short=NNMR, long=non-negative multiple regression}
\DeclareAcronym{GT}{short=GT, long=game theory}
\DeclareAcronym{LR}{short=LR, long=linear regression}
\DeclareAcronym{DPPL}{short=DPPL, long=determinantal point process learning}
\DeclareAcronym{TRPO}{short=TRPO, long=trust region policy Optimization}
\DeclareAcronym{RT}{short=RT, long=regression tree}
\DeclareAcronym{TL}{short=TL, long=transfer learning}
\DeclareAcronym{CT}{short=CT, long=classification tree}
\DeclareAcronym{REPT}{short=REPT, long=reduced error pruning tree}
\DeclareAcronym{SOHMMM}{short=SOHMMM, long=self-organizing hidden Markov model map}
\DeclareAcronym{EMA}{short=EMA, long=expectation modification algorithm}
\DeclareAcronym{FL}{short=FL, long=federated learning}
\DeclareAcronym{QNN}{short=QNN, long=Q neural network}
\DeclareAcronym{EM}{short=EM, long=expectation maximization}
\DeclareAcronym{FPGA}{short=FPGA, long=field programmable gate array}
\DeclareAcronym{FWA}{short=FWA, long=fixed wireless access}
\DeclareAcronym{AL}{short=AL, long=adaptive learning}
\DeclareAcronym{NIC}{short=NIC, long=network interface card}
\DeclareAcronym{QAM}{short=QAM, long=quadrature amplitude modulator}
\DeclareAcronym{RTS}{short=RTS, long=request to send}
\DeclareAcronym{LMT}{short=LMT, long=logistic model tree}
\DeclareAcronym{KPI}{short=KPI, long=key performance indicator}
\DeclareAcronym{ITU}{short=ITU, long=International Telecommunications Union}
\DeclareAcronym{ETSI}{short=ETSI, long=European Telecommunications Standards Institute}
\DeclareAcronym{GANA}{short=GANA, long=generic autonomic network architecture}
\DeclareAcronym{CCOD}{short=CCOD, long=centralized contention window optimization with DRL}
\DeclareAcronym{SLA}{short=SLA, long=stochastic learning automata}
\DeclareAcronym{SoA}{short=SoA, long=state-of-the-art}
\DeclareAcronym{GPU}{short=GPU, long=graphics processing unit}
\DeclareAcronym{DCC}{short=DCC, long=dilated causal convolution}

\usepackage[capitalize,noabbrev]{cleveref}
\Crefformat{figure}{#2Figure~#1#3}
\Crefformat{equation}{#2(#1)#3}

\crefname{algocf}{Algorithm}{Algorithms}
\Crefname{algocf}{Algorithm}{Algorithms}

\usepackage{pdflscape}

\usepackage[flushleft]{threeparttable} 

\definecolor{boristext}{rgb}{0.1, 0.44, 0.84}
\definecolor{boriscomments}{rgb}{0.8, 0.2, 0.04}

\begin{document}
\onecolumn
{\Large 
\noindent Please cite this paper as: \newline
S. Szott et al., ``Wi-Fi Meets ML: A Survey on Improving IEEE 802.11 Performance With Machine Learning,'' in IEEE Communications Surveys \& Tutorials, vol. 24, no. 3, pp. 1843--1893, thirdquarter 2022, doi: 10.1109/COMST.2022.3179242.



\vspace{2cm}
\begin{verbatim}
@ARTICLE{szott2022wifi,
  author={Szott, Szymon and Kosek-Szott,
  Katarzyna and Gawłowicz, Piotr and Gómez,
  Jorge Torres and Bellalta, Boris and Zubow, Anatolij 
  and Dressler, Falko},
  journal={IEEE Communications Surveys \& Tutorials},
  title={{Wi-Fi Meets ML: A Survey on Improving IEEE 802.11
  Performance with Machine Learning}},
  year={2022},
  volume={24},
  number={3},
  pages={1843--1893},
  doi={10.1109/COMST.2022.3179242}
}
\end{verbatim}
}
\twocolumn
\clearpage

\title{Wi-Fi Meets ML: A Survey on Improving IEEE 802.11 Performance with Machine Learning}

\author{%
Szymon Szott,
Katarzyna Kosek-Szott,
Piotr Gaw{\l}owicz,~\IEEEmembership{IEEE Student Member},\\
Jorge Torres Gómez,~\IEEEmembership{IEEE Senior Member},
Boris Bellalta,~\IEEEmembership{IEEE Senior Member},\\
Anatolij Zubow,~\IEEEmembership{IEEE Senior Member},
Falko Dressler,~\IEEEmembership{IEEE Fellow}
\thanks{S. Szott and K. Kosek-Szott are with AGH University of Science and Technology, Kraków, Poland, e-mail: \{szott, kks\}@agh.edu.pl.}
\thanks{P. Gaw{\l}owicz, J. Torres Gómez, A. Zubow, and F. Dressler are with the School of Electrical Engineering and Computer Science, TU Berlin, 10587 Berlin, Germany, e-mail: \{gawlowicz, torres-gomez, zubow, dressler\}@tkn.tu-berlin.de.}
\thanks{B. Bellalta is with the Department of Information and Communication Technologies, UPF Barcelona, Spain, e-mail: boris.bellalta@upf.edu.}%
\thanks{
For the purpose of Open Access, the author has applied a CC-BY public copyright licence to any Author Accepted Manuscript (AAM) version arising from this submission. 
This work was supported in part by the National Science Centre, Poland (DEC-2020/39/I/ST7/01457). 
This work was supported in part by the Federal Ministry of Education and Research (BMBF, Germany) project OTB-5G+ under grant 16KIS0985 and the 6G Research and Innovation Cluster 6G-RIC under grant 16KISK020K, as well as by the project ML4WIFI funded by the German Research Foundation (DFG) under grant number DR 639/28-1.
This work was supported in part by grants WINDMAL PGC2018-099959-B-I00 (MCIU/AEI/FEDER,UE) and SGR017-1188 (AGAUR).}
}

\maketitle

\begin{abstract}

Wireless local area networks (WLANs) empowered by IEEE 802.11 (Wi-Fi) hold a dominant position in providing Internet access thanks to their freedom of deployment and configuration as well as the existence of affordable and highly interoperable devices. The Wi-Fi community is currently deploying Wi-Fi~6 and developing Wi-Fi~7, which will bring higher data rates, better multi-user and multi-AP support, and, most importantly, improved configuration flexibility. These technical innovations, including the plethora of configuration parameters, are making next-generation WLANs exceedingly complex as the dependencies between parameters and their joint optimization usually have a non-linear impact on network performance. The complexity is further increased in the case of dense deployments and coexistence in shared bands. While classical optimization approaches fail in such conditions, machine learning (ML) is able to handle complexity. Much research has been published on using ML to improve Wi-Fi performance and solutions are slowly being adopted in existing deployments. In this survey, we adopt a structured approach to describe the various Wi-Fi areas where ML is applied. To this end, we analyze over 250 papers in the field, providing readers with an overview of the main trends. Based on this review, we identify specific open challenges and provide general future research directions. 
\end{abstract}


\IEEEpeerreviewmaketitle

\makeatletter
\def\ps@IEEEtitlepagestyle{%
  \def\@oddfoot{\mycopyrightnotice}%
  \def\@evenfoot{}%
}
\def\mycopyrightnotice{%
  {\footnotesize\hfill 
  {\color{red}
  Published in IEEE Communications Surveys \& Tutorials. DOI 10.1109/COMST.2022.3179242}
  \hfill}%
  \gdef\mycopyrightnotice{}
}
\makeatother

\section{Introduction}

\Acp{WLAN}, standardized in IEEE~802.11 and commercialized as Wi-Fi, hold a dominant position in providing wireless Internet access.
Cisco’s Visual Networking Index Forecast estimates Wi-Fi's share of Internet traffic to be 51\% in 2022~\cite{Cisco2020}.
Wi-Fi~6~\cite{khorov2019tutorial,bellalta2016ieee,afaqui2016ieee} has become state of the art for all new consumer products and Wi-Fi~7~\cite{deng2020ieee,khorov2020current,garcia-rodriguez2020wifi7} is already under development.
There are several reasons for the popularity of Wi-Fi: well-defined use cases, freedom of deployment and configuration (thanks to operating in unlicensed bands), and the existence of inexpensive in manufacturing and highly interoperable devices. 

The 802.11 protocol family has received, in recent years, regular updates leading to performance improvements and new features.
These technical innovations provide a challenge: \emph{the next generations of Wi-Fi are becoming exceedingly complex}. Specifically, each new mechanism, designed to improve network performance, comes with a plethora of parameters that have to be configured.
Additionally, there are new application requirements: Wi-Fi is no longer limited to broadband Internet access but is also being used in other situations, e.g., ultra-low latency communication for machine-to-machine communication.
This multi-modal operation needs to be supported through a proper configuration, which in most cases is left out of the standard. 
For example, depending on the combination of~\ac{RU} assignment in 802.11ax, the network throughput may vary by more than 100\%~\cite{khorov2019tutorial}. 
In most cases, multiple parameters have to be jointly optimized. This task is non-trivial  as the dependencies between parameters and their joint optimization have a highly non-linear impact on network performance. 
For example,~\textcite{wilhelmi2021spatial} show that the performance of overlapping 802.11 Wi-Fi networks does not depend linearly on sensitivity and transmission power settings. The level of complexity is further increased in the case of coexisting network technologies. 

Up till now, the goal of the mainline 802.11 amendments was to provide high throughput (802.11n, 802.11ac) and efficiency in dense environments through deterministic channel access (802.11ax). 
However, future Wi-Fi generations are anticipated to accommodate ultra-low latency and ultra-high reliability traffic (802.11be). 
Hence, the proper and timely update of the transmission settings is of key importance. Meanwhile, \emph{finding adequate configurations in an enormous search space using traditional algorithms is too time and computation resource consuming}. Additionally, new WLAN mechanisms also bring overhead in terms of additional measurements which are needed to provide input to their respective control algorithms. In the past, with only a few possible \ac{MCS} values (i.e., in early versions of 802.11), it was possible to quickly test all of them and select the best one. Currently, such a selection is practically impossible.

\subsection{Need for ML in Wi-Fi}
The increasing Wi-Fi complexity coupled with uncoordinated deployment, distributed management, and network densification may negatively impact the operation of future 802.11 networks.
A candidate approach to solving these performance-related problems is to apply \ac{ML}, a type of artificial intelligence, where ``algorithms can learn from training data
without being explicitly programmed''~\cite{wang2020thirty}.
Indeed, the IEEE 802.11 Working Group is discussing the use of \ac{ML} for improving the performance of beyond-802.11be networks
\cite{ming2022look}.
So far, \ac{ML}-based techniques have been explored for a variety of problems in networking~\cite{sun2019application,kulin2020survey,luong2019applications}.
Successful solutions are applied to fields ranging from configuring physical layer parameters to traffic prediction.

\begin{figure}
\centering
\vspace{-.5em}
\includegraphics[width=\columnwidth]{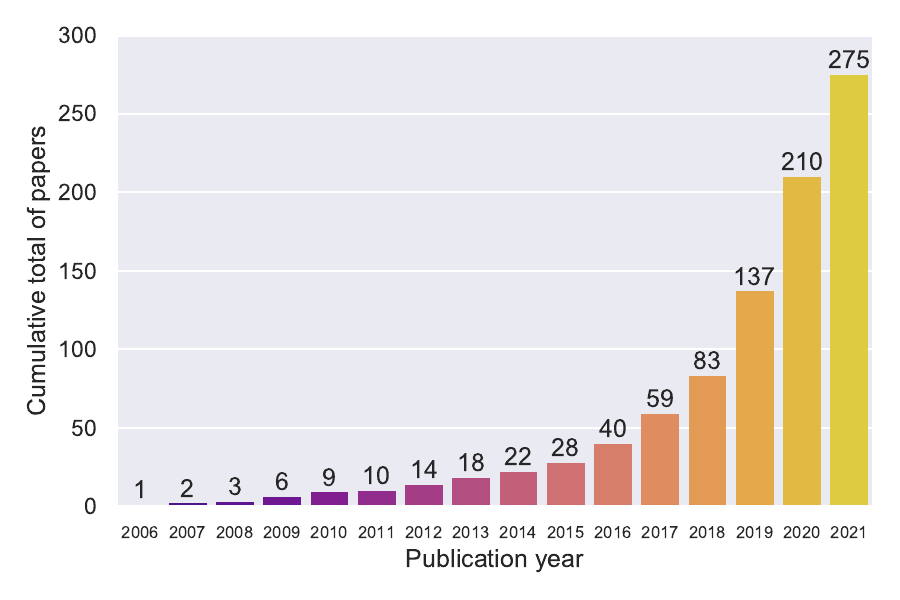}
\vspace{-1.5em}
\caption{Cumulative number of research papers published in the area of improving Wi-Fi performance using \ac{ML} and cited in this survey.
}
\label{fig_survey}
\end{figure}

Recently, \textcite{kulin2020survey} published a survey on applying \ac{ML} for general wireless networking while \textcite{zhang2019deep} reviewed almost 600 research papers on ML in 5G systems.
However, neither these nor other recent surveys (reviewed in \cref{sec:soa}) describe in detail the Wi-Fi performance improvement with \ac{ML} from different angles.
Wi-Fi is simply too complicated to be covered inside a general survey and  requires a dedicated one.

\subsection{Methodology}
\label{sec:methodology}

For the presented survey on improving Wi-Fi performance with machine learning, we started with a systematic literature review methodology~\cite{keele2007guidelines}. 
First, we searched for Wi-Fi, 802.11, and WLAN as well as machine learning in the paper abstracts in the following databases: IEEE Xplore, ACM, Elsevier, Wiley, and MDPI\footnote{We could not include SpringerLink at this stage as it does not allow to search within the abstracts of published papers. Papers from this database were added manually.}.
This yielded 1189 papers, out of which we had to remove out-of-scope papers.
Next, we added papers manually, usually found through cross-citation analysis. 
Finally, we identified (and cite in Sections~\ref{sec_core}--\ref{sec_mhop}) over 250 relevant papers in total.
Additionally, we reference over 20 survey papers in \Cref{sec:soa}.

\subsection{Survey Scope and Contributions}
The structure of the survey is depicted in \Cref{fig_MLWIFI} together with an indication of the \ac{ML} methods reported in the state-of-the-art papers, for each of the surveyed areas.
After a short summary of related surveys (\cref{sec:soa}), we first investigate core Wi-Fi features in \cref{sec_core}.
This section explores aspects such as the use of \ac{ML} for selecting PHY features, optimizing channel access, configuring frame aggregation and link parameter settings, data rate selection, as well as \ac{QoS}, admission control, and traffic classification.
In \cref{sec_advanced}, we study the benefits of using \ac{ML} to support more recent Wi-Fi features, such as channel bonding, \ac{MU-MIMO}, spatial reuse, and multi-band operation.
Wi-Fi connectivity management is discussed in \cref{sec_management}.
Here, we explore the applicability of \ac{ML} to \ac{AP} selection and association, channel and band selection, management architectures, and determining the health of Wi-Fi connections.
In \cref{sec_coexistence}, we investigate \ac{ML}-optimized coexistence of Wi-Fi with other technologies: channel sharing, network monitoring, and cross-technology signal classification.
Next, in \cref{sec_mhop}, we study \ac{ML} algorithms for multi-hop Wi-Fi deployments: ad hoc networks, mesh networks, sensor networks, vehicular networks, and relay networks.
Finally, we elaborate on future research directions in \cref{Sec:Future} and conclude the paper with \cref{Sec:Conclusion}.
Appendix~\ref{app_acronyms} contains the list of acronyms used.

\begin{figure*}
\centering
\begin{tikzpicture}

\definecolor{viridis1}{RGB}{249, 203, 53}
\definecolor{viridis3}{RGB}{173, 48, 93}
\definecolor{viridis4}{RGB}{109, 24, 110}
\definecolor{viridis5}{RGB}{68, 57, 131}

  \path[small mindmap,concept,text=black,concept color=viridis1,scale=0.8]
    node[concept] {\textbf{Improving Wi-Fi performance with ML}} [clockwise from=-85]
    child[concept color=viridis1,text=black]{
      node[concept]at(-2,-2.5){Core Wi-Fi features (Section \ref{sec_core})}[clockwise from=-145]
	child[concept color=viridis1,text=black,grow=-160]{
        node[concept,scale=1.1]{Channel access}[clockwise from = -145]
            child [concept color=viridis4,text=white]{node[concept, scale=0.85] {\textbf{RL/DRL}: QL, DQL, PDS, DQN, DDPG, multi-agent RL}}
            child [concept color=viridis3,text=white]{node[concept, scale=0.85] {\textbf{SL}: RF, Fixed-Share, DT}}
      }
      child[concept color=viridis1,text=black,grow=-110]{
        node[concept,scale=1.1]{Link configuration}[clockwise from =-110]
            child [concept color=viridis3,text=white]{node[concept, scale=0.85] {\textbf{SL/DSL}: ANN, RF, MLP, DNN}}
            child [concept color=viridis4,text=white]{node[concept, scale=0.85] {\textbf{RL/DRL}: SLA, QL, MAB, DQN, particle filter, SARSA}}
      }
      child[concept color=viridis1,text=black,grow=-15]{
        node[concept,scale=1.1]{Frame aggregation}[clockwise from = -15]
            child [concept color=viridis3,text=white]{node[concept, scale=0.85] {\textbf{SL}: RFR, RF, M5P, SVM, ANN}}
            child [concept color=viridis4,text=white]{node[concept, scale=0.85] {\textbf{RL}: $\varepsilon$-greedy policy}}
      }
      child[concept color=viridis1,text=black,grow=-65]{
        node[concept,scale=1.1]{PHY features}[clockwise from = -65]
            child [concept color=viridis3,text=white]{node[concept, scale=0.85] {\textbf{SL/DSL}: NB, NBT, J48DT, SVM, kNN, ANN, RNN, CNN, RT, GBRT, SVR}}
            child [concept color=viridis5,text=white] {node[concept, scale=0.85] {\textbf{USL}: EMA}}
      }
    }
    child[concept color=viridis1,text=black, grow=180]{
      node[concept]at(-2.5,0){Recent Wi-Fi features (Section \ref{sec_advanced})}[clockwise from=110]
      child[concept color=viridis1,text=black]{
        node[concept,scale=1.1]{MU communication}[counterclockwise from = 110]
            child  [concept color=viridis3,text=white]{node[concept, scale=0.85] {\textbf{SL/DSL}: SVM, DNN}}
            child  [concept color=viridis4,text=white]{node[concept, scale=0.85] {\textbf{RL/DRL}: MAB, QL, DQN}}
      }
      child[concept color=viridis1,text=black,grow=155]{
        node[concept,scale=1.1]{Spatial reuse}[counterclockwise from = 150]
            child [concept color=viridis3,text=white]{node[concept, scale=0.85] {\textbf{SL}: MLP}}
            child [concept color=viridis4,text=white]{node[concept, scale=0.85] {\textbf{RL}: MAB, QL}}
      }
      child[concept color=viridis1,text=black,grow=200]{
        node[concept,scale=1.1]{Channel bonding}[counterclockwise from = 185]
            child [concept color=viridis3,text=white]{node[concept, scale=0.85] {\textbf{SL}: NN, GNN}}
            child [concept color=viridis4,text=white]{node[concept, scale=0.85] {\textbf{RL/DRL}: MAB, SARSA, DQN}}
      }
      child[concept color=viridis1,text=black,grow=250]{
        node[concept,scale=1.1]{Multi-band, network MIMO, full-duplex}[counterclockwise from =240]
            child [concept color=viridis3,text=white]{node[concept, scale=0.85] {\textbf{DSL}: DNN}}
            child [concept color=viridis4,text=white]{node[concept, scale=0.85] {\textbf{DRL}: Monte-Carlo/DDPG}}
      }
       child[concept color=viridis1,text=black,grow=300]{
        node[concept,scale=1.1]{Beamforming}[counterclockwise from =280]
            child [concept color=viridis3,text=white]{node[concept, scale=0.85] {\textbf{SL/DSL}: CNN, RF, LSTM, DNN}}
            child [concept color=viridis4,text=white]{node[concept, scale=0.85] {\textbf{RL}: QL}}
      }
}
    child[concept color=viridis1,text=black, grow = -15] {
      node[concept]at(2.5,-2.5){Connectivity management (Section \ref{sec_management})}[clockwise from=95]
        child[concept color=viridis1,text=black]{
            node[concept,scale=1.1]{Channel and band selection}[clockwise from = 95]
                child [concept color=viridis4,text=white]{node[concept, scale=0.85] {\textbf{RL/DRL}: MAB, GCN, DDQN, DQL}}
                child [concept color=viridis3,text=white]{node[concept, scale=0.85] {\textbf{SL/DSL}:  LASSO, OLS, MFNN, RF, NN, LSTM}}
        }
        child[concept color=viridis1,text=black,grow=25]{
        node[concept,scale=1.1]{Traffic prediction}[clockwise from = 45]
            child [concept color=viridis3,text=white] {node[concept, scale=0.85] {\textbf{SL/DSL}: SVM, SVR, DT, DF, DRNN}}
            child [concept color=viridis5,text=white] {node[concept, scale=0.85] {\textbf{USL}: EMA}}
      }
        child[concept color=viridis1,text=black, grow=-35]{
            node[concept,scale=1.1]{Management architectures}[clockwise from =-30]
                child [concept color=viridis4,text=white]{node[concept, scale=0.85] {\textbf{DRL}: DQL}}
                child [concept color=viridis3,text=white]{node[concept, scale=0.85] {\textbf{SL}: RF}}
        }
        child[concept color=viridis1,text=black, grow=-85]{
            node[concept,scale=1.1]{Health of Wi-Fi connections}[clockwise from =-80]
                child [concept color=viridis3,text=white]{node[concept, scale=0.85] {\textbf{SL}: RF, DT, kNN, SVM, NB, NBT, MLP, REPT, ANN}}
                child [concept color=viridis5,text=white]{node[concept, scale=0.85] {\textbf{USL}: SOHMMM}}
        }
    }
    child[concept color=viridis1,text=black, grow =85] {
      node[concept]at(-1.5,2.5){Multi-hop networks (Section \ref{sec_mhop})}[clockwise from=0]
        child[concept color=viridis1,text=black, grow =60]{
            node[concept,scale=1.1]{Ad hoc networks}[clockwise from =60]
                child [concept color=viridis4,text=white]{node[concept, scale=0.85] {\textbf{RL}: QL, RL}}
        }
        child[concept color=viridis1,text=black, grow=120]{
            node[concept,scale=1.1]{Mesh networks}[clockwise from =140]
                child [concept color=viridis3,text=white]{node[concept, scale=0.85] {\textbf{SL}: SVM, NN}}
                child [concept color=viridis4,text=white]{node[concept, scale=0.85] {\textbf{RL}: QL}}
                child [concept color=viridis5,text=white]{node[concept, scale=0.85] {\textbf{USL}: k-means clustering}}
        }
        child[concept color=viridis1,text=black, grow=170]{
            node[concept,scale=1.1]{Sensor networks}[clockwise from =170]
                child [concept color=viridis3,text=white]{node[concept, scale=0.85] {\textbf{SL/DSL}: CT, DNN, SVM}}
        }
        child[concept color=viridis1,text=black, grow=235]{
            node[concept,scale=1.1]{Vehicular networks}[clockwise from =245]
                child [concept color=viridis3,text=white]{node[concept, scale=0.85] {\textbf{SL}: SVM, RF, LR}}
                child [concept color=viridis4,text=white]{node[concept, scale=0.85] {\textbf{RL, DRL}}}
        }
    }
    child[concept color=viridis1,text=black, grow =55] {
      node[concept]at(3,1){Coexistence with other technologies (Section \ref{sec_coexistence})}[clockwise from=120]
        child[concept color=viridis1,text=black]{
            node[concept,scale=1.1]{Fair channel sharing}[counterclockwise from = 110]
                child [concept color=viridis3,text=white]{node[concept, scale=0.85] {\textbf{SL/DSL}: LR, LSTM}}
                child [concept color=viridis4,text=white]{node[concept, scale=0.85] {\textbf{RL/DRL}: QL, MAB, DQN, MDP, DQL, TRPO, DDPG}}
        }
        child[concept color=viridis1,text=black, grow = 60]{
            node[concept,scale=1.1]{Network monitoring}[clockwise from =80]
                child [concept color=viridis3,text=white]{node[concept, scale=0.85] {\textbf{SL/DSL}: RF, DNN, CNN, LSTM, NNMR}}
                child [concept color=viridis4,text=white]{node[concept, scale=0.85] {\textbf{RL}: QL, fuzzy QL, double QL}}
                child [concept color=viridis5,text=white]{node[concept, scale=0.85] {\textbf{USL/DUSL}: NN, DCC}}
        }
        child[concept color=viridis1,text=black, grow=0]{
            node[concept,scale=1.1]{Signal classification}[clockwise from = 30]
                child [concept color=viridis3,text=white]{node[concept, scale=0.85] {\textbf{SL/DSL}: NN, LR, CNN, RNN, logistic regression}}
                child [concept color=viridis5,text=white]{node[concept, scale=0.85] {\textbf{USL}: K-means clustering}}
        }
        child[concept color=viridis1,text=black, grow=180]{
            node[concept,scale=1.1]{Cooperation}[clockwise from =180]
                child [concept color=viridis4,text=white]{node[concept, scale=0.85] {\textbf{RL}: TRPO, QL}}
                child [concept color=viridis3,text=white]{node[concept, scale=0.85] {\textbf{SL}: NN}}
        }
    }
   ;
  \matrix [below right] at (current bounding box.north west) {
  \node [shape=circle, draw=viridis3, fill=viridis3, line width=2,label=right:Supervised learning] {}; \\
  \node [shape=circle, draw=viridis4, fill=viridis4, line width=2,label=right:Reinforcement learning] {}; \\
  \node [shape=circle, draw=viridis5, fill=viridis5, line width=2,label=right:Unsupervised learning] {}; \\
};
\end{tikzpicture}
\caption{Organization of the survey and classification of research areas where Wi-Fi network performance is improved with \ac{ML} according to the literature.} %
 \label{fig_MLWIFI}
\vspace{-.5em}
\end{figure*}

To summarize, our contributions are the following:
\begin{itemize}
    \item \textit{A structured approach to describing the various areas of Wi-Fi performance where \ac{ML} is applied}. We elaborate on core Wi-Fi features, through recently added features to management issues as well as Wi-Fi operating in shared bands with other technologies and in multi-hop topologies.
    \item \textit{A review of over 250 papers in the field}. We provide readers with an overview of what has been done and what are the main trends of applying \ac{ML} to particular Wi-Fi performance problems.
    \item \textit{The identification of open challenges in every area of Wi-Fi performance}, given at the end of each Section~\ref{sec_core}--\ref{sec_mhop}. Additionally, we provide an overview of the general future research directions in applying \ac{ML} for improving Wi-Fi performance, to provide readers with an analysis of what remains to be done in the field.
\end{itemize}
We hope that the survey will be beneficial both for beginners\footnote{We refer readers interested in the basics of Wi-Fi to \cite{gulasckaran2021wifi} and those wanting to understand the principles of \ac{ML} to \cite{goodfellow2016deep,sutton2018reinforcement}.} and experts in the field, looking for a comprehensive summary of the latest research in the area of improving Wi-Fi performance with \ac{ML}.
We also believe that this survey will guide readers towards proposing new ideas in this developing research area.

\section{Related Surveys}
\label{sec:soa}

Many surveys address the development of \ac{ML} models to support wireless networks. 
Reported contributions consider the application \ac{ML} to both \mbox{Wi-Fi} and application-specific networks, such as \acp{WSN}, \acp{CRN}, \acp{WMN}, and \acp{HetNet}.
\mbox{Wi-Fi} also constitutes an important component of \ac{5G} and the future \ac{6G}, e.g., in the case of cellular traffic offloading. 
Due to the convergence of both technologies, not only concerning their operation but also in the context of shared unlicensed bands, \ac{5G} and \ac{6G}-related surveys provide valuable insights also for \mbox{Wi-Fi} operation.
Therefore, in this survey, we also cover some aspects and functionalities from \ac{5G} and \ac{6G} that are directly related, or equivalent, to those in the Wi-Fi area.

In terms of the direct application of \ac{ML} models in \mbox{Wi-Fi}, state-of-the-art surveys mainly focus on performance indicators and the support of a variety of applications, e.g., human activity detection, indoor localization, and network security. 
The use of \ac{ML} in application-specific wireless networks focuses on challenging functionalities like self-configuration, self-healing, and self-optimization in \acp{HetNet}, bandwidth and coverage in \ac{WMN}, and dynamic spectrum access in \ac{CRN}.
In the area of 5G, the surveys mostly focus on interference identification, link quality prediction, traffic demand estimation, and network management.
Additionally, they address the problem of unlicensed spectrum sharing between \ac{5G} and incumbent technologies, like Wi-Fi. 
All these surveys partially overlap with our literature review. However, since none of them focuses exclusively on Wi-Fi, neither the level of detail, content organization, nor the number of works covered are comparable with our survey.

\subsection{Wi-Fi-related Surveys}
Surveys of \mbox{Wi-Fi} performance-indicators cover mostly \mbox{Wi-Fi} data analytics for network monitoring~\cite{medeiros2020survey} and quality indicators accounting for user satisfaction~\cite{morshedi2020survey}.
For \mbox{Wi-Fi} analytics, the reported \ac{ML} models extract useful knowledge from big data streams produced over large-scale wireless networks \cite{medeiros2020survey}.
Additionally, \ac{ML}-based solutions to support the estimation of \ac{QoS}, \ac{QoE}, and their cross-correlation (\ac{QoS}-\ac{QoE}) as surveyed  by \textcite{morshedi2020survey}. 
For Wi-Fi-based applications, indoor localization~\cite{kunhoth2020indoor,sattarian2019indoor,liu2019survey,oguntala2018indoor} and human activity detection \cite{yousefi2017survey} are the two main covered areas.
\ac{ML}-based techniques are illustrated to detect, recognize, and categorize complex patterns to support these applications.

Security in \mbox{Wi-Fi} is also a relevant concern addressed in surveys~\cite{nivaashini2021computational,kolias2016intrusion}.
Since that \mbox{Wi-Fi} is ranked as the most deployed wireless technology, numerous attacks exploiting its vulnerabilities are observed.
In this direction, \ac{ML} models improve the autonomy and accuracy of \acp{IDS} for \mbox{Wi-Fi} networks.

\subsection{Wireless Communications-related Surveys}
\textcite{wang2020thirty} present an  interesting survey, in which the thirty-year history of \ac{ML} is reviewed. 
The survey addresses the fundamentals of \ac{SL}, \ac{USL}, and \acl{RL} (RL). 
Additionally, it summarizes the use of \ac{ML} in many compelling applications of wireless networks, e.g., \acp{HetNet}, \ac{CR}, \ac{IoT}, and \ac{M2M} communications. 
However, use of \ac{ML} in IEEE 802.11 networks is only briefly mentioned. 
Additionally, the use of \ac{ML} models for layer-specific operation is not covered.

The applications of \ac{ML} supporting the \ac{PHY}, \ac{MAC}, and network layers are also reported for wireless communication networks by \textcite{ahmad2020machine}.
Novel computing/networking concepts are also addressed like \ac{MEC} \cite{taleb2017multiaccess}, \ac{SDN} \cite{chaparadza2013sdn}, and \ac{NFV} \cite{gilherrera2016resource}. 
\ac{ML} and \ac{RL} models are also illustrated for several networks types such as \ac{5G}, \acp{LPWAN}, \acp{MANET}, and \ac{LTE} networks. 
Additionally, the survey provides a brief overview of \ac{ML}-based network security. 
However, once again, the area of 802.11 networks is only briefly touched upon.

\textcite{sun2019application} survey a variety of applications of \ac{ML} models for resource management at the \ac{MAC} layer, networking and mobility management in the network layer, and localization in the application layer.
Conditions for applying \ac{ML} models to assist developers in wireless communication systems are also identified.
The utility of \ac{ML} techniques in \mbox{Wi-Fi} scenarios is illustrated to implement power-saving mechanisms in \acp{AP} and indoor localization applications.

Several other surveys address the use of \ac{ML} models in specific wireless networks like \acp{WSN} \cite{alsheikh2014machine}, \acp{CRN} \cite{bkassiny2013survey,gavrilovska2013learning,he2010survey}, \acp{MANET} \cite{forster2007machine}, and IoT \cite{liu2021machine, ejaz2020learning}.
These surveys summarize the application of \ac{ML} models to specific-related problems on these networks like prolonged lifespan in \acp{WSN}, feature classification in \acp{CRN}, or routing in \acp{MANET}.
Only the surveys concerning \acp{CRN} discuss the applications of \ac{ML} models in \mbox{Wi-Fi} networks (e.g., coexistence, performance evaluation, channel selection, signal identification). However, specific details concerning the integration of \ac{ML} techniques and \mbox{Wi-Fi} mechanisms are only superficially covered.

Additionally, \textcite{nguyen2021transfer} provide a comprehensive survey on transfer learning for future wireless networks and \textcite{zhang2021survey} provide a general survey on federated learning, with a reference to wireless communication. Transfer learning is applied to address the problems observed for the conventional ML approaches (e.g., lack of labeled data, long-lasting training vs. varying wireless conditions, limited capacity of wireless devices vs. ML requirements) \cite{nguyen2021transfer} while federated learning is applied to ensure privacy \cite{zhang2021survey}.

\subsection{Wi-Fi and \ac{5G}-related Surveys}

In the \ac{5G} area, surveys focus mostly on the \ac{PHY}, \ac{MAC}, and network layers to account for interference identification, link quality prediction, and traffic demand estimation \cite{kulin2020survey}.  
Through \ac{ML}, patterns are automatically extracted and trends are predicted to optimize parameter settings at different protocol layers.
In this area, a variety of effective solutions are also used to: 
\begin{itemize}
    \item analyze and manage mobile networks in several directions, e.g., network state prediction, network traffic classification, call details mining, and radio-signal analysis~\cite{moysen20184g};
    \item improve the performance of mobile systems~\cite{zhang2019deep} and \ac{IoT}~\cite{luong2019applications};
    \item identify wireless modulations/technologies \cite{wu2019deep};
    \item provide fair and efficient spectrum sharing in \ac{5G} \cite{mamadou2020survey} and in future \ac{6G}~\cite{gur2020expansive} networks;
    \item maximize the potential of unlicensed bands for Industry 4.0 applications \cite{bajracharya2020future}.
\end{itemize}

\begin{table*}
  \centering
  \caption{Existing surveys concerning Wi-Fi-related topics and \ac{ML} models.}
    \begin{tabular}{clp{22.275em}p{24.82em}c}
    \toprule
    \multicolumn{1}{l}{\textbf{Network}} & \textbf{Ref.} & \multicolumn{1}{l}{\textbf{Main scope}} & {Addressed} Wi-Fi feature & \textbf{Year} \\
    \midrule
    \multirow{9}[18]{*}{\begin{sideways}Wi-Fi\end{sideways}} & \cite{medeiros2020survey} & Large-scale network monitoring & \mbox{Wi-Fi} analytics & 2020 \\
\cmidrule{2-5}          & \cite{morshedi2020survey} & Quality indicators accounting for user satisfaction & \mbox{Wi-Fi} quality indicators & 2020 \\
\cmidrule{2-5}          & \cite{kunhoth2020indoor} & \multirow{4}[8]{*}{Indoor localization} & \multirow{5}[10]{*}{Application-oriented} & 2020 \\
\cmidrule{5-5}          & \cite{sattarian2019indoor} & \multicolumn{1}{l}{} & \multicolumn{1}{l}{} & 2019 \\
\cmidrule{5-5}          & \cite{liu2019survey} & \multicolumn{1}{l}{} & \multicolumn{1}{l}{} & 2019 \\
\cmidrule{5-5}          & \cite{oguntala2018indoor} & \multicolumn{1}{l}{} & \multicolumn{1}{l}{} & 2018 \\
\cmidrule{2-3}\cmidrule{5-5}          & \cite{yousefi2017survey} & Human activity detection & \multicolumn{1}{l}{} & 2017 \\
\cmidrule{2-5}          & \cite{nivaashini2021computational} & \multirow{2}[2]{*}{Intrusion detection} & \multirow{2}[2]{*}{\mbox{Wi-Fi} security} & 2021 \\
\cmidrule{2-2}\cmidrule{5-5}          & \cite{kolias2016intrusion} & \multicolumn{1}{l}{} & \multicolumn{1}{l}{} & 2016 \\
    \midrule
    \multicolumn{1}{c}{\multirow{20}[12]{*}{\begin{sideways}\mycell{Wireless networks \\ (\ac{IoT}, \ac{CRN}, \ac{M2M}, \ac{MANET})}\end{sideways}}} 
     & \cite{liu2021machine} & Detection and identification of IoT devices & Identification of devices and security protection & 2021 \\
     \cmidrule{2-5}          & \cite{zhang2021survey} & Federated learning & Privacy protection & 2021 \\
     \cmidrule{2-5} & \cite{nguyen2021transfer} & Applications of transfer learning in wireless networks &  \multirow{8}[4]{*}{Insufficient details concerning \mbox{Wi-Fi} functionalities} & 2021 \\
    \cmidrule{2-3}\cmidrule{5-5}  & \cite{wang2020thirty} & Performance improvement in a variety of wireless networks like \acp{HetNet}, \acp{CRN}, \acp{IoT}, and \ac{M2M} & \multicolumn{1}{l}{}
     & 2020 \\
    \cmidrule{2-3}\cmidrule{5-5}          & \cite{ahmad2020machine} & Performance improvement in the \ac{PHY}/\ac{MAC}/Network layers as well as novel networking concepts (\ac{MEC}, \ac{SDN}, \ac{NFV})  & \multicolumn{1}{l}{} & 2020 \\
    \cmidrule{2-3}\cmidrule{5-5}  & \cite{ejaz2020learning} & Optimization of communication and computing technologies of IoT systems & \multicolumn{1}{l}{} & 2020 \\
\cmidrule{2-5}          & \cite{sun2019application} & \ac{ML} models to support resource management, networking and localization in wireless networks & Power saving mechanisms for \mbox{Wi-Fi} infrastructure, indoor localization mechanisms & 2019 \\
\cmidrule{2-5}          & \cite{bkassiny2013survey} & Decision making and feature classification in \acp{CRN} & {Collaborative} coexistence of \mbox{Wi-Fi} networks {with other technologies}, performance evaluation, dynamic channel selection & 2013 \\
\cmidrule{2-5}          & \cite{gavrilovska2013learning} & \ac{ML} models to support cognitive radio capabilities & {Collaborative} coexistence of Wi-Fi networks {with other technologies} & 2013 \\
\cmidrule{2-5}          & \cite{he2010survey} & \ac{ML} models to support cognitive radio capabilities & Wi-Fi signal identification & 2010 \\
    \midrule
    \multicolumn{1}{c}{\multirow{8}[10]{*}{\begin{sideways}Wi-Fi and \ac{5G}/\ac{6G}\end{sideways}}} & \cite{kulin2020survey} & Broad survey covering data science fundamentals, \ac{5G}, \mbox{Wi-Fi}, \ac{CRN} General networking concepts like interference recognition, network traffic predictions, and \ac{MAC} identification  & \multicolumn{1}{l}{Insufficient details concerning \mbox{Wi-Fi} functionalities} & 2020 \\
\cmidrule{2-5}          & \cite{mamadou2020survey} & \multirow{2}[2]{*}{Coexistence mechanisms in \ac{5G} networks} & \multirow{2}[2]{*}{Coexistence of \ac{5G} and \mbox{Wi-Fi}} & 2020 \\
\cmidrule{2-2}\cmidrule{5-5}          & \cite{bayhan2018future} & \multicolumn{1}{l}{} & \multicolumn{1}{l}{} & 2018 \\
\cmidrule{2-5}          & \cite{zhang2019deep} & Mobile and wireless networking research based on deep learning & Indoor localization applications and signal processing in \mbox{Wi-Fi} networks & 2019 \\
    \bottomrule
    \end{tabular}%
  \label{tab:addlabel}%

  \label{tab:comparison}%
\end{table*}%

\subsection{Summary}
\label{sec:soa-summary}

State-of-the-art surveys report the wide applicability of \ac{ML} models for wireless networks.
\Cref{tab:comparison} summarizes the presented surveys per addressed technology, scope, and remarking their corresponding \mbox{Wi-Fi}-related topics.

Specifically, in the \textit{\mbox{Wi-Fi} area}, the reported surveys are application-oriented, focusing on human activity detection algorithms, indoor localization mechanisms, and network security issues.
In the \textit{wireless networks area}, the surveys, in general, provide few details concerning the use of \ac{ML} models to improve the performance of the 802.11 protocol family.
The most often surveyed topics are the coexistence of \mbox{Wi-Fi} networks with other technologies, its performance evaluation, channel selection mechanism, and signal identification in the context of cognitive radio technologies.
Finally, concerning the \textit{\ac{5G} and \mbox{Wi-Fi} area}, surveys mostly cover the concept of spectrum sharing mechanisms for coexistence between the two networks.
Therefore, the lack of a dedicated Wi-Fi performance survey coupled with the variety of research papers addressing the specifics of using Wi-Fi with \ac{ML} (\Cref{fig_survey}) have motivated our work, which we hope will be valuable to the research community.

Note that there are three non-performance related areas involving both Wi-Fi and \ac{ML} which are out of the scope of this survey:
 \begin{itemize}
     \item dedicated applications of Wi-Fi (unrelated to network access), e.g., device positioning, human activity detection, 
     \item energy efficiency (e.g., power-saving protocols), and
     \item network security (e.g., detecting selfishly configured devices \cite{szott2011detecting}).
 \end{itemize}
 There has been broad adoption of \ac{ML} in these areas and they deserve literature reviews of their own, such as~\cite{kunhoth2020indoor,nivaashini2021computational}.
 Furthermore, our survey does not describe how various \ac{ML} methods operate.
 There are numerous books and research papers on this topic; we refer the reader to papers such as~\cite{kulin2020survey, wang2020thirty} for a detailed (although still wireless networking-related) discussion of these methods.

\onecolumn
\begin{landscape}
\scriptsize
\begin{xltabular}{\linewidth}{ccccccp{0.19\textwidth}*{2}{p{0.29\textwidth}}}
\caption{Summary of works on improving the performance of core Wi-Fi features with ML. The evaluation methods are theoretical (E), simulation (S), and experimental (E). The ML improvement is in comparison to \ac{SoA} methods or (if not mentioned otherwise) to IEEE 802.11.} \label{tab:core} \\
\toprule
\textbf{Area} & \textbf{Ref.} & \shortstack{\textbf{ML}\\\textbf{category}} & \shortstack{\textbf{ML}\\\textbf{mechanisms}} & \textbf{Year}&\shortstack{\textbf{Evaluation}\\\textbf{method}}  & \textbf{Application of ML} & \textbf{Novelty of approach} & \textbf{ML improvement} \\
\midrule
\endfirsthead

\multicolumn{9}{c}%
{\tablename\ \thetable{} -- continued from previous page} \\
\midrule \textbf{Area} & \textbf{Ref.} & \shortstack{\textbf{ML}\\\textbf{category}} & \shortstack{\textbf{ML}\\\textbf{mechanisms}} & \textbf{Year}&\shortstack{\textbf{Evaluation}\\\textbf{method}}  & \textbf{Application of ML} & \textbf{Novelty of approach} & \textbf{ML improvement} \\ \midrule
\endhead

\multicolumn{9}{r}{{Continued on next page}} \\
\endfoot

\endlastfoot

\multicolumn{1}{c}{\multirow{15}[0]{*}{\shortstack{Channel access \\ (Section \ref{sec_channel_access})}}} & \cite{zhu2012achieving} & RL    & QL    & 2012  & S     & Select CW update rule & Apply ABP framework for configuring DCF & Better QoS metrics for voice/video flows \\
\cline{2-9}
& \cite{amuru2015send} & RL    & PDS   & 2015  & T     & Select backoff value & Apply PDS for configuring DCF & Higher throughput, faster convergence than QL \\
\cline{2-9}
& \cite{abyaneh2019intelligent} & SL    & RF    & 2019  & S     & Select minimum CW value & Improve fairness, robust to selfish stations & Higher throughput and fairness, lower latency \\
\cline{2-9}
& \cite{ali2019deep} & RL    & QL    & 2019  & S     & Select CW value & Apply Q-learning in dense network scenario & Higher throughput \\
\cline{2-9}
& \cite{edalat2019dynamically} & SL    & fixed-share & 2019  & S     & Select CW value & Apply a fixed-share algorithm for configuring DCF & Higher throughput and fairness, lower latency \\
\cline{2-9}
& \cite{lee2020collision} & RL    & QL    & 2020  & S     & Select time slot for transmission & Stations self-organize into slot-based channel access & Higher throughput and lower latency \\
\cline{2-9}
& \cite{coronado2020improvements} & SL    & DT    & 2020  & S     & Set AIFS and CW values & Consider QoS requirements & Higher throughput for voice/video flows \\
\cline{2-9}
& \cite{ali2020performance} & RL    & QL    & 2020  & S     & Select CW values & Consider QoS requirements & Higher throughput \\
\cline{2-9}
& \cite{kihira2020adversarial} & RL    & QL    & 2020  & S     & Select time slot for transmission & Consider interference from non-ML based devices & Higher throughput than in cooperative setting \\
\cline{2-9}
& \cite{wydmanski2021contention} & RL    & DQN, DDPG & 2021  & S     & Select CW value & Apply two DRL variants for configuring DCF & Higher throughput, close to optimal \\
\cline{2-9}
& \cite{kumar2021adaptive} & RL    & QL    & 2021  & S     & Select minimum CW value & Apply DQN with rainbow agent for configuring DCF & Higher fairness, close to optimal \\
\cline{2-9}
& \cite{ali2021federated} & RL    & DQL & 2021  & S     & Select CW value & Apply FL for configuring DCF & Higher throughput than using only RL \\
\cline{2-9}
& \cite{zhang2020enhancing} & RL    & DQL, QNN & 2021  & S     & Select time slot for transmission & Apply FL for configuring slotted transmissions & Higher throughput \\
\cline{2-9}
& \cite{guo2022multi} & RL    & multi-agent RL & 2022  & S     & Select time slot for transmission & Apply multi-agent RL for random channel access scheme & Higher throughput and lower latency \\
\midrule
\multicolumn{1}{c}{\multirow{17}[0]{*}{\shortstack{Link adaptation,\\data rate selection \\ (Section \ref{sec_link})}}} & \cite{joshi2008sara} & RL    & SLA & 2008  & S+T   & Select transmission rate & Apply iterative learning for rate selection & Higher throughput than three SoA methods \\
\cline{2-9}
& \cite{punal2013rfra} & SL    & RF    & 2013  & S     & Select transmission rate & Apply the random forests method for rate selection & Higher throughput than three SoA methods \\
\cline{2-9}
& \cite{wang2013dynamic} & SL    & ANN, MLP & 2013  & S     & Select transmission rate & Use number of stations, channel conditions, and traffic intensity as input & Higher throughput than two SoA methods \\
\cline{2-9}
& \cite{karmakar2016dynamic} & RL    & MAB   & 2016  & S     & Configure link parameters & Apply MAB for link adaptation & Higher throughput, lower packet loss and delay than three SoA methods \\
\cline{2-9}
& \cite{kurniawan2019machine} & SL    & RF    & 2018  & S     & Classify channel type & Apply SL for channel classification & Higher spectral efficiency \\
\cline{2-9}
& \cite{li2020practical} & SL    & ANN   & 2020  & E     & Select transmission rate & Provide extensible rate selection framework & Higher throughput than three SoA methods \\
\cline{2-9}
& \cite{karmakar2020deep} & SL    & DNN   & 2020  & E     & Predict link-layer throughput & Apply SL for link adaptation & Higher throughput, lower packet loss and delay than three SoA methods \\
\cline{2-9}
& \cite{karmakar2020s2} & RL    & TS    & 2020  & S     & Select guard interval & Apply TS for guard interval selection & Higher throughput, lower packet loss and delay vs static settings \\
\cmidrule{2-9}          & \cite{peserico2020rate} & RL    & SARSA & 2020  & S     & Select transmission rate & Apply RL for rate selection in industrial settings & Higher throughput, lower delay than SOA method \\
\cmidrule{2-9}          & \cite{krotov2020rate} & RL    & particle filter & 2020  & S     & Select transmission rate & Apply RL for 802.11ax rate selection & Higher throughput, lower delay than two SOA methods \\
\cline{2-9}
& \cite{cho2021reinforcement} & RL    & QL    & 2021  & S     & Select transmission rate & Use packet timeouts to train RL model & Higher throughput than a SoA method \\
\cline{2-9}
& \cite{chen2021experience} & RL    & DQN   & 2021  & E     & Select transmission rate & Apply DRL for rate selection & Higher throughput than two SoA methods \\
\midrule
\multicolumn{1}{c}{\multirow{7}[0]{*}{\shortstack{Frame format,\\packet aggregation \\ (Section \ref{sec_frame})}}} & \cite{lin2009machine} & SL    & ANN   & 2009  & S+T   & Select frame size and CW & Apply ML for frame-size optimization & Higher throughput \\
\cline{2-9}
& \cite{karmakar2019online} & RL    & $\varepsilon$-greedy & 2019  & S     & Select frame size & Consider energy-consumption constraint & Better performance than three SoA methods \\
\cline{2-9}
& \cite{coronado2020adaptive} & SL    & RF    & 2020  & S     & Select frame size & Apply SL for frame-size optimization & Higher, more stable throughput \\
\cline{2-9}
& \cite{coronado2020aios} & SL    & M5P, RF & 2020  & E     & Select frame size & Apply ML-based optimization in SDN framework & Higher throughput \\
\cline{2-9}
& \cite{khastoo2020neura} & SL    & ANN   & 2020  & E     & Select frame size, transmission rate & Joint frame-size and transmission-rate optimization & Higher throughput than six SoA methods \\
\cline{2-9}
& \cite{hassani2021quick} & SL    & SVM   & 2021  & S     & Estimate aggregation level & Regulate send rates for low latency communication & Reduced latency vs two SoA transport protocols \\
\midrule 
\multicolumn{9}{c}{\multirow{3}{*}} \\ \\ \\ 
\multicolumn{1}{c}{\multirow{19}[0]{*}{\shortstack{PHY features\\ (Section \ref{sec_phy})}}} & \cite{kashyap2010deconstructing} & USL   & EMA   & 2010  & E     & Estimate probability of deferral & Infer interference relations between stations & Higher accuracy than two benchmarks \\
\cmidrule{2-9}          & \cite{lee2012dfi} & SL    & NB, NBT, J48DT, SVM & 2012  & E     & Decode frames & Alternative 802.11 protocol based on OFDM channelization & Higher throughput and fairness \\
\cmidrule{2-9}          & \cite{lee2015frequency} & SL    & NB, NBT, J48DT, SVM & 2015  & E     & Decode frames & Extension of \cite{lee2012dfi} & Higher throughput and fairness \\
\cmidrule{2-9}          & \cite{herzen2015learning} & SL    & RT, GBRT, SVR & 2015  & E     & Predict link-layer throughput & Use SL to model impact of PHY/MAC interactions on throughput & Higher accuracy (GBRT, SVR) than benchmark \\
\cmidrule{2-9}          & \cite{kim2018identifying} & SL    & kNN   & 2018  & E     & Classify signal source & Apply ML to signal identification based on channel state information & High accuracy \\
\cmidrule{2-9}          & \cite{almazrouei2019deep, almazrouei2019using} & SL    & ANN   & 2019  & S     & De-noise signals & Apply DL to improve radio signal quality & High accuracy \\
\cmidrule{2-9}          & \cite{saha2019interference} & ML    & DPP   & 2019  & T     & Estimate interference level & Apply DPP learning to characterize interference\newline{}distribution & High accuracy \\
\cmidrule{2-9}          & \cite{herath2019deep} & SL    & RNN   & 2019  & E     & Predict signal strength & Apply DL to predict radio signal quality & Higher accuracy than two benchmarks \\
\cmidrule{2-9}          & \cite{sankhe2020csiscan} & SL    & CNN   & 2020  & E     & Learn CSI & Design a CNN to determine the optimal selection of the maximum number of OFDM subcarriers and phase shifts under varying channel conditions. & Lower latency, reduced probe responses \\
\cmidrule{2-9}          & \cite{ninkovic2021deep} & SL    & CNN, DNN & 2021  & S+E   & Packet detection, carrier frequency offset estimation & Performance and complexity analysis of packet detection and CFO estimation. Investigation under which conditions the performance of the DL-based methods is superior/inferior to conventional methods. & Improved packet detection and CFO estimation \\

\bottomrule

\end{xltabular}
\end{landscape}

\twocolumn

\section{Core Wi-Fi Features}
\label{sec_core}

The new IEEE 802.11 amendments introduce augmented functionalities for ensuring robust network operation and improved user experience.
For instance, the IEEE 802.11 n/ac/ax amendments increase the data rate up to \SI{9}{\giga\bit\per\second} leveraging the increasing number of \acp{SS}, and techniques like channel bonding, multi-user transmissions, \ac{SGI}, and high modulations (up to 1024-QAM for 802.11ax) \cite{perahia2011gigabit,karmakar2017impact,karmakar2016dynamic}.
The impact of such a variety of parameters on the network performance is highly difficult to characterize considering the variability of Wi-Fi environments and the users' dynamics. 
However, the availability of performance metrics, both at the user and the \ac{AP} level, along with historical data provides a favorable environment for \ac{ML} methods to model the impact of such parameters on the network performance and optimize it.
The capability of \ac{ML}-based methods to gain knowledge, generalize, and learn from experience allows conceiving smart systems using the augmented functionalities of the IEEE 802.11 standard.

In the literature, there are many \ac{ML} solutions for 802.11's \ac{PHY} and \ac{MAC} layers to adaptively optimize the internal parameters of Wi-Fi's core features in dynamic scenarios.
As summarized in \Cref{tab:core}, contributions are reported in four areas to

\begin{itemize}
    \item reduce collisions when accessing the channel,
    \item maximize rate with the proper link configuration,
    \item find the optimum balance for the frame length with frame aggregation techniques,
    \item address interference and signal denoising at the \ac{PHY} layer.
\end{itemize}
These solutions mostly use \ac{RL} methods to adjust access parameters and \ac{SL} methods to estimate the channel condition for improved performance.

Although there is a variety of reported solutions, still more work taking into account the overall network performance is needed.
Besides, the application of \ac{ML} models is usually verified with simulations, with just a few research papers dealing with real scenarios that include, for instance, user mobility. 
In this section, we cover the core Wi-Fi features mentioned above and summarize the open challenges when conceiving \ac{ML} models in Wi-Fi environments.

\subsection{Channel Access}
\label{sec_channel_access}
\begin{figure}
\centering
\includegraphics[width=\columnwidth]{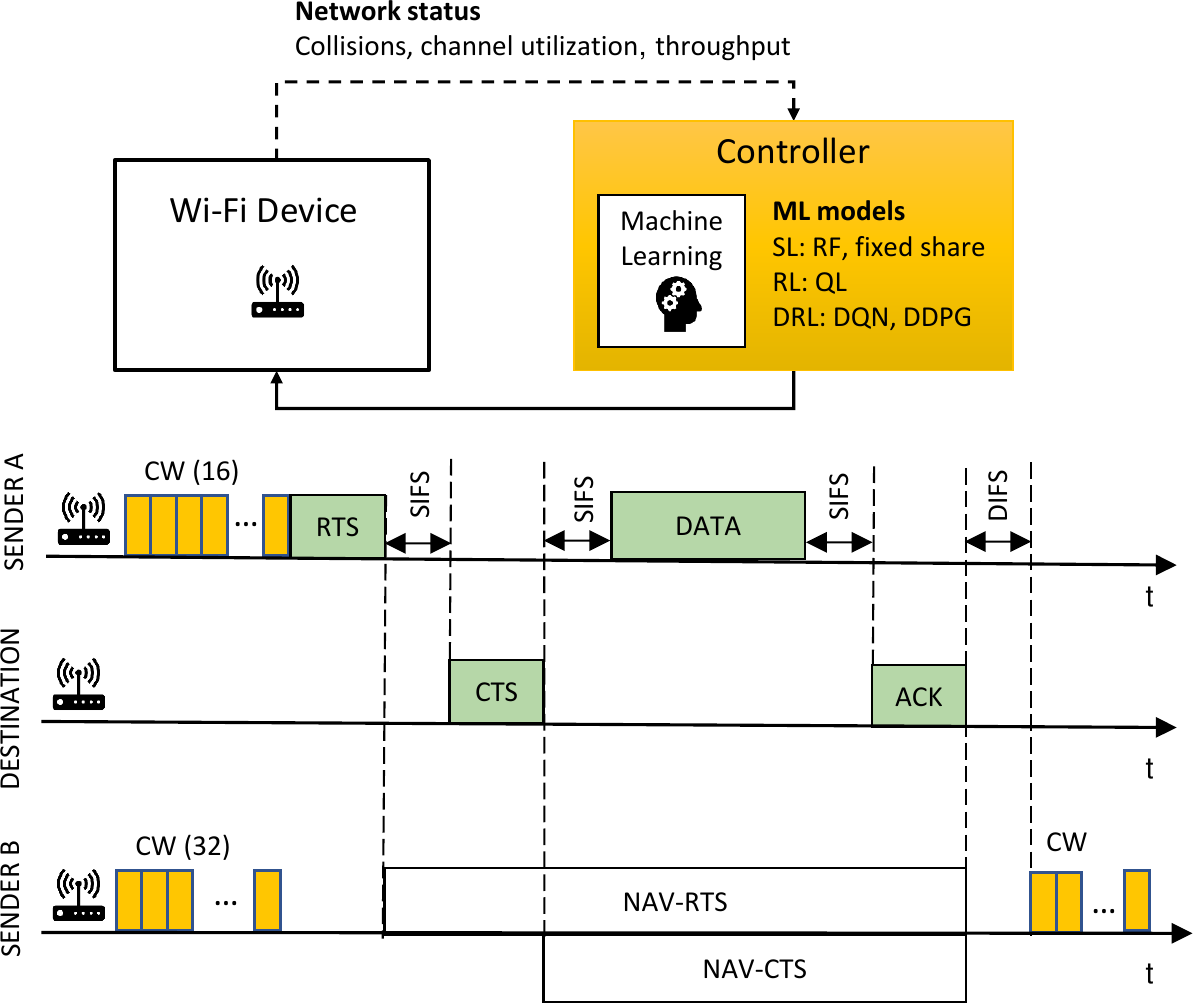}
\caption{Wi-Fi channel access (\ac{DCF}) supported by \ac{ML} models, in which network status observations are used to control the channel access settings, e.g., the contention window parameter. Dashed arrows represent observation and solid lines represent actions.}
 \label{fig_VA_1}
\end{figure}

Channel access mechanisms are perhaps the most often addressed topic concerning the improvement of Wi-Fi performance with \ac{ML}.
Proposed optimizations refer mostly to the basic 802.11 \ac{MAC} protocol, i.e., the \ac{DCF}, which is the baseline mechanism to avoid collisions among devices when accessing a common radio channel \cite{yun2012optimal}.
The main parameter responsible for the performance of \ac{DCF} is the \ac{CW}, which
defines the range from which stations randomly select their waiting periods (i.e., the backoff counter) to avoid collisions when accessing the channel.
Larger \ac{CW} values reduce collisions but increase idle times, which in turn reduces throughput.
Smaller CW values increase the chance for a station to transmit, but also increase the collision probability, thereby reducing throughput.

\begin{figure}[!t]
    \centering
    \subfloat[]{
        \includegraphics[height=5cm]{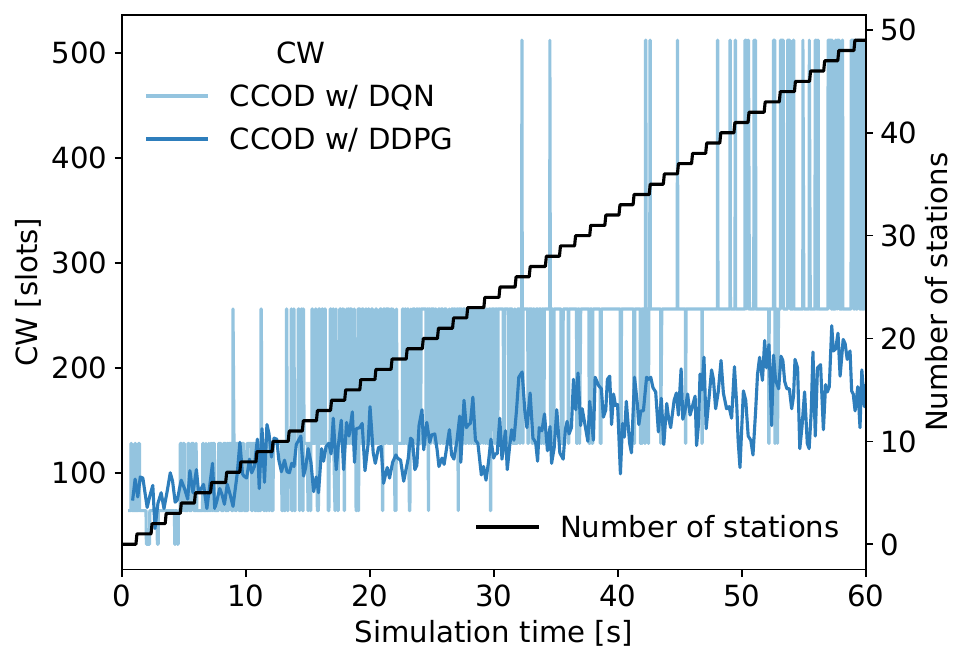}	
        \label{fig:ccod_cw_choice}
    }\\
    \subfloat[]{
        \includegraphics[height=5cm]{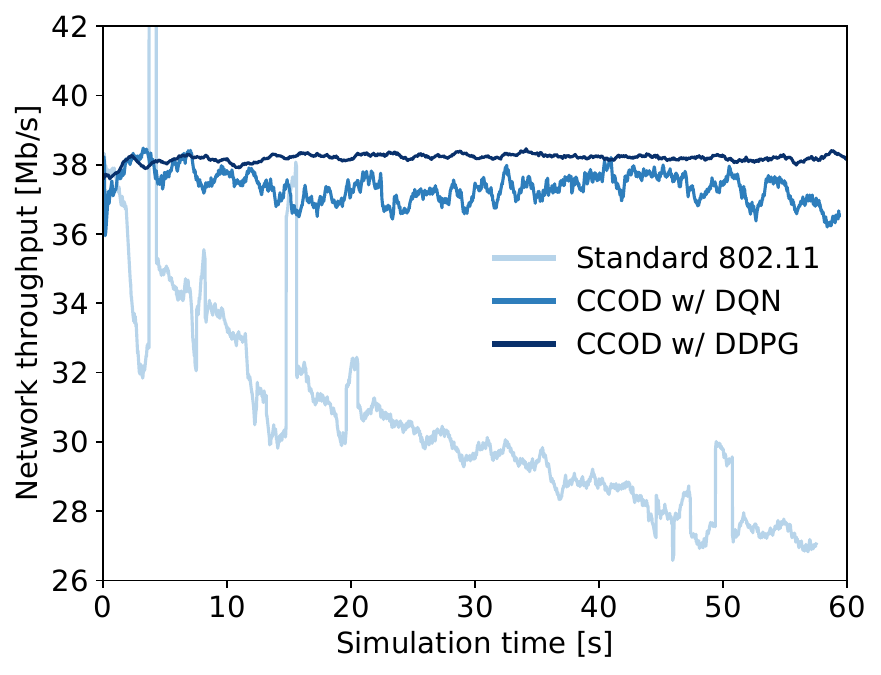}
        \label{fig:ccod_reaction}
    }
    \caption{Configuration of Wi-Fi channel access parameters with \ac{RL}: example of \ac{CW} optimization with \ac{CCOD} \cite{wydmanski2021contention}. As the number of contending stations steadily increases over time (a), the \ac{AP} monitors the collision probability and uses \ac{RL} to select the \ac{CW} value for all associated stations to maintain throughput higher than under legacy operation (b). The example shows two \ac{RL} algorithms that use different types of output (\ac{DQN} -- discrete, \ac{DDPG} -- continuous) and which explore the available parameter space in search of better \ac{CW} values (a).}
    \label{fig:ccod}
\end{figure}

Multiple studies consider the selection of \ac{CW} values to maximize throughput by reducing both collisions and idle periods.
\ac{SL} and \ac{RL} models are typically applied.
Loss functions and rewards are addressed in the form of reduced collisions \cite{ali2019deep,kumar2021adaptive}, increased difference between successful and collided frames~\cite{zhu2012achieving}, improved channel utilization~\cite{abyaneh2019intelligent}, increased successful channel access attempts \cite{edalat2019dynamically, zhang2020enhancing}, throughput~\cite{wydmanski2021contention}, network utility~\cite{moon2021neuro}, and a combination of improved throughput, reduced energy, and decreased number of collisions~\cite{amuru2015send}.
As summarized in \Cref{fig_VA_1}, \ac{SL} \cite{abyaneh2019intelligent,edalat2019dynamically}, \ac{RL} \cite{zhu2012achieving,amuru2015send,kumar2021adaptive}, \ac{DRL} \cite{ali2019deep,wydmanski2021contention,zhang2020enhancing}, and \ac{FL} \cite{zhang2020enhancing,ali2021federated} models are applied to the IEEE 802.11 standard~\cite{amuru2015send,zhang2020enhancing} and its amendments, most importantly 802.11ac \cite{abyaneh2019intelligent}, 802.11e \cite{zhu2012achieving,coronado2020improvements}, 802.11n \cite{edalat2019dynamically}, and 802.11ax~\cite{ali2019deep,wydmanski2021contention}.
We provide a summary of the major findings next, while an illustrative example of using RL to optimize Wi-Fi channel access parameters is given in Figure~\ref{fig:ccod}.

\subsubsection{Collision Reduction}
In high-density 802.11ax \acp{WLAN}, \ac{RL} with the \ac{iQRA} is considered by \textcite{ali2019deep}.
Instead of resetting the \ac{CW} value whenever the channel is idle (as in DCF), it is calculated by considering the channel collision probabilities according to the \ac{COSB} protocol~\cite{ali2018channel}.
In this direction, the cumulative reward (accounting for the probability of collisions) is minimized by optimally adjusting a policy to update the \ac{CW} size.
The \ac{iQRA} mechanism increments or decrements \ac{CW} (according to \ac{COSB}), finding a balance between optimal actions (concerning the best policy to reduce the collision probability) and exploring new actions to account for the dynamicity of Wi-Fi environments. 
Results from the ns-3 network simulator, for both small (with \num{15} stations) and dense (with \num{50} stations) networks, confirm that the solution outperforms the baseline 802.11ax protocol in terms of throughput, while the delay remains similar.

\textcite{zhu2012achieving} implement a programming paradigm called \ac{ABP}, where the reward is the difference between successful transmissions and collisions.
\Ac{ABP} optimizes the specifics of \ac{RL} for two possible actions: halve the CW size or leave CW unchanged after a successful transmission.  
Simulations performed in ns-2 with \num{20} stations show a reduction of the total number of dropped packets by four.

The \ac{RF} algorithm is applied in a supervised manner to balance the minimum \ac{CW} size among users and account for fair channel access \cite{abyaneh2019intelligent}.
The algorithm departs from monitoring channel variables (i.e., busy time, channel occupancy by the user, the number of sent frames) to build a decision tree regarding the variety of settings.
The algorithm is implemented in indoor 802.11ac scenarios with up to 8 stations.
Throughput, latency, and fairness are improved by \SI{153.9}{\percent}, \SI{64}{\percent}, and \SI{19.34}{\percent}, respectively, when compared to the 802.11ac standard. 

The size of the \ac{CW} can also be  adjusted by directly increasing the access to the channel through the fixed-share algorithm \cite{edalat2019dynamically}.
\ac{CW} is derived by weighing a set of possibilities on the \ac{CW} range predefined in advance, where the larger the weight, the larger the influence of the particular \ac{CW} value.
Whenever a successful transmission occurs, the weight of users with the largest \ac{CW} is reduced to increase the chances of transmissions and the weight of users with a lower \ac{CW} is increased. In the case of collisions, the performance is the opposite.
With this mechanism, a balance is achieved between aggressive (small \ac{CW}) and non-aggressive (large \ac{CW}) users. %
Simulations in ns-3 of %
randomly deployed senders %
show that in a heavily loaded scenario (with \num{100} users), throughput is improved by \SI{200}{\percent} and the end-to-end delay is reduced by \SI{33}{\percent} when compared to DCF.

\subsubsection{Scalability}
To address the scalability of 802.11ax networks, a \ac{DRL} model provides stable throughput under an increasing number of stations \cite{wydmanski2021contention}.
A centralized solution is applied for two trainable control algorithms: \ac{DQN} and \ac{DDPG}.
A three-phase algorithm is designed to (1) evaluate the history of collision probabilities, (2) the training of both \ac{DRL} models by maximizing the reward (throughput), and (3) their deployment in the network.
The algorithm is implemented in ns3-gym \cite{gawlowicz2019ns3gym} with a single \ac{AP} and up to 50 stations.
Compared to the 802.11ax standard, which leads to a decreased network throughput of up to \SI{28}{\percent}, the two algorithms exhibit a stable throughput value 
for an increasing number of stations.

A \ac{PDS} learning algorithm is applied by \textcite{amuru2015send} to take advantage of previous knowledge of the system components such as the \ac{CW} and the transmission buffer occupancy. 
In contrast to \ac{QL}, \ac{PDS} achieves faster convergence to optimally compute the \ac{CW} when asserting its value in specific states.
For instance, when the channel is free and the station is waiting to transmit, the \ac{CW} will certainly be reduced by one.
In such a case, the corresponding transition probabilities do not have to be learned, thereby increasing the convergence speed by eliminating exploration actions.
The solution exhibits enhanced throughput, especially with moderate network load, in comparison to Q-learning, the 802.11 standard, and alternative deterministic mechanisms like \ac{EIED}.

\subsubsection{User Fairness}
The \ac{CW} can also be  adjusted considering user fairness metrics \cite{zhang2020enhancing}.
To that end, \ac{FL} and \ac{QNN} models are implemented in \acp{AP} and stations, respectively, as a distributed method.
When each station randomly initializes its \ac{QNN} parameters, some stations will use a more aggressive strategy to  access the channel (by choosing small \acp{CW}). Such behavior, however, will block the transmissions of stations initialized with a less aggressive strategy (with large \acp{CW}).
To ensure fairness, the \ac{AP} obtains a global model of the \ac{QNN}s through \ac{FL} and later broadcasts updated \ac{CW} values to stations.
Simulation results for a single \ac{AP} and a total number of stations up to \num{50} show that throughput is improved by \SI{20}{\percent} when compared to \ac{DCF}.

An improved \ac{DQN} is trained for minimum \ac{CW} selection and deployed at stations to achieve per-user fairness \cite{kumar2021adaptive}. 
The extension of \ac{DQN} is achieved through rainbow agents \cite{hessel2018rainbow}, which incorporate six improvements: double \ac{DQN}, prioritized reply, dueling networks, multi-step learning, distributional \ac{RL}, and noisy nets.
The ns-3 simulation results, for \num{32} stations transmitting at a constant rate of \SI{1}{\mega\bit\per\second}, show that 
the solution achieves results close to optimum and it is superior to an \ac{RF}-based method.

\subsubsection{\ac{QoS}}
Driven by the need to distinguish between traffic priorities, \ac{DCF} was extended to \ac{EDCA} in the 802.11e amendment \cite{jie2005unified,szott2010ieee}.
To that end, new MAC parameters were introduced per traffic class: \ac{CW}, \ac{AIFS}, and \ac{TXOP} limit \cite{kosek2019tuning}\footnote{To support fine-grained traffic prioritization, IEEE 802.11e is extended by IEEE 802.11aa \cite{kosek2014ieee}, however, we did not find any papers devoted to ML-based optimization of 802.11aa.}.
\Ac{AIFS} together with \ac{CW} are directly responsible for the trade-off between delay and throughput.
In this direction, a three-phase scheme is implemented by \textcite{coronado2020improvements} to select the best combination of \ac{CW} and \ac{AIFS} supported by \ac{ML}. 
In the first two phases, a range of \ac{AIFS} and \ac{CW} values are selected relying on decision tree algorithms, e.g., J48 for classification and M5 for prediction. 
Then, in the third phase, the best combination for \ac{AIFS} and \ac{CW} is derived.
Simulation results exhibit high accuracy on the throughput prediction when varying the \ac{CW} range, \ac{AIFS}, and the total number of stations. 

To ensure priority-based channel access, within the \ac{EDCA} distributed scheme, a \ac{QL} model is implemented to infer network density and adjust the \ac{CW} value \cite{ali2020performance}.
In \ac{EDCA}, the \ac{CW} is set to be smaller for high-priority traffic like voice and video.
The optimal \ac{CW} value is derived for the four different traffic priorities defined by \ac{EDCA}.
Simulation results are derived in the ns-3 simulator, where the throughput per traffic type is improved in comparison with the standard \ac{EDCA} mechanism.

\subsubsection{Time-slotted Access}
Additionally, collisions are avoided in channel access mechanisms where users are scheduled per time slots \cite{lee2020collision}.
Each station stores a table consisting of the available time slots in which a given frame is to be transmitted. The available time slots are selected by an \ac{RL} method to find appropriate actions when occupying the channel.

Finally, \textcite{kihira2020adversarial} consider a channel access problem between two APs: the protagonist, which is equipped with an agent, and a second AP called the `outsider'.
Time is divided into slots, where both APs can decide to transmit independently and the goal of the agent in the protagonist AP is to find, based on learning the behavior of the outsider AP, the transmission probability that maximizes its throughput. A robust adversarial RL framework that uses game theory models the interactions between the two APs. The framework can learn the best transmission policies through Q-learning.

\subsection{Link Configuration}
\label{sec_link}

In response to growing user demands, the IEEE 802.11n/ac/ax amendments implement high-throughput wireless links through dedicated features at the \ac{PHY} and \ac{MAC} layers \cite{karmakar2020s2}.
High data rates are achieved through a variety of functionalities at the \ac{PHY} layer including channel bonding, multi-\ac{SS} transmissions, the use of \ac{SGI}, and high modulations (1024-QAM for 802.11ax) \cite{perahia2011gigabit,karmakar2017impact,karmakar2016dynamic}.
At the \ac{MAC} layer, frame aggregation and block acknowledgment are the two main features for improving the maximum link throughput. 

\begin{figure}
\centering
\includegraphics[width=\columnwidth]{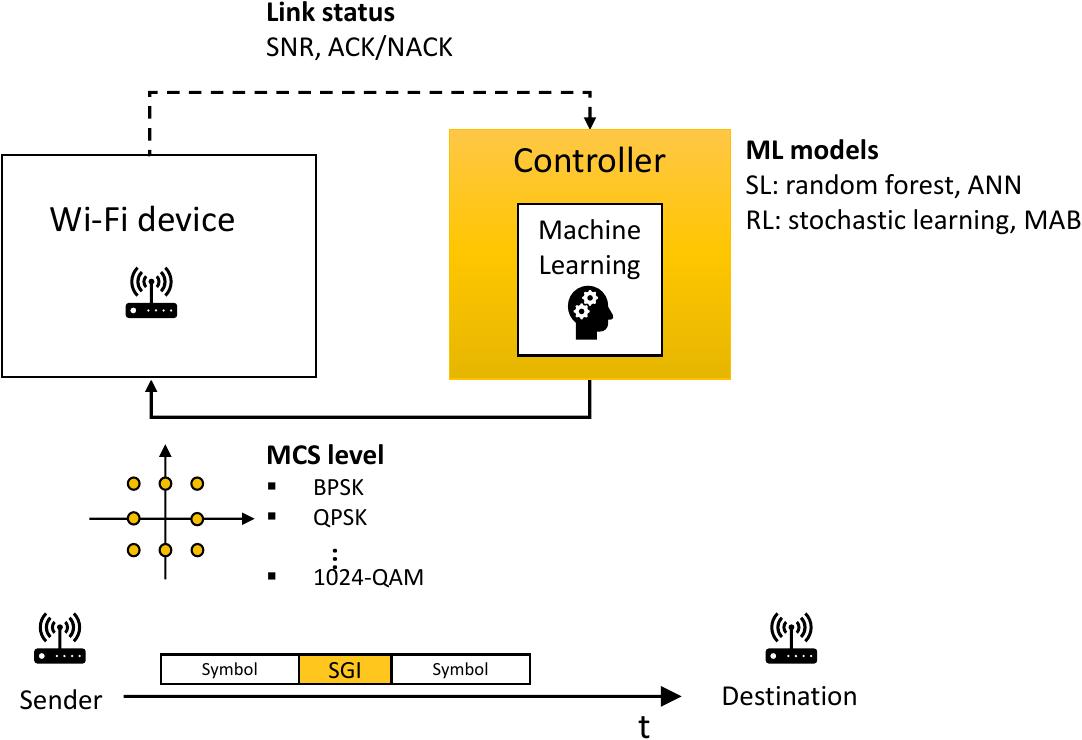}
\caption{Rate selection supported by \ac{ML} models, in which link status observation is used to adjust link configuration, e.g., MCS level. Dashed arrows represent observation and solid lines represent actions. 
}
\label{fig_VA_2}
\end{figure}

Link configuration, in the form of selecting appropriate \ac{PHY} and \ac{MAC} parameters, is required to achieve the optimum throughput for given network and channel conditions.
Rate adaptation plays an important role in link configuration, which is responsible for the selection of \ac{MCS} values for each transmission. 
In dynamic \mbox{Wi-Fi} scenarios (e.g., due to user mobility or interference), rate adaptation deals with the following counteracting mechanisms: 
\begin{itemize}
    \item high data rates may lead to high error rates when decoding the transmitted bits, thereby reducing throughput;
    \item reducing the data rate may incur poor channel utilization and thus also reduce throughput.
\end{itemize}
The trade-off between transmission errors and channel utilization can be evaluated by applying \ac{ML} models, particularly to deal with varying channel conditions.
\Cref{fig_VA_2} depicts how \ac{ML} models are used for rate selection. In the following, we summarize the contributions in the  selection of optimal \ac{MCS} and \ac{SGI} values, and a variety of trade-offs at the \ac{PHY} layer.

\subsubsection{Rate Adaptation}
Rate adaptation solutions predict the probability of successful transmissions for each \ac{MCS} candidate. Then, the data rate is selected corresponding to the \ac{MCS} with the best result.
Predictions are made based on \ac{SNR} \cite{li2020practical,punal2013rfra} or follow a cross-layer approach based on \ac{ACK} or \ac{NACK} feedback~\cite{joshi2008sara,wang2013dynamic,cho2021reinforcement}.
\ac{SNR} is preferred to timely update the channel status when dealing with station mobility, e.g., in the case of \acp{VANET}~\cite{punal2013rfra}.
However, more accurate solutions are obtained when updating the channel status based on the \ac{ACK} and \ac{NACK} feedback~\cite{punal2013rfra}.

For \ac{SNR}-based predictions, throughput is improved through a two-level data rate search algorithm based on an \ac{ANN} model \cite{li2020practical} or using an \ac{RF} algorithm \cite{punal2013rfra}.
In the former, the \ac{ANN} is implemented as a coarse estimator to find a possible set of best data rate candidates.
In the second stage, a fine-grained solution is devised to identify the best candidate from this set.
With this solution, at least a \SI{25}{\percent} improvement is reported in mobile scenarios when compared to baseline rate adaptation algorithms like Minstrel~\cite{fietkau2020minstrel}.
\textcite{punal2013rfra} implement the \ac{RF} algorithm for uplink data rate adaptation in \acp{VANET}.
The algorithm uses the position and velocity of cars to estimate the \ac{SNR} in the link between the \acp{AP} and the vehicle.
The algorithm predicts the probability of successful transmission for each possible data rate candidate and then selects the best candidate.
With this approach, the goodput is improved at least by \SI{27}{\percent} in comparison to reported solutions like \ac{CARA}.

The unpredictable impact of fast fading de-correlates, however, the correspondence between \ac{SNR} and packet loss due to their large fluctuation in short periods. 
To deal with this problem, \textcite{joshi2008sara} implement a method inspired by \ac{SLA} which does not assume any predefined relation between \ac{SNR} and packet loss.
The algorithm updates a selection probability vector in a one-to-one mapping to the available data rates.
The learning procedure is implemented to adjust this vector, with throughput being the reward function, while the \ac{ACK} frames are used as feedback to account for the channel condition.
Thus, the probability corresponding to the rate that produces the best reward is updated, 
leading to a \SI{15}{\percent} throughput improvement in comparison to other reported solutions.

Thresholds to detect successfully and non-successfully received packets are derived through \ac{ML} models to improve aggregate throughput by counting received \acp{ACK} \cite{wang2013dynamic}.
Based on the legacy \ac{ARF} algorithm, the data rate is increased or decreased when the total number of \ac{ACK} is higher than a given threshold.
Thresholds are adjusted by an \ac{ANN} when estimating their correlation with the achievable throughput considering the total number of stations, channel conditions, and traffic intensity.
Results show that the aggregated output is increased by \SI{10}{\percent} in a network of \num{10} stations.

Rate selection can also be performed by first identifying the channel condition, e.g., using \acl{SL} \cite{kurniawan2019machine} or  \mbox{Q-learning} \cite{cho2021reinforcement}. 
In the former, the channel condition is classified as residential or office environments, then the proper \ac{MCS} level is selected.
The model is trained based on selected characteristics of an 802.11 frame's preamble. 
In the \mbox{Q-learning} model, the \ac{MCS} level is adjusted based on the total number of received \acp{ACK}.
Observation of the network state is conceived through timeout events, which are referred to as the total number of missing \acp{ACK}.
Simulations in ns3-gym \cite{gawlowicz2019ns3gym} consider a dynamic scenario, where the receiver station moves away from the sender at a speed of $\SI{80}{\meter\per\second}$ with throughput comparable to Minstrel.

Alternatively, \ac{MCS} is selected considering also the available bandwidth and selected spatial streams.
\textcite{chen2021experience} apply the \ac{DDQN} model using goodput as a reward and include further learning techniques like prioritized training, history-based initialization, and adaptive training interval.
Results show that this method, implemented in hardware, significantly outperforms default mechanisms.

\subsubsection{SGI Adaptation}
The selection of the \ac{SGI} values is another link configuration mechanism that is supported by \ac{ML} models.
The \ac{SGI} assumes two (802.11ac) or three  (802.11ax) different values.
The selection between them is implemented through \ac{TS} by \textcite{karmakar2020s2}. Such an online learning mechanism deals with the fluctuation of channel quality (signal interference, signal fading, and attenuation).
The \ac{TS} model is evaluated with simulations in ns-3 for an 802.11ac network with up to 40 stations.
For \ac{SNR} varying randomly in the range of \SIrange{20}{60}{\decibel}, the results show a slight throughput improvement compared to the static \ac{SGI} settings.

\subsubsection{PHY Layer Trade-offs}
\label{sec:multilayer}
There are a variety of trade-offs inherent to the \ac{PHY} layer:
wider channels versus more interference, \ac{MCS} versus required \ac{SNR}, frame aggregation versus packet loss, etc.
These trade-offs may be jointly addressed to optimize the overall performance using \ac{ML} methods such as \ac{MAB} \cite{karmakar2020online,karmakar2016dynamic,karmakar2017ieee,karmakar2017smarla} and \ac{DL} \cite{karmakar2020deep}.

Karmakar et al. design an online learning-based mechanism based on the \ac{MAB} framework for link configuration in 802.11ac networks \cite{karmakar2020online,karmakar2016dynamic,karmakar2017ieee,karmakar2017smarla}. 
This solution considers both network load and channel conditions and uses a \ac{MAB}-based \ac{AL} (i.e., the $\varepsilon$-greedy algorithm) along with fuzzy logic.
Through this approach, the network performance is improved thanks to the ability to explore multiple configurations.
The resulting implementation exhibits increased throughput (up to \SI{358}{\percent}) when compared to existing solutions.

\textcite{karmakar2020deep}  improve throughput with a two-step algorithm that considers several parameters from the \ac{PHY} and \ac{MAC} layers simultaneously (channel bonding, \ac{MCS}, and frame aggregation settings).
First, a \ac{DNN} estimates throughput assuming different link parameter settings. 
Then, a predictive control-based search algorithm finds the optimal parameter values which maximize throughput.
Experimental results are obtained through IEEE 802.11ac client boards installed on laptops.
Results exhibit superior performance concerning delay and throughput in comparison to three baseline algorithms.

Rate adaptation algorithms are also designed for specific applications in industrial networks \cite{peserico2020rate}.
An \ac{RL}-based mechanism solves the trade-off between reduced packet loss and increased transmission rate.
The learning procedure is implemented through the \ac{SARSA} algorithm. The balance between exploration and exploitation is conceived through the $\varepsilon$-greedy algorithm.
With this approach, packet losses are reduced by \SI{6}{\percent} when compared to non-\ac{RL}-based algorithms.

\subsection{Frame Aggregation}
\label{sec_frame}

Frame aggregation directly impacts the communication efficiency in terms of useful transmitted data and overhead~\cite{skordoulis2008ieee}. %
Efficiency is analyzed in terms of errors produced during packet decoding:
larger frames can lower the impact of overhead, but they are also more susceptible to transmission errors. 
This trade-off is addressed by frame aggregation techniques to derive the optimum frame size to maximize efficiency.
The 802.11 standard introduces two basic aggregation methods: the \ac{A-MSDU} and the \ac{A-MPDU} \cite{khorov2019tutorial}. These aggregations can also be used together \cite{kosek2019tuning}.

The \ac{A-MSDU} method is more efficient but more prone to errors than \ac{A-MPDU} since it contains only one \ac{FCS} accounting for all aggregated frames.
The \ac{A-MPDU} method is more robust but introduces more overhead as it generates several \acp{FCS}, one per each subframe.
However, their dynamic adjustment in the 802.11 standard is not designed to deal with the varying \ac{CSI} in wireless links. 

\begin{figure}
\centering
\includegraphics[width=.9\columnwidth]{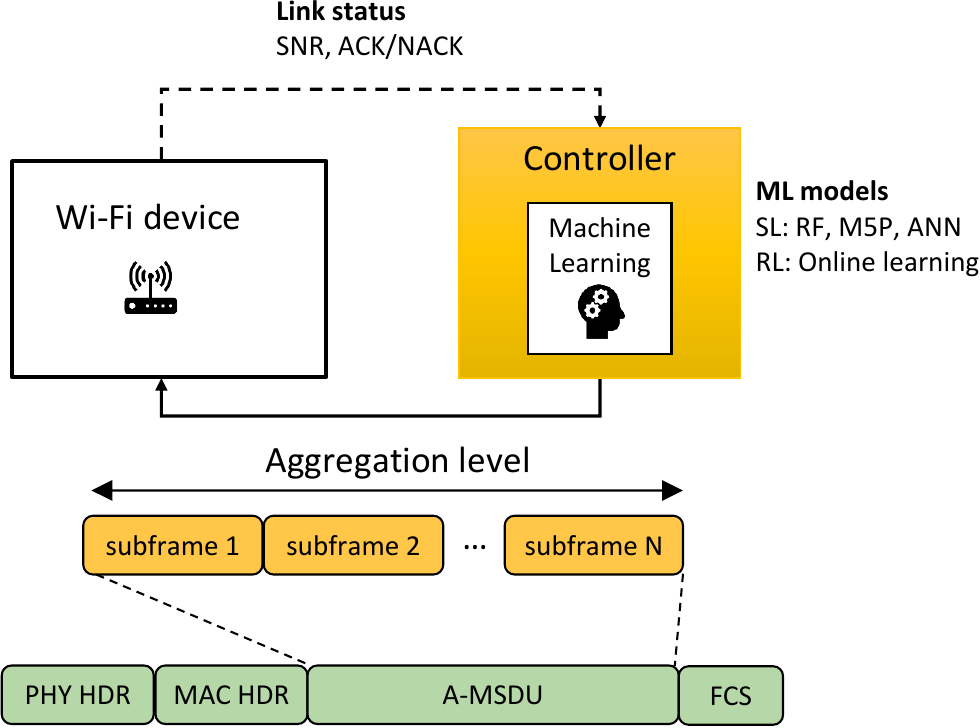}
\caption{Frame aggregation supported by \ac{ML} models, in which link status observation is used to adjust the number of aggregated subframes. Dashed arrows represent observation and solid lines represent actions. 
}
\label{fig_VA_3}
\end{figure}

To optimally select the frame size under dynamic conditions, \ac{ML} techniques are used (\Cref{fig_VA_3}), including \ac{SL} \cite{coronado2020adaptive,coronado2020aios,hassani2021quick, lin2009machine,khastoo2020neura} and \ac{RL} \cite{karmakar2019online}.
Their use is reported for generic 802.11 networks to maximize throughput \cite{lin2009machine}, for 802.11n to maximize goodput \cite{coronado2020adaptive,coronado2020aios} and for 802.11ac to address the energy-throughput trade-off \cite{karmakar2019online} as well as to estimate the aggregation levels in 802.11ac \cite{hassani2021quick}.

\textcite{coronado2020adaptive} implement a low computational complexity technique for the downlink direction.
A \ac{RFR} model configures both the aggregation and \ac{MCS} settings. %
Results are obtained for small and medium-sized networks (up to 20 stations).
This solution lowers the rate of retransmission resulting in goodput improved by \SI{18.36}{\percent} when compared to legacy 802.11 aggregation mechanisms.

Aggregation methods supported by \ac{ML} are also designed for \acp{SD-WLAN} as an \ac{AI}-based operating system \cite{coronado2020aios}.
The M5P and the \ac{RFR} models are implemented due to their low computational complexity. 
Intended to provide a frame length that maximizes goodput for each user, their training is performed with real measurements in a \mbox{Wi-Fi} scenario with up to \num{10} stations.
Here, the \ac{RFR} model presents the highest goodput improvement (\SI{55}{\percent}) when compared to the \ac{A-MSDU} mechanism.

The \ac{MCS} level can also be  predicted through an \ac{ANN} \cite{khastoo2020neura}.
The model is trained in a client device by receiving packets from an \ac{AP} using all available rates within a \SI{1}{\second} time window.
Estimated rates are then used to compute the best aggregation level using a previously designed  (non-\ac{ML}) method \cite{abedi2020practical}.
The implemented solution outperforms baseline algorithms by at least \SI{13}{\percent} in terms of throughput.

Aggregation level estimators can help in queue backlogging.
\textcite{hassani2021quick} use \ac{ML} techniques on obtained hardware-level timestamps to determine the aggregation level implemented at a given \ac{AP}.
A logistic regression estimator model provides an accurate aggregation level estimator with low computational complexity.
This solution is implemented in non-rooted hardware as client devices, where the achieved accuracy to determine the proper aggregation level is close to \SI{100}{\percent}.

Frame aggregation settings can also consider the associated energy costs \cite{karmakar2019online}.
Based on the channel condition (given by the \ac{SNR} value), the aggregation level is selected as the one with the smallest \ac{FER} to reduce the energy costs caused by retransmissions.
The solution combines an online learning algorithm to define a set of suitable aggregation levels and fuzzy logic to select the most suitable level from that set,
by estimating which frame size would have the lowest \ac{FER}.
With this approach, the resulting energy efficiency with \num{10} stations is \SI{14}{\percent} better when compared to the standard use of \ac{A-MSDU} and \ac{A-MPDU} mechanisms.

Finally, channel condition and impact of collisions are jointly addressed by \textcite{lin2009machine} to adjust both the frame size and \ac{CW}. %
An \ac{ANN} model is trained with frame size-throughput patterns to provide a gradient indicating the direction of the optimal frame and the \ac{CW} sizes.
Simulation results, provided for \num{10} mobile users, show that 
throughput is improved when compared to the case when only the frame size is optimized (i.e., without additionally considering the optimal \ac{CW}).

\subsection{PHY Features}
\label{sec_phy}
\begin{figure}
\centering
\includegraphics[width=\columnwidth]{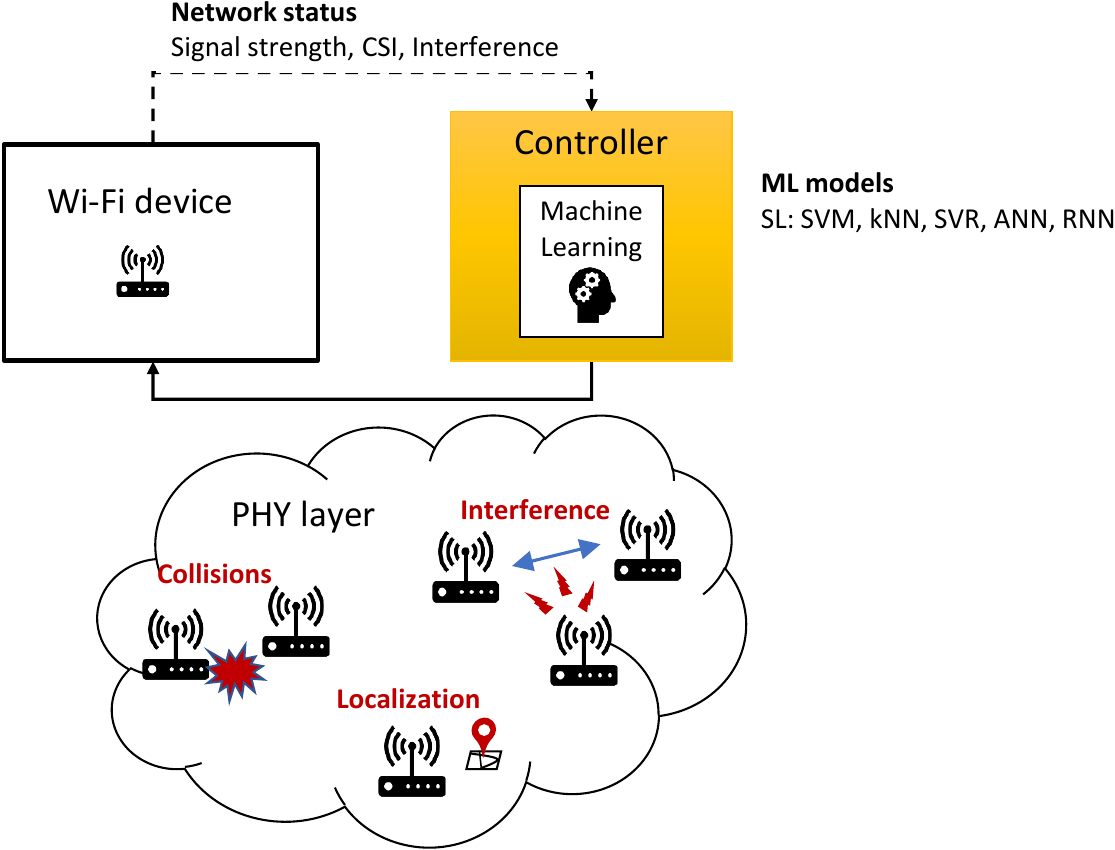}
\caption{\ac{ML} models mitigate negative PHY effects by observing network status and avoiding interference. Dashed arrows represent observations, solid lines represent actions.}
 \label{fig_VA_4}
\end{figure}

At the \ac{PHY} layer, a variety of actions are supported by \ac{ML} techniques to improve the performance of \mbox{Wi-Fi} networks.
Issues that are addressed include:
\begin{itemize}
    \item collision detection characterization \cite{kashyap2010deconstructing} and its mitigation \cite{lee2015frequency,lee2012dfi},
    \item interference power-level characterization \cite{saha2019interference} and its mitigation \cite{zhao2019joint},
    \item signal de-noising~\cite{almazrouei2019deep}, source detection to improve spectral efficiency~\cite{kim2018identifying},
    \item prediction of signal strength variability \cite{herath2019deep},
    \item the enhanced modeling of the \ac{PHY} and \ac{MAC} layer interactions to improve throughput \cite{herzen2015learning}.
\end{itemize}
As depicted in \Cref{fig_VA_4}, a variety of \ac{ML} models are available to deal with these problems, which we describe next. 

\subsubsection{Collision Reduction}
To estimate the number of collisions in the channel, the activity of stations in the network is modeled as a \ac{HMM} \cite{kashyap2010deconstructing}.
The approach is to use \ac{RL} techniques to learn the parameters of such models, then to mathematically evaluate the probability of collisions.
The transition probabilities are assessed through the \ac{EMA}.  
Based on the derived transition probabilities, the probability of collisions is directly computed based on the estimated total number of senders that simultaneously transmit.
Results are provided for seven \acp{AP} deployed with an equal number of clients over two floors of a building.
The estimated deferring probabilities exhibit a good correspondence with the real condition scenario.

To improve the decoding of \ac{RTS} frames during collisions, Lee et al. \cite{lee2015frequency,lee2012dfi} implement an \ac{ML} model.
A Bloom filter decodes the \ac{RTS} frames, and a supervised \ac{ML} technique solves the inherent ambiguity with an accuracy larger than~\SI{99}{\percent}.
The \ac{ML} is implemented through a variety of algorithms such as naive Bayes, naive Bayesian tree, J48 decision tree, and \ac{SVM}.
Additionally, this solution is connected to a second, $k\varepsilon$-greedy algorithm for channel allocation.
The integration of both algorithms improves the performance~\num{3.3} times over legacy 802.11 operation.

\subsubsection{Interference Estimation}
The interference level is estimated by modeling the network through a \ac{DPP} \cite{saha2019interference}.
An \ac{SL}-based process is implemented to evaluate the total number of active transmitters that may interfere with each other and learn their locations.
Interference is evaluated by providing the \ac{CDF} for the total number of active users.
This evaluation is then used when modeling the power of the interference signals through a path-loss model for each link.
Results illustrate a good match with a theoretical model %
regarding the \ac{CDF} of interference levels.

\subsubsection{Signal Quality Estimation and Management}
The received signal strength is predicted through deep learning techniques by \textcite{herath2019deep}.
In a \ac{RNN} model, encoder and decoder components are implemented to capture the \ac{CSI} and predict its variability, respectively.
The model is trained according to three different schemes to balance the trade-off between convergence speed and performance: 
\begin{itemize}
    \item guided training which uses current measured signal strength (resulting in faster convergence),
    \item unguided training which uses predicted signal strength (resulting in better prediction performance), 
    \item curriculum training which combines both previous methods to balance the speed and prediction performance.
\end{itemize}
With the curriculum training scheme, the resulting prediction accuracy of the signal strength is improved when compared to linear regression and auto-regression methods.

The quality of the received signal can also be improved at the \ac{PHY} layer using \ac{DL} techniques \cite{almazrouei2019deep}.
With an \ac{ANN}, the preamble of the 802.11 family protocols is de-noised by unfolding the useful signal from noise in the spectrogram domain (i.e., time-frequency domain).
The spectrogram is processed as an image, where the \ac{ANN}, used as a convolutional de-noising auto-encoder, estimates the originally emitted patterns.
With this approach, the derived reconstruction accuracy is around \SI{85}{\percent}.

The spectral efficiency of \mbox{Wi-Fi} transmissions can also be  improved when avoiding the exposed terminal problem.
To that end, senders are identified according to their \ac{CSI} to later predict whether they will interfere with each other \cite{kim2018identifying}.
To implement such an identification mechanism, a model is trained through \ac{kNN} and \ac{ANN} with \num{20} wireless stations in indoor scenarios, where an accuracy of \SI{90}{\percent} is achieved with at least \num{30} samples per station.
In the case of reduced total samples, better performance is obtained with the \ac{kNN} model.

\subsubsection{Interaction with the MAC Layer}
The \ac{PHY} layer can also be  modeled in unison with the \ac{MAC} layer to characterize the impact of different features on observed throughput \cite{herzen2015learning}.
The selected input features are received power, channel width, spectral separation between users, traffic load, and physical rates.
The idea is to find a mathematical function that maps input features to throughput values supported by supervised \ac{ML}.
This mathematical function then becomes a black box representation of a given link to later optimize throughput.
The learning phase, which is used to obtain this function, is  derived through regression techniques: regression tree, \ac{GBRT}, and \ac{SVR}.
In particular, simulation results show that \ac{GBRT} and \ac{SVR} provide the most accurate results in comparison to a benchmark.

\subsection{Open Challenges}
From the multitude of papers addressing core Wi-Fi \ac{PHY} and \ac{MAC} features, we identify several open challenges related to:
\begin{itemize}
    \item studying more realistic settings (including user mobility), 
    \item removing common simplifying assumptions,
    \item improving \ac{ML}-based solutions.
\end{itemize}
We describe these challenges below.

First, there is a need for more \emph{realistic simulations}.
Several reports address the intention to provide simulation testbeds with less simplifying assumptions.
For instance, the inclusion of more realistic channel and traffic models, variable channel conditions per user, dense networks, or the addressing of multi-hop networks are some remarked requirements to conduct further research as remarked in \cite{zhu2012achieving, ali2019deep,edalat2019dynamically, jie2005unified, kosek2019tuning, wydmanski2021contention,joshi2008sara,peserico2020rate}.

Second, studies are needed to consider \emph{overall network performance}.
Currently, papers address specific optimization parameters under specific conditions. Although some work is reported to simultaneously address a variety of parameters of \mbox{Wi-Fi} networks (cf.\ \Cref{sec:multilayer}), an overall perspective of network functioning, which would account for optimization criteria in several layers simultaneously, has been not conducted yet.
While improved performance is achieved when addressing cross-layer designs \cite{xiaojun2006tutorial}, solutions to posed problems in this direction are rather difficult to solve by analytical means due to the variety of related parameters.
As yet unexplored, this constitutes a promising research direction to address by \ac{ML} models.
    
Third, only a few papers provide details on the \emph{impact of user mobility} on communication performance~\cite{cho2021reinforcement,lin2009machine}. 
However, considering the growing number of mobile \mbox{Wi-Fi} devices (e.g., phones, tablets, even vehicles), further insights can be provided to better characterize the influence of their movement on the network performance.

Finally, many reported works remark future directions concerning the \emph{improvement of \ac{ML}-based solutions} to:
\begin{itemize}
    \item provide accurate \ac{ML} models (additional loss functions) \cite{edalat2019dynamically},
    \item reduce the coordination overhead between agents in decentralized solutions \cite{amuru2015send},
    \item further study the impact of network status parameters on traffic prediction \cite{karmakar2016dynamic,karmakar2017ieee}, and
    \item increase the complexity of \ac{ML} models to better characterize network functioning \cite{kashyap2010deconstructing,saha2019interference,coronado2020adaptive,cho2021reinforcement}.
\end{itemize}

\section{Recent Wi-Fi Features}
\label{sec_advanced}

In a push for higher and more efficient performance levels, recent Wi-Fi amendments such as 802.11ac~\cite{ong2011ieee}, 802.11ax~ \cite{bellalta2016ieee}, and 802.11be~\cite{lopez2019ieee} introduce new advanced and complex techniques such as multi-user communications (OFDMA, MU-MIMO)~\cite{bellalta2019ap}, spectrum aggregation and opportunistic spectrum access (channel bonding~\cite{barrachina2021wi}, multi-link operation~\cite{lopez2022multi,naik2020next,carrascosa2021experimental}), spatial reuse~\cite{wilhelmi2021spatial}, and multi-AP coordination~ \cite{deng2020ieee,nunez2021txop}. All these techniques promise high-performance gains in both throughput and latency but also open new challenges. These challenges are solved, or at least alleviated, using ML methods (see Table \ref{tab:recent}) as we show in this section.

\subsection{Beamforming}
\label{sec_mmWave}

Transmissions in the \ac{mmWave} 60~GHz shared band are a specific Wi-Fi use case aimed at greatly increasing the transmission rate in \ac{LoS} communication scenarios, both short-range (indoor) and long-range (outdoor), the latter known at \ac{FWA}~\cite{aldubaikhy2020mmwave}. 
To cope with the increased attenuation in this band, beamforming of transmissions is required. 
This functionality was first introduced to Wi-Fi in 802.11ad and later extended in 802.11ay.

\begin{figure}
\centering
\includegraphics[width=\columnwidth,trim=50 50 20 0, clip]{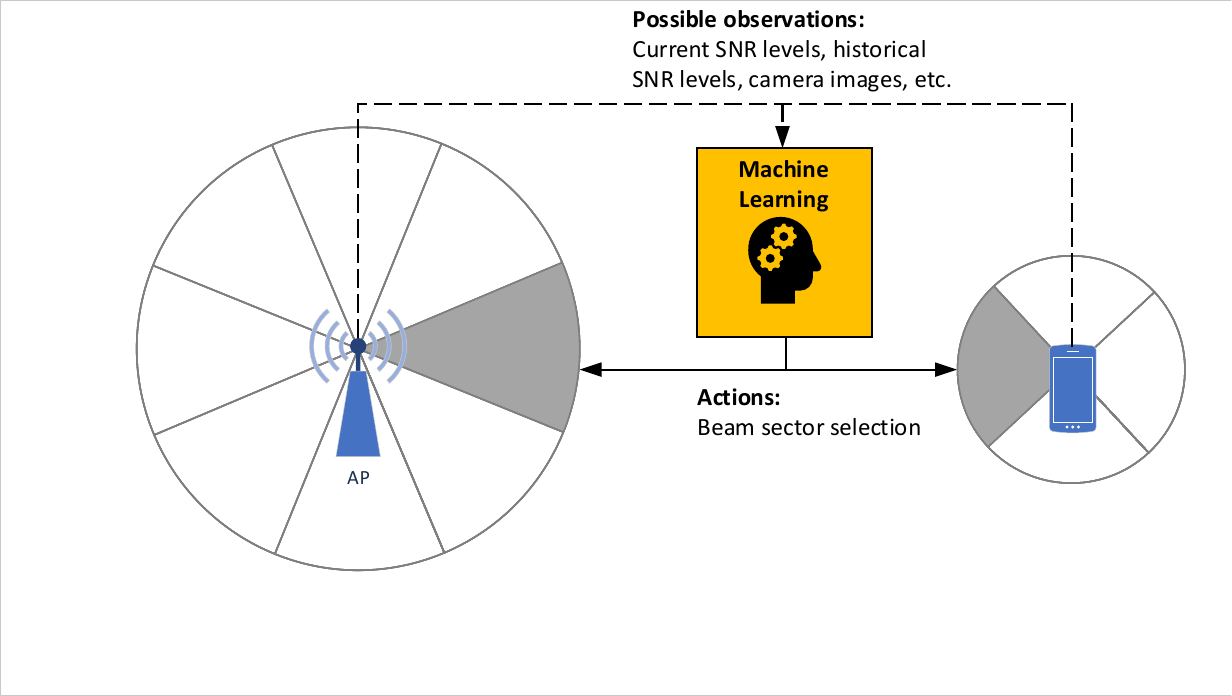}
\caption{Example of beam sector alignment in IEEE 802.11ad/ay networks: 8-sector \ac{AP} and 4-sector client station.}
 \label{fig_beamforming}
\end{figure}

A key problem of 802.11ad/ay networks, which are solved using \ac{ML}, is finding the optimum beam sector pairs (i.e., beam alignment) between transmitter and receiver (\cref{fig_beamforming}). 
Alignment is derived from a beam sweeping procedure, which can take up to tens of milliseconds and needs to be periodically repeated.
To facilitate the beam sector pair selection, \textcite{chang2019learning} replace the standard method of an exhaustive beam search with one of three \ac{NN}-based algorithms to predict the optimal beam sector, including with historical data. 
This work is extended by \textcite{shen2021design} with the training duration reduced through a combination of \ac{SL}-based feature extraction and \ac{RL}-based training beam selection.
Meanwhile, \textcite{polese2021deepbeam} develop DeepBeam, a framework for beam selection that replaces the time-consuming beam sweeping procedure with inferring the beam sector to use through deep learning based on passive listening to other transmissions.

Alternatively, improved \ac{ML}-based beam alignment predictions are performed with camera images.
\textcite{salehi2020machine} show that visual information can significantly reduce the time required to establish the best beam pairs.
\textcite{nishio2019proactive} also apply \ac{ML} camera images to accurately and rapidly predict received power, which is the necessary information needed to find beam sectors.
Additionally, camera-based predictions of link outage with \ac{DRL} lead to improved handovers in \ac{mmWave} networks \cite{koda2020handover}.

Since the range of mmWave bands is short, 802.11ad/ay APs may need to be densely deployed for certain use cases.
Under such network densification, beam coordination, and interference management become necessary. 
\textcite{mohamed2017coordinated} reduce cross-beam interference by applying statistical learning to construct a radio map of the network environment,
which serves as input for beam selection.
In this scenario, signaling is carried over the Wi-Fi network in the 5~GHz band through a centralized AP controller.
\textcite{zhou2019deep} optimize the beams in a centrally-managed deployment with a \ac{DNN}-based solution. 
Their solution achieves nearly the same performance as an optimization algorithm at a fraction of the computational time.

A related problem in dense deployment scenarios is the association between user stations and \acp{AP}, especially since
next-generation stations will have multi-homing capabilities (i.e., methods allowing sustained connectivity to multiple \acp{AP}).
This leads to an interesting user-to-multiple \acp{AP} association problem, which is solved using \ac{ML} methods.
\textcite{dinh2021deep} consider a generic \ac{WLAN} where users can autonomously learn, using their own \ac{DQN}, which \acp{AP} to connect to and using which band (sub-6 GHz or \ac{mmWave}).

\onecolumn
\begin{landscape}
\scriptsize
\begin{xltabular}{\linewidth}{ccccccp{0.19\textwidth}*{2}{p{0.29\textwidth}}}
\caption{Summary of works on improving the performance of recent Wi-Fi features with ML. The evaluation methods are theoretical (E), simulation (S), and experimental (E). The ML improvement is in comparison to SoA methods or (if not mentioned otherwise) to IEEE 802.11.} \label{tab:recent} \\
\toprule
\textbf{Area} & \textbf{Ref.} & \shortstack{\textbf{ML}\\\textbf{category}} & \shortstack{\textbf{ML}\\\textbf{mechanisms}} & \textbf{Year}&\shortstack{\textbf{Evaluation}\\\textbf{method}}  & \textbf{Application of ML} & \textbf{Novelty of approach} & \textbf{ML improvement} \\
\midrule
\endfirsthead

\multicolumn{9}{c}%
{\tablename\ \thetable{} -- continued from previous page} \\
\midrule \textbf{Area} & \textbf{Ref.} & \shortstack{\textbf{ML}\\\textbf{category}} & \shortstack{\textbf{ML}\\\textbf{mechanisms}} & \textbf{Year}&\shortstack{\textbf{Evaluation}\\\textbf{method}}  & \textbf{Application of ML} & \textbf{Novelty of approach} & \textbf{ML improvement} \\ \midrule
\endhead

\multicolumn{9}{r}{{Continued on next page}} \\
\endfoot

\endlastfoot

    \multicolumn{1}{c}{\multirow{24}[0]{*}{\shortstack{Beamforming \\ (Section \ref{sec_mmWave})}}} & \cite{mohamed2017coordinated} & SL    & Statistical learning & 2017  & T     & Select AP and beam & Reduce cross-beam interference & Higher throughput, lower packet loss and delay \\ \cline{2-9}
          & \cite{kurniawan2017machine} & SL    & RF    & 2017  & S     & Classify channel type & Classify channel using received preamble & Faster classification than SoA methods \\ \cline{2-9}
          & \cite{bai2018predicting} & SL    & CNN   & 2018  & T+E   & Predict channel characteristics & Apply prediction to massive MIMO channels & Good fitting between predicted and real channel statistical characteristics \\ \cline{2-9}
          & \cite{zhou2019deep} & SL    & DNN   & 2019  & S     & Select beam directions, beamwidths, and transmit power & Perform joint beam management and interference coordination & Reduced complexity vs classical algorithms \\ \cline{2-9}
          & \cite{chang2019learning} & SL    & DNN   & 2019  & S     & Select beam pair & Reduce number of input sectors & Higher throughput, lower beam training latency \\ \cline{2-9}
          & \cite{nishio2019proactive} & SL    & CNN, RF & 2019  & S+E   & Predict received power & Use camera imagery & Predict link blockage 0.5 s in advance \\ \cline{2-9}
          & \cite{azzino2020scheduling} & RL    & QL    & 2020  & S     & Select contention-free period duration & Optimize 802.11ad MAC with ML & Guarantee throughput and delay with lower channel occupancy \\ \cline{2-9}
          & \cite{salehi2020machine} & SL    & CNN   & 2020  & E     & Select beam pair & Use camera imagery & Reduce exploration time \\ \cline{2-9}
          & \cite{lin2020location} & SL    & DNN   & 2020  & S     & Estimate channel frequency response & Use transceiver location information & Decrease number of required pilot signals \\ \cline{2-9}
          & \cite{aggarwal2020experimental} & SL    & DT, RF, SVM & 2020  & S+E   & Select transmission rate & Use PHY layer features for rate selection & Higher throughput \\ \cline{2-9}
          & \cite{aggarwal2020learning} & SL    & RF    & 2020  & S     & Select transmission rate and beam pair & Joint transmission rate and beam pair selection & Higher throughput than two simple heuristics \\ \cline{2-9}
          & \cite{koda2020handover} & RL    & DRL   & 2020  & E     & Predict link outage & Use camera imagery & Improved handovers \\ \cline{2-9}
          & \cite{dinh2021deep} & RL    & DRL   & 2021  & S     & Select AP and band & Consider multiple APs and bands & Higher throughput, lower outage probability than two benchmarks \\ \cline{2-9}
          & \cite{shen2021design} & SL, RL & CNN, DRL & 2021  & S     & Select beam pair & Combine feature extraction with training beam selection & Higher throughput, shorter training than two benchmarks \\ \cline{2-9}
          & \cite{wang2021deep} & SL    & DNN   & 2021  & S     & Predict link-level performance & Apply DNN for link-level error performance prediction with transfer learning & High accuracy \\ \cline{2-9}
          & \cite{polese2021deepbeam} & SL    & CNN   & 2021  & E     & Select beam pair & Passively listen to ongoing transmissions & Higher beam selection accuracy, lower beam training latency \\ %
\midrule          
    \multicolumn{1}{c}{\multirow{18}[0]{*}{\shortstack{Multi-user\\communication\\(Section \ref{sec_multi-user})}}} & \cite{rico2014learning} & SL    & SVM   & 2014  & S     & Select users, spatial mode, transmission rate & Address MU-MIMO challenges & Higher throughput \\ \cline{2-9}
          & \cite{karmakar2019intelligent} & RL    & MAB   & 2019  & E     & Select users and configure links & Joint parameter optimization & Higher throughput, improved fairness vs three SoA methods \\ \cline{2-9}
          & \cite{su2019client} & RL    & QL    & 2019  & E     & Predict benefits from participating in MU-MIMO & Address MU-MIMO station mobility & Higher throughput \\ \cline{2-9}
          & \cite{balakrishnan2019deep} & SL    & DNN   & 2019  & S     & Schedule RUs & Apply DRL to resource allocation in downlink OFDMA & Higher throughput than heuristic \\ \cline{2-9}
          & \cite{sangdeh2020lb} & SL    & DNN   & 2020  & E     & Compress CSI & Minimize CSI feedback airtime without reducing accuracy & Higher throughput \\ \cline{2-9}
          & \cite{kotagiri2020multi} & RL    & DQN   & 2020  & S     & Schedule RUs & Apply DRL to resource allocation in uplink OFDMA & Higher throughput, lower latency \\ \cline{2-9}
          & \cite{kotagiri2021distributed} & RL    & DQN   & 2021  & S     & Schedule RUs & Improved Q-value estimation & Faster learning rate than \cite{kotagiri2020multi} \\ \cline{2-9}
          & \cite{pasandi2021poster} & RL    & N/A   & 2021  & E     & Select MU-MIMO group & RL-based MU-MIMO grouping and mode selection in the MAC layer & Improved video streaming \\ \cline{2-9}
          & \cite{su2021data} & RL    & QL    & 2021  & E     & Select MU-MIMO group & Use ML to improve MU-MIMO performance & Higher throughput \\ \cline{2-9}
          & \cite{sangdeh2021deepmux} & SL    & DNN   & 2021  & E     & Compress CSI, schedule RUs, allocate power & Joint MU-MIMO and OFDMA optimization & Reduced channel sounding overhead, higher throughput \\ \cline{2-9}
          & \cite{kotera2021lyapunov} & RL    & NN    & 2021  & S     & Schedule RUs & Queue-based resource allocation & Lower latency, improved fairness \\ %
\midrule          
\multicolumn{9}{c}{\multirow{3}{*}} \\ \\ \\ 
    \multicolumn{1}{c}{\multirow{16}[0]{*}{\shortstack{Spatial reuse \\ (Section \ref{sec_spatial_reuse})}}}
          & \cite{timmers2009spatial} & RL    & QL    & 2009  & S     & Select carrier sense threshold, transmission rate and power & Joint optimization of these parameters & Higher throughput \\ \cline{2-9}
          & \cite{nguyen2016enhancing} & RL    & MAB   & 2016  & E     & Select antenna orientation & Directional transmissions for multi-AP indoor scenario & Higher throughput \\ \cline{2-9}
          & \cite{wilhelmi2017implications} & RL    & QL    & 2017  & S     & Select channel and transmit power & Dynamic channel assignment in a fully decentralized scenario & Higher throughput, improved fairness \\ \cline{2-9}
          & \cite{wilhelmi2019potential} & RL    & MAB   & 2019  & S     & Select channel, transmit power and sensitivity threshold & Improve spatial reuse in dense and uncoordinated WLANs & Higher throughput, improved fairness \\ \cline{2-9}
          & \cite{wilhelmi2019collaborative} & RL    & MAB   & 2019  & S     & Select channel and transmit power & Assess new action-selection strategies vs \cite{wilhelmi2017implications} & Higher throughput than static approach \\ \cline{2-9}
          & \cite{lopez2019combining} & RL    & MAB   & 2019  & S     & Select channel and transmit power & Deploy spatial reuse methods in a SDN framework & Higher throughput, improved fairness \\ \cline{2-9}
          & \cite{zhao2019joint} & RL    & QL    & 2019  & S     & Select channel and transmit power & Apply QL with event-driven training & Higher throughput than three SOA methods \\ \cline{2-9}
          & \cite{ak2020fsc} & SL    & MLP   & 2020  & S     & Select carrier sensitivity threshold & Combine local and global operations & Higher throughput than five benchmarks \\ \cline{2-9}
          & \cite{yin2019learning} & RL    & QL    & 2020  & S     & Decide whether to transmit concurrently & Per-frame adaptive CCA & Higher throughput \\ \cline{2-9}
          & \cite{bardou2021improving} & RL    & MAB   & 2021  & S     & Select transmit power and sensitivity threshold & Solve dimensionality problem by subsampling state space & Higher throughput and fairness than three benchmarks \\ %
\midrule
    \multicolumn{1}{c}{\multirow{16}[0]{*}{\shortstack{Channel bonding \\ (Section \ref{sec_channel_bonding})}}} & \cite{ayush2017supporting} & RL    & MAB   & 2017  & T+E   & Identify stations under starvation & Investigate fairness of dynamic bandwidth channel access & Higher throughput, improved fairness \\ \cline{2-9}
          & \cite{karmakar2019learning} & RL    & SARSA & 2019  & E     & Select channel bandwidth & Packet scheduler supporting dynamic bandwidth allocation and QoS & Higher throughput, lower packet losses and delay than two benchmarks \\ \cline{2-9}
          & \cite{khan2019interactive} & RL    & MAB   & 2019  & T+S   & Select channel and bandwidth & Consider non-continuous channels & Stations reach a Nash equilibrium \\ \cline{2-9}
          & \cite{qi2020on-demand} & RL    & DQN   & 2020  & S     & Select channel and bandwidth & Apply DL to channel bonding problem & Lower latency than three benchmarks \\ \cline{2-9}
          & \cite{luo2020learning} & RL    & DQN   & 2020  & S     & Select channel and bandwidth & Consider spatio-temporal changes in traffic demands & Higher user satisfaction than in five benchmarks \\ \cline{2-9}
          & \cite{karmakar2020smartbond} & RL    & MAB   & 2020  & E     & Select channel and bandwidth & Consider hidden stations & Higher throughput than three benchmarks \\ \cline{2-9}
          & \cite{han2020deep} & RL    & DNN   & 2020  & S     & Select channel and bandwidth & Consider per-channel traffic load & Higher throughput than four heuristics \\ \cline{2-9}
          & \cite{kanemasa2021dynamic} & RL    & MAB   & 2021  & E     & Select channel and bandwidth & Use laser oscillations to guide lasers to action space exploration & Higher throughput than two benchmarks \\ \cline{2-9}
          & \cite{wilhelmi2021machine} & SL    & NN    & 2021  & S     & Predict throughput & Consider impact of channel bonding & Performance comparison of five competing models \\ \cline{2-9}
          & \cite{soto2021atari} & SL    & GNN   & 2021  & S     & Predict throughput & Apply GNN to performance prediction & Better prediction than five benchmarks \\ \cline{2-9}
          & \cite{barrachina2021stateless} & RL    & MAB   & 2021  & S     & Select channel and bandwidth & Apply stateless RL to channel allocation & Higher throughput \\ \cline{2-9}
          & \cite{barrachina2021multi} & RL    & MAB   & 2021  & S     & Select channel and bandwidth & Propose general RL-based spectrum management framework & Higher throughput satisfaction and fairness than two benchmarks \\ %
\midrule
    \multicolumn{1}{c}{\multirow{7}[0]{*}{\shortstack{Multi-band,\\network MIMO,\\and full-duplex \\ (Section \ref{sec_multi-band})}}} & \cite{yano2019achievable} & SL    & DNN   & 2019  & S     & Predict channel idle state & Apply PNN to multi-band Wi-Fi network & Near-optimal throughput \\ \cline{2-9}
          & \cite{krishnan2019optimizing} & RL    & \shortstack{Monte-Carlo policy gradient,\\DDPG} & 2019  & S     & Select channel and cluster APs & Joint parameter optimization for distributed MU-MIMO networks & Higher throughput than two heuristics \\ \cline{2-9}
          & \cite{zhang2020pointer} & SL    & DNN   & 2020  & S     & Find pairs for concurrent transmissions & Apply ML to orchestrating full-duplex transmissions & Higher spectral efficiency than two heuristics \\ \cline{2-9}
          & \cite{wang2021intelligent} & SL    & DNN   & 2021  & S     & Select band for retransmission & Retransmissions solution in multi-band Wi-Fi network & Higher throughput, lower latency \\
\bottomrule

\end{xltabular}
\end{landscape}

\twocolumn

Once appropriate 802.11ad/ay beam sectors have been found, rate adaptation is required.
\Ac{MCS} selection for mmWave transmissions relies on appropriate channel classification, i.e., determining whether a channel is \ac{LoS} or \ac{NLoS}.
This classification is augmented with \ac{ML}, as shown by \textcite{kurniawan2017machine}, where classification is done with the \ac{RF} technique.
The prediction of statistical characteristics of a channel can also be useful and many papers focus on the \ac{PHY} layer (regardless of the wireless technology).
For example, \textcite{bai2018predicting} use a trained \ac{CNN} to predict the statistical characteristics of a channel for any given (indoor) location for technologies using massive \ac{MIMO}.

Alternatively, rate adaptation can be based on typical metrics available in \ac{COTS} devices. 
\textcite{aggarwal2020experimental} predict optimal \ac{MCS} settings using three ML models: \ac{DT}, \ac{RF}, and \ac{SVM}.
They conclude that \ac{RF} provides the best results and outperformes \ac{SNR}-based rate selection strategies.
This approach is extended in the \ac{LiBRA} framework \cite{aggarwal2020learning}, 
where the same \ac{ML}-based classification methods determine which of the two adaptation methods (rate selection or beam selection) gives better performance for a given link.

The data rate of \ac{mmWave} links can be improved by better channel estimation techniques. 
\textcite{lin2020location} combine transceiver location information with a \ac{DNN} to evaluate the channel frequency response. 
This approach decreases the number of transmitted pilot signals, leaving more room for user data.

Finally, in terms of channel access, 802.11ad introduces a new hybrid MAC with contention-free and contention-based periods.
The specifics of the resource scheduler are out of the standard scope and remains an open research challenge \cite{mohebi2020challenges}.
\textcite{azzino2020scheduling} find the optimal duration of the contention-free period by observing the time-varying network load and using an \ac{RL}-based approach. 
Their scheme preserves throughput for allocated streams while leaving more resources for contention-based traffic.

\subsection{Multi-user Communication}
\label{sec_multi-user}

With the IEEE 802.11ac amendment, and its support for downlink \ac{MU-MIMO} transmissions, Wi-Fi opened the door to support multi-user transmissions, i.e., simultaneously transmitting to different stations in the same \ac{TXOP} using spatial multiplexing. IEEE 802.11ax extends IEEE 802.11ac \ac{MU-MIMO} features and provides support to both downlink and uplink, as well as \ac{OFDMA}. \ac{OFDMA}, which could be considered as the most disruptive novelty introduced in IEEE 802.11ax, divides the available bandwidth into different sub-channels, called \acp{RU}, which is then allocated to different users. Both \ac{MU-MIMO} and \ac{OFDMA} will also play an important role in future IEEE 802.11be networks. In 802.11be, beyond extending Wi-Fi capabilities by using \SI{320}{\mega\hertz} channels and up to 16 spatial streams, some improvements such as the allocation of multiple \acp{RU} to the same user and support for implicit channel sounding will be introduced~\cite{lopez2019ieee}.

\begin{figure}
\centering
\includegraphics[width=\columnwidth]{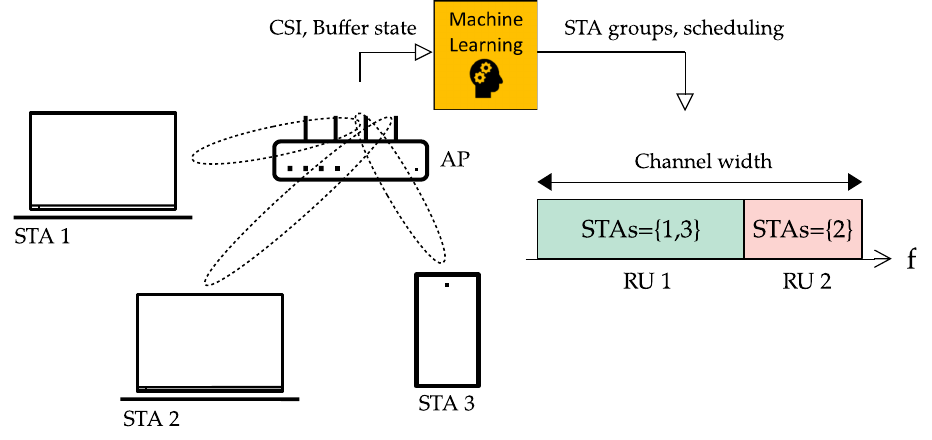}
\vspace{-1em}
\caption{The ML agent is in charge of finding groups of compatible stations and scheduling transmissions based on experience.}
\label{Fig:FigMU_ML}
\vspace{-.5em}
\end{figure}

The most significant challenge in multi-user communications is identifying and creating groups of compatible stations that, when simultaneously scheduled, result in higher network performance. The problem is both complex and non-linear so suitable to be tackled using \ac{ML} techniques due to the need to choose a particular group of stations and configure their link parameters with only partial information in a rapidly changing environment. \Cref{Fig:FigMU_ML} shows the case where an \ac{AP} empowered with an \ac{ML} agent is in charge of taking these scheduling decisions. First, it must learn that \ac{STA}~1 and \ac{STA}~3 can belong to the same \ac{MU-MIMO} group. Then, given the \ac{AP} has data to transmit to all three stations, the \ac{ML} agent has to decide how to allocate the different available \acp{RU} to them. In this example, it has agreed to allocate a larger \ac{RU} to \ac{STA}~1 and \ac{STA}~3 for a \ac{MU-MIMO} transmission, and a smaller one to \ac{STA}~2.

Several papers address the problems of user selection, link adaptation, and channel sounding overhead reduction in \ac{MU-MIMO}-enabled \acp{WLAN} using a variety of \ac{ML} strategies. \textcite{karmakar2019intelligent} implement an $\varepsilon$-greedy strategy to find the best configuration (group and link parameters) using experience. 
\textcite{rico2014learning} use an \ac{SVM} classifier to develop a robust \ac{MCS} selection procedure. Reducing the channel sounding overheads using \acp{DNN} to compress \ac{CSI} at each station and decompress \ac{CSI} at the \ac{AP} is presented by \textcite{sangdeh2020lb}. 
Finally, a different approach is considered by Su at al. \cite{su2019client,su2021data}, where a policy gradient technique determines if a certain client will benefit from participating in \ac{MU-MIMO} transmissions. The policy function is represented by a neural network consisting of two convolutional layers. In all analyzed cases, results show significant network throughput improvements.

For \ac{OFDMA}, in the case of \ac{AP}-initiated transmissions, either in downlink or uplink, the AP must determine the group of stations scheduled at each \ac{TXOP}, and which is the best \ac{RU} allocation to them. Alternatively, in the case of uplink transmissions, stations may be allowed to select the \ac{RU} that they will use for transmission \cite{kosek2022efficient}. These problems are considered using \ac{DRL} techniques \cite{kotagiri2020multi,kotagiri2021distributed, balakrishnan2019deep}. 
Kotagiri et al. \cite{kotagiri2020multi,kotagiri2021distributed} focus on the uplink case and use a decentralized \ac{RU} selection method with \ac{DRL} (i.e., a \ac{CNN}-based \ac{DQN}) that provides higher gains when compared against the case when \acp{RU} are selected randomly. 
\textcite{balakrishnan2019deep} consider the opposite case, i.e., only \ac{AP}-initiated downlink transmissions, using \ac{DRL}-based scheduling . 
Per-station channel quality and traffic information are the inputs for different objective policies. Results confirm the potential of \ac{ML} for scheduling in \ac{OFDMA} systems. 
\textcite{kotera2021lyapunov} use a deep deterministic policy gradient (DDPG) algorithm to solve the OFDMA resource allocation problem formulated as a \ac{MDP} using Lyapunov optimization. 
Results show that this solution can meet the system latency requirements in situations where other baseline solutions fail, while also improving fairness.

Additionally, \textcite{sangdeh2021deepmux} address joint MU-MIMO and OFDMA optimization by using \ac{DSL}. The solution, called DeepMux, is executed at the APs and relies on \acp{DNN} to minimize the impact of channel sounding and find a near-optimal resource allocation policy. Experimental results show gains of up to 50\% in throughput using DeepMux.

Importantly, \ac{OFDMA}-based channel access is a common feature for both Wi-Fi and \ac{5G} and it was adopted in \acp{WLAN} after it was successfully applied in the cellular domain. In the following papers, \ac{ML} addresses several problems in \ac{OFDMA}-based cellular networks: fair scheduling \cite{comsa2019comparison, comcsa2019enhancing}, \ac{CFO} estimation for uplink transmissions \cite{li2018carrier, li2019unsupervised}, inter-network interference control \cite{galindo2010distributed}, and resource allocation \cite{yang2010resource}. 
These works implement \ac{RL} \cite{comsa2019comparison, comcsa2019enhancing, galindo2010distributed}, supervised deep learning \cite{li2018carrier}, unsupervised deep learning \cite{li2019unsupervised}, and a genetic learning algorithm \cite{yang2010resource} to support performance optimization. We believe that these papers may provide interesting insights and guidelines for researchers working in the \mbox{Wi-Fi} domain.

\subsection{Spatial Reuse}
\label{sec_spatial_reuse}

The IEEE 802.11ax amendment first introduced~\ac{SR} to Wi-Fi networks \cite{wilhelmi2021spatial}. The main objective of this mechanism is to support concurrent transmissions between devices that belong to different~\acp{BSS}. When a device detects an ongoing transmission, it must first decide whether another concurrent transmission is possible, and in case it is, which transmission power to use to avoid disrupting the ongoing one.  
IEEE 802.11ax \ac{SR} offers good performance gains, despite its conservative rule-based design. In such a context, \ac{ML} techniques make such a mechanism adaptive to different scenarios, and decide when and how a device detecting an ongoing transmission can benefit from a spatial reuse opportunity, should result in even higher throughput and latency gains. It is expected that IEEE 802.11be will further extend Wi-Fi \ac{SR} capabilities by allowing neighboring APs to coordinate their transmissions. \ac{ML} techniques can contribute to improve coordinated \ac{SR} by solving the challenge of identifying which devices can transmit at the same time by combining the collected \ac{CSI} information from different devices.

\begin{figure}
\centering
\includegraphics[width=\columnwidth]{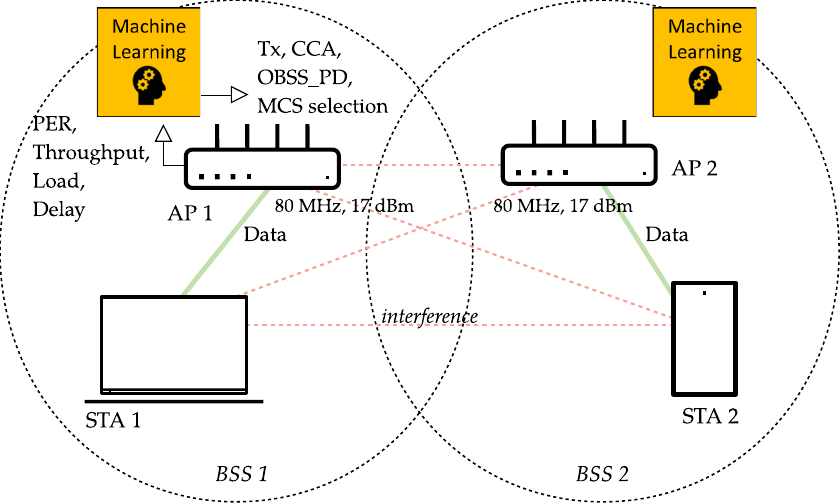}
\vspace{-1em}
\caption{An \ac{AP} empowered with \ac{ML} can learn from experience which is the best \ac{SR} configuration to maximize its own, or the overall network, performance.}
\label{Fig:FigSR_ML}
\vspace{-.5em}

\end{figure}

\Cref{Fig:FigSR_ML} shows the case of two neighboring \acp{AP} that, empowered by \ac{ML} agents, find a suitable configuration  that gives them the highest possible throughput while sharing spectrum resources. In this case, we assume both prefer to use the same 80~MHz channel but transmitting at lesser power. With this configuration, the \acp{AP} maximize mutual spatial reuse opportunities in front of other options such as using non-overlapping channels (less bandwidth) or transmitting at high power (with higher \acp{MCS}) but causing the other \ac{BSS} to defer.

The use of \ac{ML} solutions to tackle the \ac{SR} problem has raised some attention in recent years. Most of the works implement \ac{RL} techniques for learning the best configuration for each \ac{ML} agent-empowered \ac{AP} online. 
Popular methods include Q-learning  \cite{timmers2009spatial,yin2019learning} and \acp{MAB}  \cite{wilhelmi2017implications,wilhelmi2019collaborative,wilhelmi2019potential,lopez2019combining,nguyen2016enhancing}. 
All these papers share the concept of multiple agents that either do not share information or they share it only partially (i.e., the action performed and the obtained reward) and learn by interacting through the environment. 
The results show that in multi-agent scenarios where the agents compete with each other without collaborating, convergence may be hard or impossible to achieve. There are also papers using \ac{SL} techniques, such as \acp{NN} \cite{ak2020fsc,jamil2016novel}, to  select proper \ac{SR} parameters (transmission power and sensitivity levels) given that the characteristics of the scenario are known.

In the following, we overview some of these papers, as they are illustrative to understand how \ac{ML} can improve \ac{SR} operation in \mbox{Wi-Fi}.
\textcite{timmers2009spatial} use a Q-learning algorithm to optimize power, transmission rate, and \ac{CCA}. States are defined as a combination of transmission power, interference, and the \ac{MCS} used, and actions consist of changing the transmission power and \ac{MCS}. Agents are placed at every device and act selfishly. 
\textcite{yin2019learning} use Q-learning to improve 802.11ax's \ac{SR} mechanism: the agent learns the best decision (i.e., transmit concurrently or wait) assuming knowledge of current interferers. 
For non-stationary scenarios, the learning rate of the less frequently chosen actions is increased to ensure rapid adaptation to environmental changes.

Wilhelmi et al. address the problem of channel selection and transmission power allocation with a stateless Q-learning solution \cite{wilhelmi2017implications} and different \acp{MAB} action-selection strategies ($\varepsilon$-greedy, EXP3 \cite{auer1995gambling}, \ac{UCB} \cite{auer2002finite}, and \ac{TS} \cite{thompson1933likelihood})   \cite{wilhelmi2019collaborative}. 
Two approaches are also examined: 1) concurrent -- all networks take actions simultaneously, and 2) sequential -- only one network changes its configuration at a time. Results show that optimal proportional fairness is achieved even if the different networks operate selfishly (i.e., they aim to maximize their throughput) without sharing information. 
Meanwhile, sequential action taking between actors reduces the throughput variability at the different \acp{BSS}. 
However, this comes at the expense of lower throughput values. 
\textcite{wilhelmi2019potential} also use \acp{MAB} for improving decentralized \ac{SR} decisions. 
If the different \ac{ML} agents can communicate and share the performance obtained when playing a certain action, it is possible to apply utility functions in the online optimization process that directly target network fairness, such as max-min, effectively reducing cases where some \acp{BSS} are starved due to the selfish operation of others. 
Meanwhile, \textcite{bardou2021improving} also consider \acp{MAB} in a centralized solution to dynamically change spatial reuse parameters. 
The reward function prevents starvation by using \ac{TS} to select the best configuration. 
The dimensionality problem is solved by subsampling the state space. Simulation results show the ability of this solution to improve the performance of dense WLANs with multiple interacting BSSs.

Supervised learning techniques such as \ac{MLP} and \acp{DT} are considered by \textcite{ak2020fsc} to select \ac{SR} parameters at both the \ac{AP} and stations. The models are trained offline using a dataset that covers multiple scenarios and configurations. A different approach is considered by \textcite{jamil2016novel}, where a centralized \ac{NN} configures all \acp{BSS} so that spatial reuse is maximized. The \ac{NN} considers the correlation function between the throughput achieved by the different devices in the network and their associated link layer parameters. 

Interference can also be mitigated by jointly optimizing the transmitted power of \acp{AP} and the channel allocation policies~\cite{zhao2019joint}.
A Q-learning model maximizes throughput in dense \acp{WLAN}.
The model is trained through a learning process of reduced total iterations driven by an event-triggered mechanism. Whenever the network status changes due to the mobility of users, the learning process is called again to optimize power and channel allocation policies.
Results are derived based on the deployment of \num{15} \acp{AP}, where a \SI{16}{\percent} throughput improvement is obtained in comparison to \ac{SoA} power and channel allocation mechanisms.

Lastly, a completely different approach to achieve \ac{SR} with directional transmissions is taken by \textcite{nguyen2016enhancing}. 
The selection of the antenna orientation is tackled as a non-stationary \ac{MAB} problem. 
Results from an \ac{SDR} implementation show the correct operation and resilience to co-channel interference.

\subsection{Channel Bonding}
\label{sec_channel_bonding}

The option to enable channels wider than 20~MHz was introduced in IEEE 802.11n, where up to \SI{40}{\mega\hertz} channels were supported. The IEEE 802.11ac and IEEE 802.11ax amendments further increased the maximum channel width to \SIlist{80;160}{\mega\hertz}, respectively. IEEE 802.11be will continue to increase the channel width, with up to \SI{320}{\mega\hertz} channels. Wider channels allow higher transmission rates and therefore higher performance. However, in dense scenarios, it may notably increase contention between neighboring \acp{BSS}, which may cause the opposite result. Therefore, correctly deciding when to use a wider channel, what should be its size, and which particular channels to use is necessary for successfully improving WLAN performance. Unfortunately, there is no single answer to the previous question but rather it depends on each specific scenario, including the number and position of contending devices, the load of each BSS, and the available channels. 

\begin{figure}
\centering
\includegraphics[width=\columnwidth]{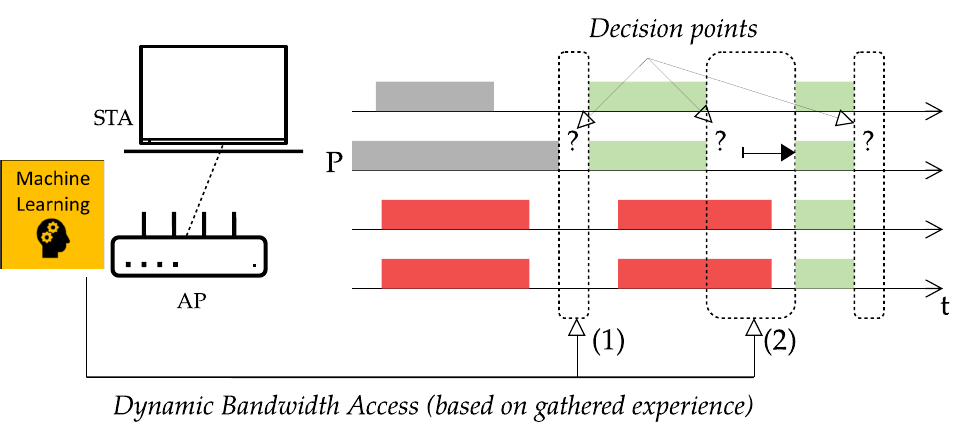}
\vspace{-1em}
\caption{An ML-enabled AP learns which is the best set of actions that maximize its performance. The primary channel is identified as P. \textcolor{black}{Grey rectangles illustrate idle channels, red -- busy channels, green -- AP transmissions.}}
\label{Fig:FigML_CB}
\vspace{-.5em}
\end{figure}

\ac{ML} techniques can solve such a situation by learning the best channel allocation and bonding configurations in a given scenario. Online learning seems a natural option in this case, especially if RL techniques and prediction models are combined to foster a rapid convergence \cite{barrachina2021stateless}. For example, \cref{Fig:FigML_CB} shows the case where an agent learns from experience which actions to perform given that the environment is found in a particular state (i.e., the state may be defined by the occupancy of the different \SI{20}{\mega\hertz} channels) every time the primary channel becomes idle. In this case, the agent has learned that the best action when all four \SI{20}{\mega\hertz} channels are idle is to transmit in the first \SI{40}{\mega\hertz} primary channel, but not in the secondary \SI{40}{\mega\hertz} channel. Similarly, when the secondary \SI{40}{\mega\hertz} channel is busy, the AP has learned the best action is to wait until the \SI{40}{\mega\hertz} secondary channel becomes idle to perform an \SI{80}{\mega\hertz} transmission. 

Out-of-the-box \acp{MAB} mainly decide which are the best channel widths to be used when no further information, neither from the network nor from user requirements, is considered. The goal is to maximize WLAN performance \cite{karmakar2019learning,khan2019interactive,kanemasa2021dynamic}. When traffic loads and other performance metrics are considered, such as delay and throughput, \ac{DRL} techniques are successfully applied \cite{qi2020on-demand,luo2020learning}.

\textcite{karmakar2019learning} show that the default dynamic channel bonding operation is improved by considering the individual needs of each station, as well as the \ac{AC} they are using, selecting the most appropriate channel widths to use. With that goal in mind, a \ac{MAB} algorithm, \ac{UCB}, learns when the use of secondary channels is required. Testbed results show that this solution provides gains higher than 100\% in some cases. Similarly, \textcite{khan2019interactive} apply learning from a trial and error perspective (i.e., exploring) which are the best channels and bonding strategies to use, including both contiguous and non-contiguous 20~MHz channels. The mechanism, called \ac{ITE}, includes different states depending on both the actions taken and the reward obtained. Exploration is implemented in \ac{ITE} using an $\varepsilon$-greedy strategy. The mechanism is implemented in WARP nodes. Results show that \ac{ITE} outperforms the default $\varepsilon$-greedy mechanism and improves the performance of \ac{SBCA} and \ac{DBCA} thanks to its availability to select the channel width properly. Lastly,  \textcite{ayush2017supporting} introduce \ac{HA-DBCA} to solve the starvation problem that affects some \ac{DBCA} devices. 

\ac{HA-DBCA} uses a polling-based adaptive mechanism for contention-free access and \ac{UCB} to identify the stations that are starving, and so allow them to transmit their data during the contention-free access. The channel bonding problem is also modelled as a MAB by \textcite{kanemasa2021dynamic}: chaotically oscillating waveforms generated by semiconductor lasers guide the exploration of the different available actions. Then, dynamically adapting the different thresholds used to select one or another action based on the amplitude of the generated waveform at sampling instants shows that such a technique can outperform default MABs such as UCB and $\varepsilon$-greedy in terms of throughput. Finally, \textcite{barrachina2021multi} justify model-free \ac{RL} techniques to address the channel bonding problem, design a complete \ac{RL} framework and call into question whether complex \ac{RL} algorithms allow rapid learning in realistic scenarios. 
Through extensive simulations, results show that a stateless \ac{RL} in the form of lightweight \acp{MAB} is an efficient solution for rapid adaptation, avoiding the definition of broad and/or meaningless states. In summary, lightweight MABs are an appropriate alternative to the cumbersome and slowly convergent methods such as Q-learning, and especially, deep reinforcement learning.

\Ac{DRL} is considered for configuring channel bonding. 
\textcite{qi2020on-demand} address the channel allocation problem (i.e., group of selected channels and position of the primary channel) in a scenario with multiple BSSs. 
The channel allocated to each BSS should depend on its expected load and performance. Then, considering the goal of minimizing latency, a \ac{DCB} algorithm that uses \ac{DRL}, along with a \ac{MADDPG} for training, to find suitable channel allocations. Results show that by reducing the channel width in \acp{AP} with low traffic demands, the delay in the overall network is improved as the channel access contention is reduced. A similar problem is considered by \textcite{luo2020learning}, where \ac{DRL} tackles the channel assignment problem in \acp{WLAN} with channel bonding while considering spatio-temporal changes in traffic demands. Therefore, the \ac{DRL} solution (i.e., a \ac{DQN})  learns to adapt to offer satisfactory service. The agent in each AP learns from historical traffic loads when more or fewer channels should be bonded together, trying to minimize the interactions with other \acp{BSS} when not required. 
An opportunistic contiguous and non-contiguous channel aggregation scheme for 802.11ax WLANs is presented by \textcite{han2020deep}. 
Since the default strategy of aggregating all available channels may degrade network performance due to the inter-WLAN contention,
an efficient probabilistic channel aggregation scheme should consider the traffic load of secondary channels. 
To adjust aggregation probabilities of the secondary channels, a \ac{DRL} strategy is used. Results confirm that this strategy outperforms others based on predefined rules such as aggregating all channels and aggregating one or two channels randomly selected.

The problem of throughput prediction in dense WLANs supporting channel bonding is considered by \textcite{wilhelmi2021machine}, where several predictors are built using \ac{SL} techniques that include \acp{ANN}, \acp{GNN}, \ac{RF} regression, and gradient boosting. Both training and validation are performed on an open dataset generated using the IEEE 802.11ax-oriented Komondor network simulator \cite{barrachina2019komondor}. While the accuracy achieved by the methods demonstrates the suitability of ML for predicting the throughput performance of complex WLANs, more importantly, this work can be easily extended by considering other approaches. The same dataset is used by \textcite{soto2021atari} to predict Wi-Fi performance using a \ac{GNN} model that incorporates the deployment's topology information. 
Finally, the problem of collisions with hidden stations when channel bonding is used is described by \textcite{karmakar2020smartbond}. 
APs use a recursive neural network, namely a \ac{MH-GAN}, to predict the activity of neighbouring \acp{BSS}. Results confirm that the presented solution, called Smart Bond, can reduce the probability of suffering transmission errors due to hidden stations.

\subsection{Multi-link Operation, Network MIMO, and Full-duplex}
\label{sec_multi-band}

\Ac{ML} techniques also improve the operation of a wide variety of advanced mechanisms that include multi-band WLAN operation \cite{yano2019achievable}, multi-AP coordination for network MIMO \cite{krishnan2019optimizing}, and in-band full-duplex \cite{zhang2020pointer}. Both \ac{RL} and \ac{SL} techniques are used. For example, \ac{DRL} considers both channel allocation and \ac{AP} clustering to maximize the performance of distributed MIMO transmissions \cite{krishnan2019optimizing}. Similarly, \acp{NN} predict channel states and so improve the performance of multi-band WLANs \cite{yano2019achievable} and to find groups of stations that enable full-duplex communication at the \acp{AP} \cite{zhang2020pointer}. In the following, we overview these solutions in more detail.

\begin{figure}
\centering
\includegraphics[width=\columnwidth]{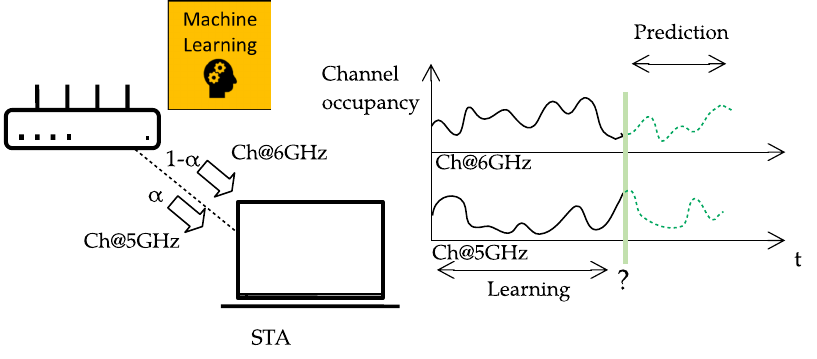}
\vspace{-1em}
\caption{An ML-enabled multi-band AP determines the best traffic balancing policy based on predicting channel occupancy values. In the figure, $\alpha$ represents the fraction of the traffic directed to the 5~GHz interface.}
\label{Fig:FigML_MLO}
\end{figure}

\Cref{Fig:FigML_MLO} shows the case where an ML-enabled multi-band AP has to decide how to distribute traffic to a given station between the 5~GHz and 6~GHz bands. The agent predicts channel occupancy values on both bands to decide how to use the two interfaces better. For instance, if the occupancy predicted at the 5~GHz band is high, it may choose to turn off such an interface and use only the one working at the 6~GHz band, thus saving energy.

\subsubsection{Multi-link Operation}
Multi-link operation enables a WLAN device to simultaneously tranmit over multiple interfaces on the same or different bands. 
This feature is currently under development in the IEEE 802.11be task group.
In the synchronous version, when any of the active backoff instances reach zero for a given interface, the state of the other interfaces is checked, and those idle are bonded together to support the subsequent transmission. However, if other interfaces are currently busy but some may become idle soon, it may be more efficient to wait and then aggregate these links instead of immediately transmitting using only a single interface. \textcite{yano2019achievable} solve this uncertainty by learning and predicting when a given interface will become idle using a \ac{PNN}. Multi-band operation is also considered by \textcite{wang2021intelligent} in combination with HARQ to improve packet retransmissions' efficiency. \ac{SL} decides if a packet retransmission should use the same band as the previous transmission, a different one, or all available bands simultaneously by sending multiple copies. A K-means algorithm and a \ac{DNN} discern which is the best operation mode. Results show that a more efficient network utilization is obtained, leading to higher throughput values.

\subsubsection{Network MIMO}
\textcite{krishnan2019optimizing} consider the joint problem of channel allocation and AP clustering in distributed \ac{MU-MIMO} for Wi-Fi networks. \ac{DRL} solves these two problems by maximizing per-user throughput. Since both underlying problems are NP-hard, only heuristic solutions exist in the literature. The \ac{DRL} framework consists of an agent, implemented using a \ac{DNN}, and a distributed MIMO Wi-Fi simulator. Although not explicitly specified, the solution is implemented in a central controller. Results show that using the \ac{DRL} framework, a 20\% improvement in user throughput is achieved. Also, the \ac{DRL} framework can attain multiple objectives, such as maximizing throughput and fairness simultaneously. 

\subsubsection{Full-duplex Communication}
Full-duplex (in-band) communication allows a device to transmit and receive simultaneously, thus `doubling' the channel capacity. In WLANs, a key challenge to solve is the user pairing problem: finding groups of different stations that allow the AP to transmit to one while receiving from another. To solve this combinatorial problem, which becomes impractical when the number of stations is high, \textcite{zhang2020pointer} use a \ac{DSL} architecture introduced in \cite{vinyals2015pointer}. The main benefit of this solution is that the \ac{NN} does not need to be re-trained when the length of input (e.g., the number of users) changes within an expected range. 
Results  confirm that the \ac{DSL}-based solution outperforms two  low-complexity methods called greedy assignment and random assignment.

\subsection{Open Challenges}

This section has covered recent and advanced Wi-Fi features such as beamforming, multi-user communications, channel bonding, spatial reuse, and multi-band.
Although quite different, in all of them, ML techniques are used mainly either:
\begin{itemize}
    \item to adapt to the environment through selecting the most proper actions at the right moment,
    \item for system-level performance predictions, or
    \item to improve the operation of specific mechanisms by completing unavailable data.
\end{itemize}

Since most of these features are recent, complex, and in development, many aspects are still not considered or considered only superficially. Therefore, there is room for future work in this area either by addressing the problems listed in previous subsections with different ML techniques or by simply picking some of the still uncovered aspects. In the following, we detail some open aspects in the different categories.

First, the success of using \emph{beamforming} in indoor Wi-Fi scenarios will be based on the ability to properly perform beam sector alignment (\Cref{fig_beamforming}). 
Research has shown that \ac{ML} methods can have a positive impact, but robust solutions available for \ac{COTS} devices are required, e.g., to minimize latency \cite{aldubaikhy2020mmwave}.
For outdoor scenarios, beamforming-aware resource allocation (intra-\ac{AP}) and resource coordination (inter-\ac{AP}) methods based on \ac{ML} need to be updated to the recently released 802.11ay amendment, where \ac{FWA} is an important use case, which has so far not been researched in depth.
	
In the area of \emph{multi-user communication}, more works focusing on ML solutions for allocating spatial streams and RUs to active stations are required, especially when mixed with realistic traffic patterns and \ac{QoS} requirements. 
In terms of future traffic estimates, contending devices and environmental conditions may improve the Wi-Fi response to sensitive traffic, improving criteria such as worst-case latency by pre-reserving resources. Moreover,  predictions using ML techniques can improve how channel sounding is implemented, as only stations that are likely to be scheduled will be requested to provide such information. 
	
Many works in the area of \emph{spatial reuse} have considered BSSs operating in a completely decentralized way, so using a spatial reuse opportunity depends only on each individual's observed inputs. This situation justifies that many papers have considered ML techniques such as \acp{MAB} or Q-learning to infer which is the best action in a particular situation. However, with IEEE 802.11be, \ac{TXOP} sharing and cooperative schemes may be enforced, thus requiring a different approach, augmented with \ac{ML} techniques to optimize its operation.
	
The case of \emph{channel bonding} has been addressed using \ac{RL} and \ac{SL}. Both techniques capture the interactions between BSSs that appear when channel widths change dynamically. Further work is required to test and compare these results with each other. Furthermore, an exciting aspect is to couple channel aggregation techniques with \ac{OFDMA} \ac{RU} allocation, for which complex \ac{DRL} techniques may be well suited. 
	
Finally, a disruptive new feature introduced by IEEE 802.11be is \emph{multi-link operation}. This will open several exciting challenges, such as which channels to use and how to distribute the different flows between links. ML techniques can learn, for example, when is the best moment to perform a channel switch, which link occupancy patterns favor a particular traffic pattern, and how to allocate or distribute flows to links.

\onecolumn
\begin{landscape}
\scriptsize
\begin{xltabular}{\linewidth}{ccccccp{0.19\textwidth}*{2}{p{0.29\textwidth}}}
\caption{Summary of works on improving Wi-Fi management with ML. The evaluation methods are theoretical (E), simulation (S), and experimental (E). The ML improvement is in comparison to SoA methods or (if not mentioned otherwise) to IEEE 802.11.} \label{tab:management} \\
\toprule
\textbf{Area} & \textbf{Ref.} & \shortstack{\textbf{ML}\\\textbf{category}} & \shortstack{\textbf{ML}\\\textbf{mechanisms}} & \textbf{Year}&\shortstack{\textbf{Evaluation}\\\textbf{method}}  & \textbf{Application of ML} & \textbf{Novelty of approach} & \textbf{ML improvement} \\
\midrule
\endfirsthead

\multicolumn{9}{c}%
{\tablename\ \thetable{} -- continued from previous page} \\
\midrule \textbf{Area} & \textbf{Ref.} & \shortstack{\textbf{ML}\\\textbf{category}} & \shortstack{\textbf{ML}\\\textbf{mechanisms}} & \textbf{Year}&\shortstack{\textbf{Evaluation}\\\textbf{method}}  & \textbf{Application of ML} & \textbf{Novelty of approach} & \textbf{ML improvement} \\ \midrule
\endhead

\multicolumn{9}{r}{{Continued on next page}} \\
\endfoot

\endlastfoot

\multicolumn{1}{c}{\multirow{20}[27]{*}{\shortstack{Channel and\\band selection \\ (Section \ref{sec_channel_band_selection})}}} & \cite{liu2010cognitive} & SL    & MFNN  & 2010  & E     & Select channel & Apply traffic prediction methods for channel allocation & Higher throughput \\
\cmidrule{2-9}          & \cite{bojovic2011supervised} & SL    & RF    & 2011  & E     & Select AP for association & Apply SL for AP selection & Higher throughput than three benchmarks \\
\cmidrule{2-9}          & \cite{song2017leveraging} & SL    & LR, DT & 2017  & E     & Select AP for association & Use frame aggregation characteristics to infer expected throughput & Increased AP selection accuracy \\
\cmidrule{2-9}          & \cite{pei2017why} & SL    & RF    & 2017  & S     & Select AP for association & Consider connection establishment & Reduced connection failure, improved setup time \\
\cmidrule{2-9}  &   \cite{jeunen2018machine} & SL    & LASSO, OLS & 2018  & E     & Select channel & Avoid interfering devices & Increased QoE \\
\cmidrule{2-9}     & \cite{feltrin2018machine} & SL    & NN    & 2018  & S     & Handover decision & Monitor RSSI patterns for upcoming handovers & Low prediction error \\
\cmidrule{2-9}          & \cite{carrascosa2019decentralized} & RL    & MAB   & 2019  & S     & Select AP for association & Apply MAB for AP selection & Higher throughput \\
\cmidrule{2-9}          & \cite{kafi2019online} & RL    & QL    & 2019  & S     & Select AP for association & Centralized, SDN-based solution & Higher throughput than three benchmarks \\
\cmidrule{2-9}          & \cite{han2019artificial} & RL    & DQN   & 2019  & S     & Handover decision & Consider spatial and temporal characteristics of the wireless signal & Higher throughput than three benchmarks \\
\cmidrule{2-9}          & \cite{zeljkovic2019abraham} & SL    & LSTM  & 2019  & E     & Handover decision & Predict user location, AP load, and signal strength to preserve QoS during handovers & Increased QoS for mobile users \\
\cmidrule{2-9}          & \cite{nakashima2020deep} & RL    & GCN, DDQN & 2020  & S     & Select channel & Apply DL for channel allocation & Higher throughput than two benchmarks \\
\cmidrule{2-9}          & \cite{lopez2020concurrent} & RL    & MAB   & 2020  & S     & Select channel and AP & Joint channel allocation and AP selection & Higher channel utilization and fairness \\
\cmidrule{2-9}          & \cite{carrascosa2020multi} & RL    & MAB   & 2020  & S     & Select AP for association & Halt exploration of $\varepsilon$-greedy AP selection to reduce network dynamics & Higher throughput \\
\cmidrule{2-9}          & \cite{dinh2021distributed} & RL    & DQL   & 2021  & S     & Select APs for association & Apply DQN to multi-AP association & Higher aggregate throughput \\
\midrule
\multicolumn{1}{c}{\multirow{2}[4]{*}{\shortstack{Management\\architectures \\ (Section \ref{sec_management_arch})}}} & \cite{bast2019deep} & RL    & DQL   & 2019  & S     & Optimize slice configuration & Apply DRL to network slicing & Dynamically find optimum configuration \\
\cmidrule{2-9}          & \cite{lyu2019largescale} & SL    & RF    & 2019  & E     & Predict AP load & Dynamically enable APs in large-scale network based on load predictions & Reduced power consumption \\
\midrule
\multicolumn{1}{c}{\multirow{8}[6]{*}{\shortstack{Traffic prediction \\ (Section \ref{sec_traffic})}}} & \cite{feng2006svm} & SL    & SVM   & 2006  & E     & Predict traffic intensity & Apply SVM to Wi-Fi traffic prediction & Better prediction than three SOA methods \\
\cmidrule{2-9}          & \cite{thapaliya2018predicting} & SL, USL & SVR, EM & 2017  & E     & Predict network congestion & Apply both regression and clustering to predict congestion & Provide expected level of congestion \\
\cmidrule{2-9}          & \cite{khan2020real} & SL    & MLP, SVR, DT, DF & 2020  & E     & Predict transmission throughput & Apply ML for Wi-Fi throughput & Better prediction performance than four SOA methods \\
\cmidrule{2-9}          & \cite{chen2021flag} & SL    & DRNN  & 2021  & E     & Predict user load & Apply DRRN with separate RNNs (encoder and decoder) for AP-level user load prediction & Better prediction than two SOA methods \\
\midrule
\multicolumn{9}{c}{\multirow{6}{*}} \\ \\ \\ \\ \\ \\ 
\multicolumn{1}{c}{\multirow{10}[11]{*}{\shortstack{Predicting\\health of\\Wi-Fi connections \\ (Section \ref{sec_predicting_health})}}} & \cite{gallo2018widia} & SL    & DT    & 2018  & S+E   & Classify Wi-Fi issues & Provide real-time diagnostics for two Wi-Fi issues & High classification accuracy \\
\cmidrule{2-9}          & \cite{syrigos2019employment} & SL, USL & DT, RF, SVM, kNN & 2019  & E     & Classify Wi-Fi issues & Provide real-time diagnostics for five Wi-Fi issues & High classification accuracy (for kNN) \\
\cmidrule{2-9}          & \cite{trivedi2020winetsense} & SL    & kNN, NB & 2020  & E     & Classify and predict link status & Provide sensing and analysis framework for user-centric decisions & High classification and prediction accuracy \\
\cmidrule{2-9}          & \cite{allahdadi2021hidden} & USL   & SOHMMM & 2021  & S     & Classify Wi-Fi issues & Provide real-time diagnostics for four Wi-Fi issues & Higher classification accuracy than two benchmarks \\
\cmidrule{2-9}          & \cite{morshedi2021estimating} & SL    & LMT, MLP, NBT, REPT & 2021  & E     & Estimate QoE of video streaming & Apply ML to estimate QoE by monitoring 802.11 parameters only & High estimation accuracy (for LMT) \\
\cmidrule{2-9}          & \cite{morshedi2021conferencing} & SL    & LMT, MLP, NBT, REPT & 2021  & E     & Estimate QoE of video conferencing & Apply ML to estimate QoE by monitoring 802.11 parameters only & High estimation accuracy (for REPT) \\
\cmidrule{2-9}          & \cite{scanzio2022predicting} & SL    & ANN   & 2022  & E     & Predict channel gain & Apply ML to data gathered through periodic probing & Higher prediction accuracy than benchmark \\

\bottomrule

\end{xltabular}
\end{landscape}

\twocolumn

\section{Connectivity Management}
\label{sec_management}

Connectivity management is an important task in Wi-Fi networks that includes, among others, channel allocation, band selection, and AP selection. The task is complex and challenging as the configuration change of a single link affects not only its performance but  often (e.g., in densely deployed networks) the performance of all neighboring networks. In this section, we present ML-based approaches for solving the connectivity management subtasks. Moreover, we cover also ML-based approaches for predicting future traffic load as well as the health of Wi-Fi link connections. These techniques allow to prepare and update the network configuration in advance (e.g., before a rapid change of communication conditions), which helps to minimize outage probability and improve user QoE. Table \ref{tab:management} presents a summary of works augmenting Wi-Fi with \ac{ML} in this area, which we present next.

\subsection{Channel and Band Selection}
\label{sec_channel_band_selection}

Channel allocation is an important problem in dense Wi-Fi networks, where a limited set of available channels has to be shared by a large number of co-located Wi-Fi BSSs. Poor channel allocation causes substantial contention among the APs and stations, hence reducing the throughput of each station. Typically, in the proposed solutions, the research goal is to assign channels in a way that
\begin{itemize}
    \item the \acp{AP} using the same channel do not interfere with each other (e.g., they are out of each other's interference range) and/or
    \item highly loaded \acp{AP} are not allocated the same channel  (i.e., a form of load balancing).
\end{itemize}   
Note that in the case of variable traffic load, channel allocation has to be performed periodically.

\begin{figure}
\centering
\includegraphics[width=0.7\columnwidth]{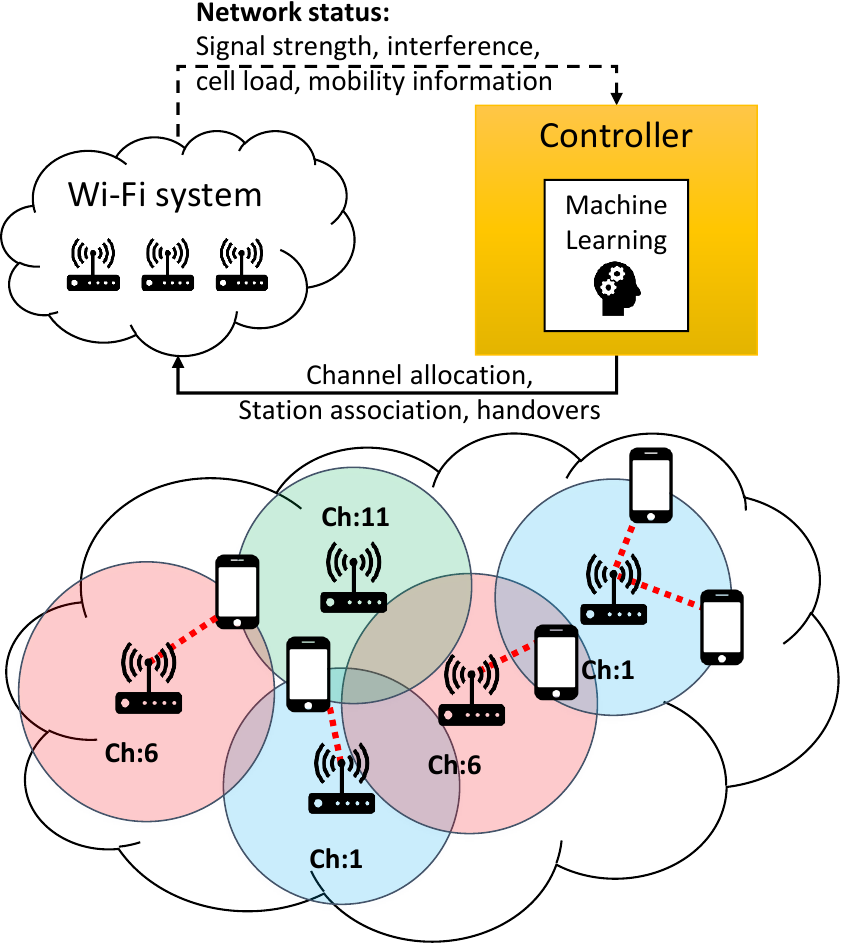}
\caption{An ML-based network controller determines channel allocation, station association, and handovers. Dashed arrows represent observations, solid lines represent actions.}
\label{fig_VC_1}
\end{figure}

As depicted in~\Cref{fig_VC_1}, \Ac{ML}-based algorithms can solve the problem of channel and band selection. They provide models that may consider changing interference relations (e.g., due to station mobility) and variable traffic loads (e.g., as a result of stations becoming active or passive).

\textcite{nakashima2020deep} use a \ac{DRL}-based channel allocation scheme to maximize throughput in densely deployed multi-\ac{BSS} WLANs. 
A central controller is aware of the global system state and able to control all APs. 
The interactions between \acp{AP}, under a certain channel allocation, are represented through contention graphs (i.e., channel adjacency matrices) to extract the features of carrier sensing relationships among the APs (i.e., topology information) using \acp{GCN}. The learning algorithm is \ac{DDQN} with a dueling network and prioritized experience replay. In addition, to prevent overfitting, selective observation data buffering is used, i.e., experiences are filtered to reduce the duplication of data for learning, which can often adversely influence the generalization performance. The simulation results demonstrate that the method enables the allocation of channels in densely deployed WLANs such that the system throughput increases.

\textcite{jeunen2018machine} introduce a framework able to passively monitor dense Wi-Fi environments, compute overlapping airtime periods, and detect networks which are the main cause of performance degradation in a \ac{WLAN}. A centralized (\ac{SDN}-based) network architecture is assumed. To implement the framework, different \ac{ML} techniques are used, e.g., \ac{LASSO}, \ac{OLS}. The extraction and ranking of relevant features from the gathered data is done using, e.g., \ac{LPA} and \ac{GNA}. Results show that the presented framework can find a new channel allocation that solves the interference problems. 

\subsubsection{AP Selection and Association}
\label{sec_ap_selection}

The proliferation and densification of Wi-Fi networks often lead to the existence of multiple spatially overlapping Wi-Fi cells. Hence, a station has to choose which of the discovered \acp{AP} to connect to.
The association method of 802.11 has stations select the \ac{AP} that provides the strongest signal. Unfortunately, in many cases, this simple approach leads to the under-utilization of some \acp{AP} while overcrowding others.
Consequently, \ac{AP} selection and load balancing approaches have been extensively studied as a way to improve network throughput.
For example, Carracscosa et  al. \cite{carrascosa2019decentralized,carrascosa2020multi} use a decentralized \ac{AP} selection procedure where stations employ an \ac{MAB}-based approach to dynamically learn the optimal mapping between \acp{AP} and stations. This procedure distributes the stations evenly among the available \acp{AP}. Specifically, each station independently explores the different \acp{AP} inside its coverage range and selects the one that better satisfies its needs. A novel opportunistic $\varepsilon$-greedy approach with stickiness halts the exploration when a suitable \ac{AP} is found. Then, the station remains associated to that same \ac{AP} while it is satisfied, only resuming the exploration after several unsatisfactory association periods. Results show that this approach increases the number of satisfied stations and the aggregated network throughput by up to 80\% in the case of dense \ac{AP} deployments.

Similarly, \textcite{lopez2020concurrent} study \ac{MAB}-based solutions for the decentralized channel allocation and \ac{AP} selection problems in enterprise WLAN scenarios. \acp{AP} and stations use agents that, through a Thompson sampling algorithm, explore and learn:
\begin{itemize}
    \item at the AP side: which is the best channel to use, and
    \item at the station side: which is the best \ac{AP} to associate with.
\end{itemize}  
Results from a custom-built simulator, called Neko\footnote{https://github.com/wn-upf/Neko}, show that the learning-based approach outperforms the static one, regardless of the network density and  traffic requirements. 

\textcite{bojovic2011supervised} propose a cognitive \ac{AP} selection scheme, where a station selects an \ac{AP} that is expected to yield the best throughput according to past experienced performance. The scheme belongs to the \ac{SL} family and uses a \ac{MFNN} to learn the correlation between the observed environmental condition (e.g., \ac{SNR}, probability of failure, beacon delay) and the obtained performance (i.e., throughput). The results from an 802.11 testbed show that the approach effectively outperforms legacy \ac{AP} selection strategies in a variety of scenarios.
\textcite{liu2010cognitive, karaca2016loadaware} use a similar approach of predicting performance under \ac{AP} selection constraints.

An interesting \ac{RL}-based scheme of user-to-multiple AP association is presented by~\textcite{dinh2021distributed}. Two distributed association methods based on \ac{DQL} enable stations to learn their best set of APs to connect to, using only local knowledge of the wireless environment or with a limited feedback from the APs.
Note that each device is equipped with multiple wireless interfaces. The objective is to maximize the long-term sum-rate subject to multiple constraints (e.g., AP load or application QoS constraints). A numerical evaluation reveals that the algorithms improve the targeted objectives and enhance fairness among applications.

A centralized approach is proposed by \textcite{kafi2019online}: an \ac{RL}-based client-\ac{AP} association algorithm to enhance the aggregated throughput in dense Wi-Fi networks. The Q-learning-based algorithm is deployed centrally in an \ac{SDN}-controller and controls the associations of new users, as well as performs re-associations of connected stations. As simulation results show, the approach outperforms the standard 802.11 association procedure when the distribution of users is not uniform and performs similarly when it is uniform.

\textcite{pei2017why} determine, through large-scale measurements, which factors affect the Wi-Fi connection set-up process. The analysis of 0.4~billion Wi-Fi sessions collected using the \textit{Wi-Fi Manager} mobile app from 5~million mobile devices
shows that 45\% of Wi-Fi connection attempts fail and about 5\% of attempts consume more than 10 seconds. 
Based on this analysis, the developed \ac{SL}-based \ac{AP} selection algorithm significantly improves Wi-Fi connection set-up performance. The algorithm uses \ac{RF} to classify candidate \acp{AP} into \textit{slow} or \textit{fast} sets by taking the following features as an input: hour of the day, \ac{RSSI}, mobile device model, \ac{AP} model, encryption enabled. Based on the classification, a station avoids connecting to \acp{AP} in the \textit{slow} set. The evaluation results show that the described approach reduces connection failures to 3.6\% and improves the connection set-up time over 10 times.

As shown by \textcite{song2017leveraging}, frame aggregation offers an efficient representation of expected throughput for improving \ac{AP} selection. Specifically, the characteristics of subframes during frame aggregation can uniquely embody the utilization, interference, and backlog traffic pressure for an \ac{AP}. With an \ac{SL}-based approach, simple regression models (based on linear regression and \ac{DT} regression) predict the \ac{AP} expected throughput for better AP selection. The results show a prediction accuracy above 80\%.

\subsubsection{Station Handovers}
In mobile scenarios, it frequently happens that a station leaves the coverage area of one \ac{AP} and enters an area covered by another \ac{AP}. In such a case, the station has to perform a handover from the old \ac{AP} towards the new \ac{AP}. The decision about a potential handover operation should be made early enough to avoid low data rate periods or even connectivity outage. \ac{ML} methods can predict network conditions and hence make correct handover decisions. 
For example, \textcite{feltrin2018machine} employ \ac{ML} to predict upcoming handover by making an AP monitor the \ac{RSSI} of connected stations and use a \ac{NN} for specific pattern recognition in the \ac{RSSI} evolution. This technique provides good prediction accuracy and is resilient to noise, speed, and fading phenomena. 

ABRAHAM (mAchine learning Backed multi-metRic Handover AlgorithM)~\cite{zeljkovic2019abraham} is an \ac{ML}-based proactive handover algorithm that uses multiple metrics to predict the future location of stations and the future \ac{AP} load. Additionally, using \ac{LSTM}, it predicts the future \ac{RSSI} values. These predictions are used to optimize the load on the APs by handing over stations to \acp{AP} to preserve QoS and QoE metrics. 
ABRAHAM achieves 139\% higher overall throughput compared to the legacy 802.11 handover algorithm. 

\textcite{han2019artificial} describe a handover management scheme for dense WLAN networks, which uses \ac{DRL}, specifically a deep Q-network. The scheme enables the \ac{NN} to learn from user behavior and network status, adapting its learning in time-varying dense WLANs. The handover decision is modeled as an \ac{MDP} leveraging the temporal correlation property, while the scheme depends on real-time network statistics to make decisions. Simulation results show that this solution can effectively improve the data rate during the handover process and outperform the 802.11 handover scheme.

\subsection{Management Architectures}
\label{sec_management_arch}
Management of Wi-Fi networks is a complex task as it requires tuning a plethora of parameters across distributed devices. Here, we present management frameworks that facilitate this task by providing an AI-based control plane. 

aiOS~\cite{coronado2020aios} is an \ac{AI}-based operating system for \acp{SD-WLAN} (i.e., the control plane). This system embeds state-of-the-art \ac{ML} toolboxes to provide a global intelligence platform, which is at the same time driven by \ac{AI} and designed to drive future \ac{AI}-powered applications and services. A proof-of-concept implementation of aiOS validates it by using several low-complexity \ac{ML} models for adaptive frame length selection in 802.11-based \acp{SD-WLAN}. The approach improves the aggregated network throughput by up to 55\% as evaluated in a real-world testbed.

\textcite{bast2019deep} use \ac{DRL} to dynamically optimize network slice configuration in Wi-Fi networks. A slice configuration consists of multiple parameters, e.g., \ac{CCA} sensitivity level, \ac{MCS}, and  transmit power level. Therefore, the action search space grows with the number of active slices in the network.
Interestingly, in the approach the selected action does not consist of absolute configuration values, but the increasing or decreasing current parameter values. 
A simple \ac{DQN} agent is enhanced with \ac{DDQN}, experience replay, and fitted Q-learning to improve convergence speed and stability. 
Results from the ns-3 simulator show that the solution can achieve the same optimal performance as found with an exhaustive search. Finally, \ac{DDQN} can optimize at run-time, without the need for \ac{AP} deployment information or knowledge about coexisting networks.

\textcite{lyu2019largescale} use large-scale AP usage data from a university campus Wi-Fi system with over 8,000 APs and more than 40,000 active users. 
An extensive spatio-temporal analysis of the data set includes the following:
\begin{itemize}
    \item AP load, i.e., the number of associated users,
    \item AP traffic throughput, i.e., the amount of traffic consumption within a period.
\end{itemize} 
A so-called \textit{idle phenomenon} prevails throughout the whole trace. Specifically, multiple APs remain unused (i.e., without any user association).
Second, the AP load follows a long tail distribution (i.e., most APs serve only a few users, while a small number of APs serve hundreds of users), hence, the per-AP utilization is imbalanced. 
Therefore, a new management system, named LAM (large-scale AP management), has the unused APs switched off intelligently according to the underlying user association conditions.
LAM leverages an ML-based algorithm to predict the AP load over time based on historical AP association records.
Results for diverse algorithms (including \ac{RF}, \ac{SVM}, \ac{kNN}, and \ac{DT}) show that the load prediction accuracy can reach as high as 90\%. In addition, more than 70\% of power energy is saved, with over 92\% of Wi-Fi coverage guaranteed. These savings translate to \$59,000 per year in the aforementioned university Wi-Fi system.

An SDN-based Wi-Fi control system is considered to manage a group of APs by \textcite{patro2015outsourcing}. The central controller  configures channel and transmission power settings for the  APs in the network. Decisions on how to configure the network are taken after learning from the collected data. A set of ML-based techniques are used, for example, \acp{REPT} -- to predict Wi-Fi and non-Wi-Fi activity (such as microwave ovens) so that better configurations can be deployed. The framework reduces channel congestion by up to 47\%. 

\subsection{Traffic Prediction}
\label{sec_traffic}
Network management operations are assisted by traffic prediction techniques for better short- and long-term planning.
Proper planning, using methods such as traffic forecasting, congestion control, power saving, bandwidth allocation, and buffer management, leads to improved user \ac{QoE}.
For instance, based on the predicted traffic, \acp{AP} can improve load balancing and admission control.

\begin{figure}
\centering
\includegraphics[width=\columnwidth]{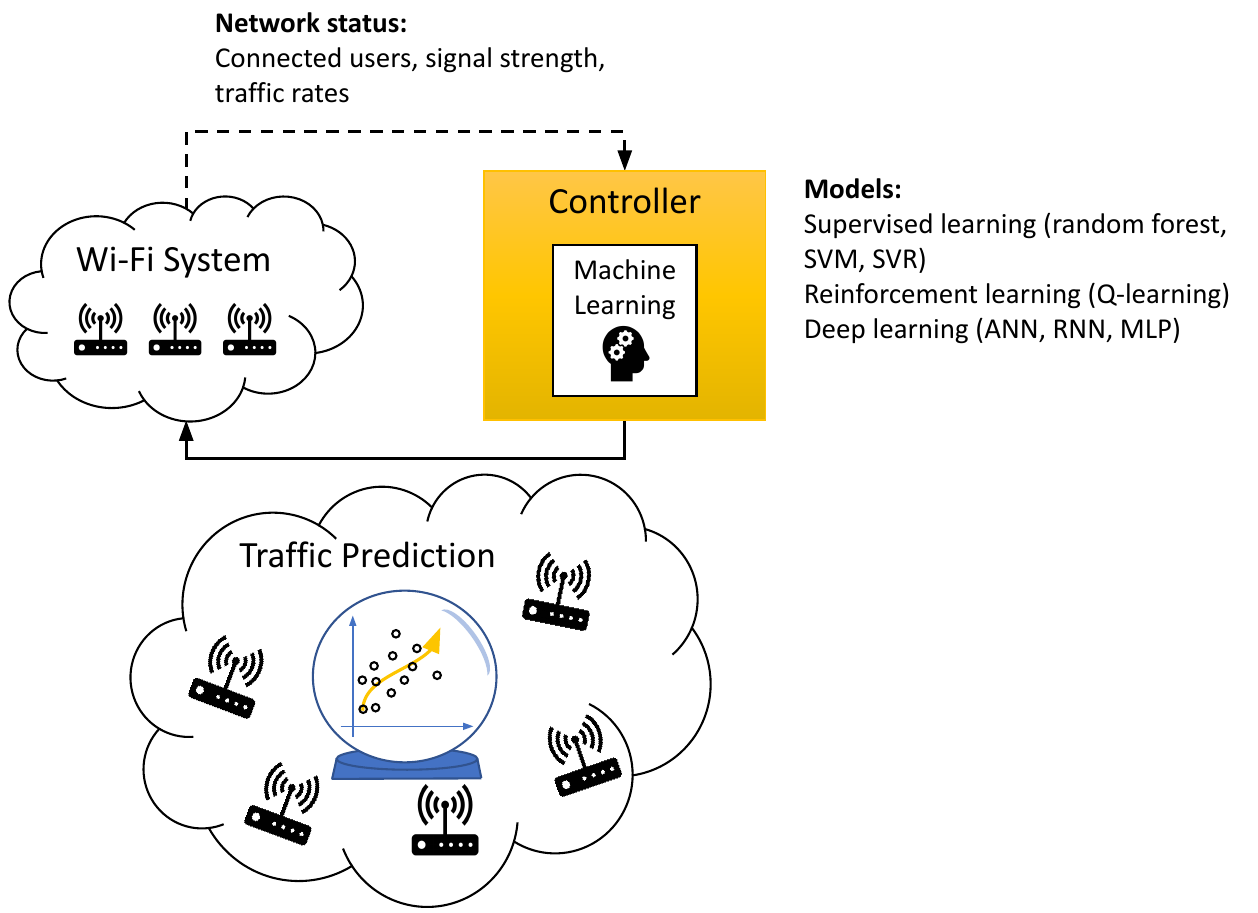}
\vspace{-1em}
\caption{Illustration of the traffic prediction capabilities of \ac{ML} models. Dashed arrows represent observation and solid lines represent actions.
}
\label{fig_VA_5}
\end{figure}

Real-time traffic prediction becomes a challenging problem in \mbox{Wi-Fi} networks due to varying channel conditions, changing network topologies, and random user traffic.
Traffic estimation is also dependent on several other parameters, such as the total number of users in the network, the \ac{SNR} on the link, or the communication capabilities of users and \acp{AP} \cite{khan2020real}.
In such scenarios, \ac{ML} models deal with the diverse conditions of \mbox{Wi-Fi} networks, otherwise intractable through analytical methods. 
As summarized in \Cref{fig_VA_5}, \ac{ML} models augment legacy 802.11 devices through \ac{SVM}~\cite{feng2006svm}, \ac{RNN}, \ac{MLP}, \ac{SVR}, and polynomial regression \cite{thapaliya2018predicting}, or \ac{DT} and \ac{RF} \cite{khan2020real}. %

Specifically, the solution proposed by \textcite{feng2006svm} trains an \ac{SVM} to predict the traffic evolution one step ahead.
Besides, by recursively applying the one-step-ahead solution, traffic estimation for $l$-step-ahead is also conceived.
The \ac{SVM} model is implemented as a Gaussian radial basis function and trained with \num{100} samples to predict the next \num{100} samples.
Through the \ac{SVM} model, the error to predict the upcoming traffic is reduced at least by \SI{33}{\percent} when compared to the performance of the \ac{ANN}.

\textcite{khan2020real} analyze the most suitable \ac{ML} models to predict traffic among \ac{MLP}, \ac{SVR}, \ac{DT}, and \ac{RF}.
To train these models, several features are extracted from simulation and real data (i.e., Wireshark network trace). In particular, the number of connected users, signal strength, modulation scheme, data rate, inter-arrival time, packet arrival rate, number of re-transmissions, and several other channel parameters are extracted.
The solutions are implemented in a \mbox{Wi-Fi} network consisting of \num{10} users and a single 802.11 \ac{AP}.
The reported prediction accuracy presents a maximum value of \SI{96.2}{\percent}, \SI{94.5}{\percent}, \SI{93.3}{\percent}, and \SI{91}{\percent} using \ac{MLP}, \ac{DT}, \ac{RF}, and \ac{SVR}, respectively.
The study also analyses the complexity of these mechanisms in real-time schemes by reporting the time elapsed for each model.
The highest time-consuming model is \ac{MLP} followed by \ac{RF}, \ac{SVR}, and \ac{DT}.

Finally, Thapaliya et al. apply both \ac{SL} and \ac{USL} models~\cite{thapaliya2018predicting} to predict network congestion levels.
Based on captured data attributes (the number of clients, throughput, frame retry rate, and frame error rate), \ac{SVR} and polynomial regressor models predict the same values for a certain location, day, and time.
These predicted values are then fed to an \ac{EM} algorithm to predict congestion levels by forming three different clusters.
Each cluster is identified with high, medium, and low congestion levels based on the numeric value of the clustered samples. 
The obtained accuracy is \SI{24}{\percent}, \SI{50}{\percent}, and \SI{26}{\percent}, for a low, medium, and high level of network area congestion.

\subsection{Predicting the Health of Wi-Fi Connections}
\label{sec_predicting_health}
Unlicensed bands are becoming crowded with dense and uncontrolled deployments of Wi-Fi networks, generally managed by different users. These environments exacerbate the effects of well-known pathological conditions such as hidden terminals, flow starvation, and performance anomaly. Unfortunately, these problems become increasingly difficult to detect in real world scenarios. Specifically, while performance degradation is a common symptom of these unwanted conditions, different causes require different solutions. ML seems to be the right toolset for the detection of individual impairments,  as it can handle a large amount of raw measurement data and learn to deduce the current operation regime (e.g., using classification methods).
Therefore, \textcite{gallo2018widia} provide an automatic diagnostic tool, Wi-Dia, for detecting the causes of performance impairments by recognizing the wireless operating context. Wi-Dia follows a data-driven approach and exploits ML methods for classifying Wi-Fi problems (e.g., hidden stations and flow starvation). 
It uses features related to network topology and measures channel utilization without impacting regular network operations. The classifier is jointly trained using simulated and experimental data, thus taking advantage of the flexibility of network simulators and the realistic details of wireless testbeds. As the results show, Wi-Dia achieves high detection accuracy of pathological Wi-Fi conditions in real-world scenarios.

Similarly, \textcite{syrigos2019employment} detect the causes of Wi-Fi under-performance, e.g., high contention with other Wi-Fi and non-Wi-Fi devices, operation in low \ac{SNR} region, hidden terminal, or capture effect. A centralized Wi-Fi network controller collects two performance metrics from connected \acp{AP} (i.e., those exposed by the ath9k driver):
\begin{itemize}
    \item \ac{NCA}, i.e., the ratio between channel access attempts per second and the maximal channel access attempts per second as calculated with analytical 802.11 models); and
    \item \ac{FDR}, i.e., the ratio between successful transmissions per second and channel access attempts per second.
\end{itemize}  
The classification is preceded by data modeling and feature extraction and performed with four diverse algorithms: \ac{DT}, \ac{RF}, \ac{SVM}, and \ac{kNN}. After fine-tuning the algorithms' parameters, the results show a remarkable detection accuracy of 99.2\% with the \ac{kNN} algorithm.

\textcite{trivedi2020winetsense} propose WiNetSense, a centralized sensing framework, which collects the Wi-Fi link quality statistics (e.g., \ac{RSSI}) from network devices and uses this information to build the global network topology and instantaneous network health information. Furthermore, the collected data is analyzed using \ac{ML} algorithms such as \ac{kNN} and \ac{NB} to predict the health of wireless links. This knowledge can trigger specific decisions regarding load balancing, smooth handovers, or dynamic power control.

An anomaly-detection approach that uses a \ac{SOHMMM} is considered by~\textcite{allahdadi2021hidden}. The self-organizing map is an artificial \ac{NN} that is trained through a \ac{USL} process. The \ac{SOHMMM} shows improved anomaly detection accuracy and sensitivity, compared to other HMM-based approaches, as tested in a simulated environment.

Morshedi and Noll propose a novel ML-based approach for estimating the perceived \ac{QoS} of video streaming~\cite{morshedi2021estimating} and video conferencing~\cite{morshedi2021conferencing} using only 801.11-specific network performance parameters collected from \ac{AP}. The studies use datasets comprising 802.11n/ac/ax specific network performance parameters in the form of mean opinion scores. These datasets train multiple ML algorithms, i.e., \ac{LMT}, \ac{REPT}, \ac{NBT}, \ac{MLP}, and achieve a 93-99\% accuracy of estimating the perceived QoS classes. 
\ac{LMT} and \ac{REPT} are the most suitable algorithms to estimate the perceived QoS of video streaming and conferencing, respectively, in terms of accuracy, interpretability, and computational cost criteria. 
Additionally, the generated ML model can be transferred to an \ac{AP} as a lightweight script to continuously monitor QoS. 

\subsection{Open Challenges}

While most of the presented \ac{ML}-based solutions for cross-network optimization (e.g., channel allocation) feature centralized operation, we believe that distributed approaches are better suited for the unplanned and random nature of Wi-Fi deployments. Moreover, we cannot assume the existence of a centralized controller that manages co-located but separately owned Wi-Fi networks (e.g., in typical residential Wi-Fi deployments). Note that the potential operation of such a central controller poses a significant privacy threat, as it might require the collection of sensitive user data (e.g., the traffic volume of individual stations).
Therefore, we argue that there is an increasing need for research in the scope of a decentralized and distributed \ac{ML}-based optimization. Particularly, multi-agent \ac{RL}-based schemes seem to be a fit: a set of agents (e.g., one at each \ac{AP}) interact and share limited information to collaboratively optimize wireless resources while also preserving privacy.
Finally, an open management challenge that has not yet been addressed using \ac{ML} methods is improving the coexistence of modern and legacy Wi-Fi devices \cite{kosek-szott2017coexistence}.

\onecolumn
\begin{landscape}
\scriptsize
\begin{xltabular}{\linewidth}{ccccccp{0.19\textwidth}*{2}{p{0.29\textwidth}}}
\caption{Summary of works on improving wireless network coexistence with ML. Papers indicated with an asterisk (*) implement Wi-Fi agents; other papers deploy agents only on the competing technology side. The evaluation methods are theoretical (E), simulation (S), and experimental (E). The ML improvement is in comparison to SoA methods.} \label{tab:coexistence} \\
\toprule
\textbf{Area} & \textbf{Ref.} & \shortstack{\textbf{ML}\\\textbf{category}} & \shortstack{\textbf{ML}\\\textbf{mechanisms}} & \textbf{Year}&\shortstack{\textbf{Evaluation}\\\textbf{method}}  & \textbf{Application of ML} & \textbf{Novelty of approach} & \textbf{ML improvement} \\
\midrule
\endfirsthead

\multicolumn{9}{c}%
{\tablename\ \thetable{} -- continued from previous page} \\
\midrule \textbf{Area} & \textbf{Ref.} & \shortstack{\textbf{ML}\\\textbf{category}} & \shortstack{\textbf{ML}\\\textbf{mechanisms}} & \textbf{Year}&\shortstack{\textbf{Evaluation}\\\textbf{method}}  & \textbf{Application of ML} & \textbf{Novelty of approach} & \textbf{ML improvement} \\ \midrule
\endhead

\multicolumn{9}{r}{{Continued on next page}} \\
\endfoot

\endlastfoot

    \multicolumn{1}{c}{\multirow{27}[52]{*}{\shortstack{Fair channel sharing \\ with cellular networks \\ (Section \ref{sec_fair})}}} & \cite{rupasinghe2015reinforcement} & RL    & QL    & 2015  & S     & Select duty cycle duration & Apply QL to configure LAA transmission gaps & Higher capacity than with fixed duty cycles \\
\cmidrule{2-9}          & \cite{xu2016dynamic}* & RL    & QL    & 2016  & S     & Calculate maximum transmission duration & Apply QL to set limit on maximum transmission duration & Higher throughput and fairness \\
\cmidrule{2-9}          & \cite{lin2017fairly} & RL    & QL    & 2017  & S     & Coordinate channel access & Apply Markov game framework for Wi-Fi/LTE-U coexistence & Higher throughput and fairness \\
\cmidrule{2-9}          & \cite{liu2017dynamic} & RL    & QL    & 2017  & S     & Allocate blank subframes & Apply QL for blank subframe allocation & Improved spectrum utilization \\
\cmidrule{2-9}          & \cite{bairagi2018qoe} & RL    & QL    & 2018  & S     & Select resource allocation scheme & Apply QL for LTE-U resource allocation & Higher Wi-Fi throughput and coexistence fairness \\
\cmidrule{2-9}          & \cite{tan2018learning} & RL    & QL    & 2018  & S     & Select LAA transmission time & Apply QL for LAA channel access & Higher throughput \\
\cmidrule{2-9}          & \cite{haider2018enhanced} & RL    & QL    & 2018  & S     & Select LTE-U defer period & Apply QL for LTE-U channel access & Lower packet loss, lower Wi-Fi delay \\
\cmidrule{2-9}          & \cite{maglogiannis2018q} & RL    & QL    & 2018  & S     & Select duty cycle parameters & Apply QL to configure LAA transmission durations and gaps & Higher coexistence fairness \\
\cmidrule{2-9}          & \cite{challita2018proactive} & SL    & LSTM  & 2018  & S     & Select channel and learn channel access probability, toggle carrier aggregation & Apply ML to joint channel selection and channel access & Higher coexistence fairness \\
\cmidrule{2-9}          & \cite{su2018lte} & RL    & QL    & 2018  & S     & Select duty cycle & Apply QL for duty cycle optimization & Higher throughput and coexistence fairness \\
\cmidrule{2-9}          & \cite{tian2019traffic} & RL    & DQL & 2019  & S     & Predict Wi-Fi traffic patterns, allocate LTE-U resources & Apply two-layer network approach to predict and decide & Higher throughput, reduced delay \\
\cmidrule{2-9}          & \cite{zhou2019lwcq} & RL    & QL    & 2019  & S     & Select LAA scheduling and transmission duration & Apply QL to LAA channel access control & Higher system throughput and coexistence fairness \\
\cmidrule{2-9}          & \cite{kushwaha2019novel}* & RL    & QL    & 2019  & S     & Select LTE-U frame structure and transmit power & Apply QL for dynamic frame selection & Higher coexistence fairness \\
\cmidrule{2-9}          & \cite{mosleh2020dynamic}* & RL    & QL    & 2020  & S     & Select subchannel & Apply RL for subchannel selection considering PHY and MAC parameters & Near-optimal system throughput \\
\cmidrule{2-9}          & \cite{han2020reinforcement}* & RL    & MAB   & 2020  & S     & Select CW & Apply RL for CW selection of both Wi-Fi and LAA & Higher system throughput and coexistence fairness \\
\cmidrule{2-9}          & \cite{tan2020intelligent} & RL    & DRL, MDP, DQN & 2020  & S     & Learn Wi-Fi traffic demands, configure LTE-U duty cycle & Apply DRL for optimizing LTE-U transmission duration & Higher LTE-U throughput while protecting Wi-Fi \\
\cmidrule{2-9}          & \cite{neto2020multi} & RL    & QL    & 2020  & S     & Select duty cycle parameters & Apply ML in multi-cell scenario & Higher system throughput \\
\cmidrule{2-9}          & \cite{ali2020relbt} & RL    & QL    & 2020  & S     & Select CW & Apply RL to LBT & Higher efficiency and coexistence fairness \\
\cmidrule{2-9}          & \cite{tang2020almost} & RL    & QL    & 2020  & S     & Allocate blank subframes & Apply QL to learn Wi-Fi transmission patterns & Higher system throughput \\
\cmidrule{2-9}          & \cite{yu2020non} & RL    & DRL, TRPO & 2020  & S     & Access the channel & Apply DRL for coexisting with unknown technologies & Higher throughput or fairness \\
\cmidrule{2-9}          & \cite{hirzallah2021sense}* & RL    & MAB & 2021  & S     & Select sensing thresholds & Apply MAB for sensing threshold adaptation & Higher throughput \\
\cmidrule{2-9}          & \cite{kishimoto2021reinforcement} & RL    & QL    & 2021  & S     & Select channel and number of subframes & Apply RL for joint parameter optimization & Improved coexistence fairness, lower packet loss \\
\cmidrule{2-9}          & \cite{pei2021q} & RL    & QL    & 2021  & S     & Select energy threshold & Apply QL for energy threshold optimization & Improved coexistence fairness, lower Wi-Fi collision probability and transmission delay \\
\cmidrule{2-9}          & \cite{kala2021optimizing} & SL    & Linear regression & 2021  & E     & Infer relationship between network features and coexistence performance & Apply SL to LTE/Wi-Fi network feature relationships & Higher network capacity and signal strength \\
\cmidrule{2-9}          & \cite{girmay2021coexistence} & RL    & QL    & 2021  & S     & Allocate blank subframes & Apply experience replay for LTE/Wi-Fi coexistence & Higher throughput and fairness \\
\cmidrule{2-9}          & \cite{wang2021joint} & RL    & DDPG  & 2021  & S     & Allocate bandwidth and transmission opportunities & Apply DDPG to joint bandwidth and transmission opportunity allocation & Higher NR-U throughput \\
\cmidrule{2-9}          & \cite{pei2021qlearning} & RL    & QL    & 2021  & S     & Allocate power and duty cycle & Apply QL for resource allocation of D2D-Unlicensed & Higher throughput and fairness \\
\cmidrule{2-9}          &     \cite{pei2021deep} & RL    & DQL   & 2021  & S     & Select access time and transmission duration & Apply DQL for unlicensed spectrum access & Higher throughput and fairness \\
\cmidrule{2-9}          & \cite{naveen2021coexistence}* & RL    & QL    & 2021  & S     & Schedule transmissions & Apply QL to solve LTE/Wi-Fi airtime allocation & Improved efficiency \\

\midrule

\multicolumn{1}{c}{\multirow{16}[14]{*}{\shortstack{Network monitoring \\ (Section \ref{sec_net_monitoring})}}} & \cite{galanopoulos2016efficient} & RL    & QL, double QL & 2016  & S     & Estimate channel occupancy time, select power level & Apply double QL for joint channel occupancy learning and power level adaptation & Improved throughput \\
\cmidrule{2-9}          & \cite{fakhfakh2017incentive} & RL    & QL    & 2017  & S     & Select base station or AP & Apply QL to Wi-Fi offloading decisions & Improved Wi-Fi offloading efficiency \\
\cmidrule{2-9}          & \cite{fakhfakh2017optimised} & RL    & QL    & 2017  & S     & Select base station or AP & Use QL for Wi-Fi offloading decisions (more inputs) & Improved Wi-Fi offloading efficiency \\
\cmidrule{2-9}          & \cite{yang2018channel} & RL    & fuzzy QL & 2018  & S     & Adapt LAA's alignment reference interval & Apply fuzzy QL to frequency reuse & Improved system capacity and coexistence fairness \\
\cmidrule{2-9}          & \cite{yang2019machine}* & SL    & DNN, CNN, LSTM & 2019  & S+E   & Predict contending devices & Apply pretrained DNN model for channel allocation & Improved system throughput \\
\cmidrule{2-9}          & \cite{pulkkinen2020understanding} & SL    & CNN, NNMR & 2020  & E     & Detect and classify interference & Apply DL to interference detection & Improved interference detection accuracy \\
\cmidrule{2-9}          & \cite{el-sha2021machine} & SL    & RF    & 2021  & S     & Select LTE's channel access method & Consider multiple LTE channel access methods & Improved LTE throughput without Wi-Fi throughput degradation \\
\cmidrule{2-9}          &     \cite{kuzdeba2021transfer} & USL   & DNN & 2021  & E     & Learn network features and detect new devices & Apply TL to transfer features learned on synthetic data to real data & Retraining of features improves TL accuracy \\
\cmidrule{2-9}          & \cite{yin2021learning} & USL   & NN    & 2021  & S     & Estimate active Wi-Fi users & Use USL to filter collision probability measurements & Higher estimation accuracy than Kalman filter \\

\midrule

\multicolumn{1}{c}{\multirow{10}[14]{*}{\shortstack{Signal classification \\ (Section \ref{sec_sig_classification})}}} & \cite{olbrich2017wiplus} & USL   & K-means clustering & 2017  & E     & Detect signal clusters & Apply K-means clustering for interference detection & Accurate estimation of remaining Wi-Fi link capacity \\
\cmidrule{2-9}          & \cite{sathya2020machine} & USL, SL & NN, DCNN & 2020  & S     & Detect contending Wi-Fi networks & Observe energy values during LTE-U OFF time & Higher detection accuracy than two benchmarks \\
\cmidrule{2-9}          & \cite{gu2020deep} & SL    & CNN/RNN & 2020  & S     & Identify LTE-U and Wi-Fi signals & Apply DL to classify RAT signals & High recognition accuracy and robustness \\
\cmidrule{2-9}          & \cite{mosleh2020novel} & SL    & NN, logistic regression & 2020  & S     & Estimate throughput and probability of coexistence & Apply SL to estimate probability of coexistence & High throughput tracking accuracy \\
\cmidrule{2-9}          & \cite{yang2019blind} & SL    & CNN   & 2020  & S     & Identify LTE-U and Wi-Fi signals & Apply DL to classify RAT signals using in-phase and quadrature samples & High recognition accuracy and robustness \\
\cmidrule{2-9}          & \cite{bhatti2021shared} & SL    & CNN   & 2021  & E     & Classify signals & Apply CNN to detect ten signal classes & High classification accuracy \\
\cmidrule{2-9}          & \cite{fonseca2021radio} & SL    & CNN, TL & 2021  & S+E   & Classify signals & Apply CNN to spectrogram images & High classification accuracy \\
\cmidrule{2-9}          & \cite{girmay2021machine} & SL    & CNN   & 2021  & S     & Determine Wi-Fi network saturation & Apply CNN to analyze inter-frame spacing statistics & High classification accuracy \\

\midrule

\multicolumn{1}{c}{\multirow{9}[10]{*}{\shortstack{Cooperative coexistence \\ (Section \ref{sec_hybrid})}}} & \cite{ahmad2020load} & RL    & TRPO  & 2020  & S     & Select AP & Apply RL in hybrid Li-Fi/Wi-Fi networks & Higher throughput, user satisfaction, and outage probability than benchmark \\
\cmidrule{2-9}          & \cite{ahmad2020reinforcement} & RL    & TRPO  & 2020  & S     & Select AP & Consider user satisfaction and fairness in RL-based load balancing & Higher throughput and fairness than three benchmarks \\
\cmidrule{2-9}          & \cite{wu2020novel} & SL    & NN, backpropagation & 2020  & S     & Handover decision & Apply ANN to Li-Fi/Wi-Fi handovers & Higher throughput than two benchmarks \\
\cmidrule{2-9}          & \cite{sanusi2020development} & SL    & NN with fuzzy logic & 2020  & S     & Handover decision & Apply fuzzy logic and ANN to Li-Fi/Wi-Fi handovers & Higher handover decision accuracy than benchmark \\
\cmidrule{2-9}          & \cite{alenezi2020reinforcement} & RL    & QL    & 2020  & S     & Select AP & Apply QL for Wi-Fi offloading & Higher throughput than two benchmarks \\
\cmidrule{2-9}          & \cite{alenezi2021reinforcement} & RL    & QL    & 2021  & S     & Select AP & AP assignment based on requested data rate & Higher fairness and user satisfaction than benchmark \\

\bottomrule

\end{xltabular}
\end{landscape}

\twocolumn

\section{Coexistence Scenarios} %
\label{sec_coexistence}

The coexistence of Wi-Fi and cellular technologies is currently a popular and attractive research area\footnote{Channel sharing with other technologies is described in \cref{sec_mhop} where, among others, we address sensor and vehicular networks. Additionally, we refer the readers to \cite{ejaz2020learning}, in which different learning paradigms for \ac{IoT} communication and computing technologies are surveyed, and to \cite{liu2021machine}, in which ML-supported detection and identification of IoT devices is surveyed.}. 
These technologies are already advanced and their newest generations provide peak data rates in the order of $\si{\giga\bit\per\second}$. However, under coexistence scenarios in unlicensed bands (e.g., with LTE-LAA), they still rely on rather primitive coexistence schemes based on energy sensing and hence suffer from frequent collisions and significant throughput degradation of up to 90\%~\cite{chai2016lte,abinader2014coexistence}.
The coexistence schemes need to tackle the problem of heterogeneity of the underlying technologies: they implement different \ac{MAC} and \ac{PHY}, they are usually managed by separate operators, and they do not natively support inter-technology communication for spectrum sharing. Therefore, fair sharing of unlicensed radio resources is still an open challenge \cite{gur2020expansive}.

Several research papers address the problem of the coexistence of multiple \acp{RAT} in unlicensed bands, e.g., \cite{li2016modeling, chen2016space, li2016share, chen2017dynamic,wszolek2021revisiting}.
In this survey, we describe only those which 
propose \ac{ML}-based solutions (Table \ref{tab:coexistence}). Both centralized and decentralized approaches are considered, together with both offline and online training. The proposed mechanisms appear in the following main areas:
\begin{itemize}
    \item fair channel sharing,
    \item network monitoring,
    \item signal classification, and
    \item cooperative networking.
\end{itemize}
The majority of the proposed mechanisms are based on reinforcement learning (mostly Q-learning) and deep supervised learning (mostly \acp{CNN}). Often, the $\varepsilon$-greedy policy is used for Q-learning to balance exploration and exploitation.

\begin{figure}
\centering
\includegraphics[width=\columnwidth]{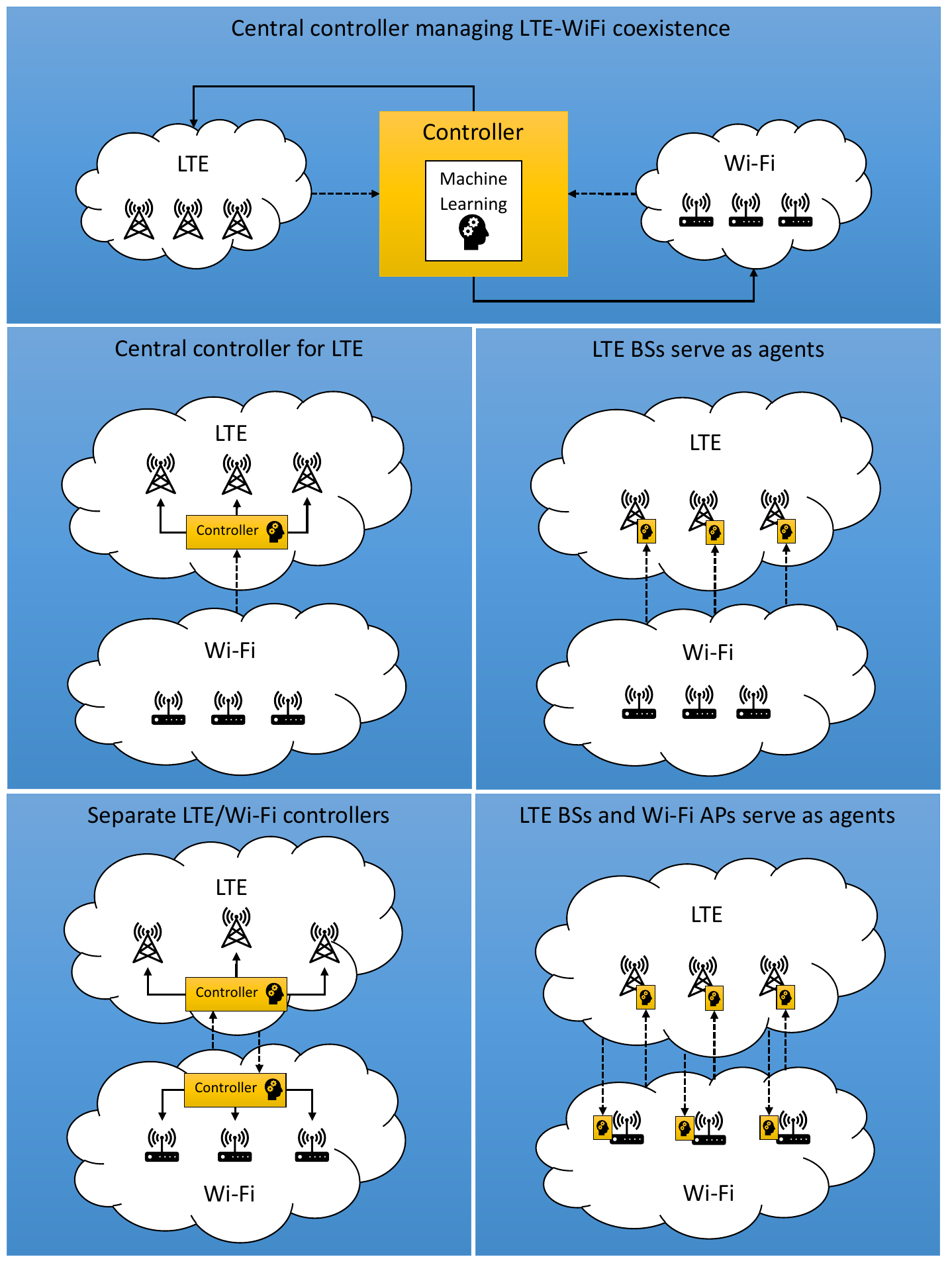}
\vspace{-.5em}
\caption{Types of ML implementations in LTE/Wi-Fi coexistence scenarios. Dashed arrows represent observations, solid lines represent actions, and blue rectangles represent interference domains.
}
\label{fig_coexistence}
\end{figure}

Most analyzed papers optimize LTE behavior (i.e., the newcomer to the unlicensed bands) so that Wi-Fi performance is not degraded~\cite{zinno2018fair}. In some cases, however, it is proposed that both technologies implement some sort of \ac{ML} to improve the coexistence of both technologies. \Cref{fig_coexistence} presents different approaches considered by researchers: from a central controller implemented for both technologies up to separate \ac{ML} agents installed in LTE \acp{BS} and Wi-Fi \acp{AP}, which independently observe the environment (i.e., perform local observation) and take actions. Note that the state of the environment depends on the joint action of all agents, which may be unaware of individual decisions.
Additionally, in the reviewed papers, typically only downlink LTE transmissions interfere with either uplink or downlink Wi-Fi transmissions, while LTE uplink traffic is scheduled in the licensed band.  

\subsection{Fair Channel Sharing with Cellular Networks}
\label{sec_fair}
Several papers propose to adjust \ac{LTE-U} behavior, by either a central controller or by distributed learning. Their main goal is to intelligently avoid interference with incumbent technologies, like Wi-Fi, as a solution to the problem of the negative impact of periodic LTE transmissions on channel utilization and channel access fairness \cite{tinnirello2018impact}. 

\begin{figure}
\centering
\includegraphics[width=\columnwidth]{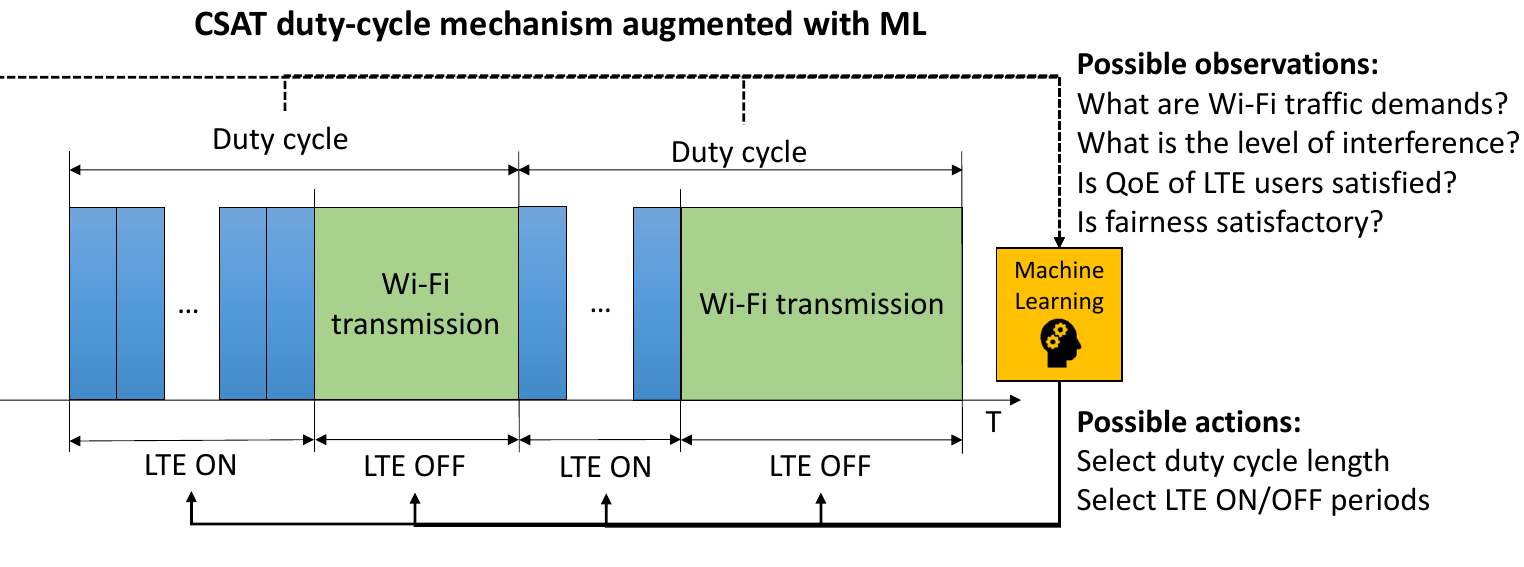}
\vspace{-1em}
\caption{LTE-U \ac{CSAT} coexistence mechanism. Blue rectangles represent LTE-U subframes of a length of 1~ms and green rectangles represent LTE-U OFF periods, during which Wi-Fi transmissions can occur.}
\label{fig_lteu_csat}
\end{figure}

\begin{figure}
\centering
\includegraphics[width=\columnwidth]{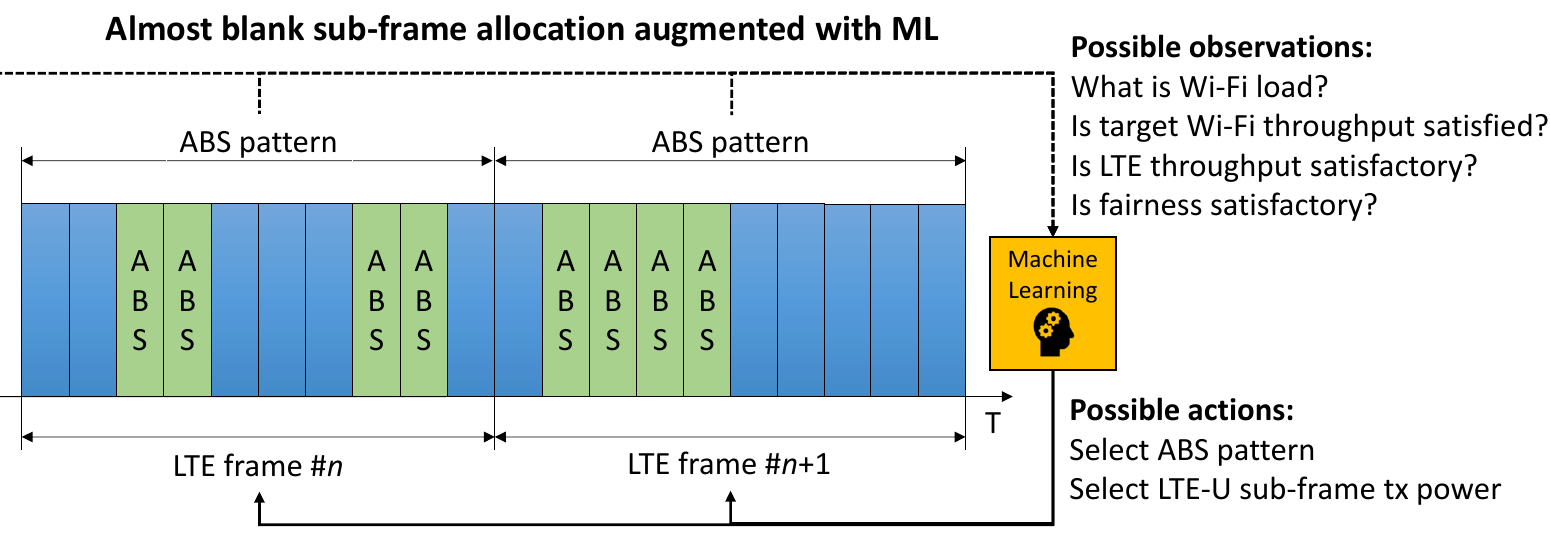}
\vspace{-1em}
\caption{LTE-U \ac{ABS} coexistence mechanism. Blue rectangles represent LTE-U subframes of a length of 1~ms and green rectangles represent \acp{ABS}, during which Wi-Fi transmissions can occur.}
\label{fig_lteu_abs}
\end{figure}

Many papers implement Q-learning and modify the \ac{DCM} function, which is a part of the \ac{CSAT} algorithm (\cref{fig_lteu_csat}), or to the \ac{ABS} allocation mechanism \cite{huang2017recent, tian2019traffic}. The \ac{ABS} mechanism is normally used to avoid co-channel cross-tier interference in the case of heterogeneous cellular scenarios, e.g., in scenarios composed of macro and small cells (Figure~\ref{fig_lteu_abs}). The main goal of these modifications is to improve coexistence and channel sharing efficiency by intelligently disabling LTE transmissions in certain subframes to allow Wi-Fi transmissions and outperform legacy \ac{DCM}.

Centralized \ac{LTE-U}/Wi-Fi channel access management is proposed in the following papers. \textcite{naveen2021coexistence} model the traffic load of each system as an M/M/1 queue and Q-learning is used by a central controller to adjust the allocation of LTE subframes in the \ac{CSAT} duty cycles. 
\textcite{kushwaha2019novel} use an inter-\ac{RAT} controller with Q-learning to improve Wi-Fi/\ac{LTE-U} coexistence fairness by considering the Wi-Fi load. In particular, the controller selects the optimum subframe configurations out of the ones defined by 3GPP. Additionally, it reduces \ac{LTE-U} subframe transmission power to limit interference to co-channel users and increase the overall channel utilization. 
Similar approaches are used elsewhere: 
\begin{itemize}
    \item an agent controls \ac{DCM} to maximize LTE-U throughput while protecting  Wi-Fi transmissions, based on observing Wi-Fi traffic demands and using \ac{DRL} \cite{tan2020intelligent};
    \item a centralized \ac{RL}-based DCM learns from measured interference \cite{neto2020multi}; and
    \item a centralized Q-learning-based mechanism of blank subframe allocations improves the overall utility function, i.e., considering target Wi-Fi throughput and satisfactory LTE throughput and delay \cite{liu2017dynamic}. 
\end{itemize} 

Decentralized channel access management for \ac{LTE-U}/Wi-Fi coexistence is proposed in the following papers. \textcite{rupasinghe2015reinforcement} use Q-learning for distributed control of duty cycle periods by \ac{LTE-U} \acp{BS}, while considering the beaconing mechanism of 802.11n. %
Additionally, \textcite{haider2018enhanced} apply a Q-learning based \ac{LBT} for \ac{LTE-U} downlink transmissions\footnote{\ac{LBT} is commonly used in case of LTE-LAA, however, it was proposed in the literature also for LTE-U \cite{yin2016lbt}.}. 
\ac{LTE-U} devices are treated as secondary users that need to protect Wi-Fi transmissions. 
Therefore, \ac{LTE-U} users for which defer periods increase in case of increasing Wi-Fi backoff timers (i.e., when Wi-Fi defer periods increase) are rewarded. 
Simulation results show improved throughput and decreased delay of Wi-Fi stations in comparison to legacy \ac{LBT}. 
Wi-Fi protection is also considered by \textcite{bairagi2018qoe}.
A \ac{VCFG} is used and an optimization problem is defined within each virtual coalition composed of Wi-Fi \acp{AP} and \ac{LTE-U} \acp{SBS} operating in the same unlicensed band. Then, (i) Kalai-Smorodinsky bargaining is used for fair time-sharing between \ac{LTE-U} and Wi-Fi and (ii) Q-learning is used for resource allocation for \ac{LTE-U}. 
Each \ac{SBS} maximizes the sum of QoE for its users under the constraint of protecting Wi-Fi APs. QoE is measured in terms of the \ac{MOS} which is mapped to the transmission characteristics of the following applications: web browsing, file downloading, and video streaming. This approach provides higher throughput for Wi-Fi than standard \ac{LBT}.

\textcite{lin2017fairly} implement adaptive learning to improve coexistence fairness for  \ac{LTE-U} \acp{BS}.
Bianchi's Markov model \cite{bianchi2000performance} is embedded in a sequential game to describe the contention nature of Wi-Fi networks. The time-slotted behavior of LTE-U devices is also modeled as a sequential game. These two processes are combined to form a Markov game. Each \ac{LTE-U} \ac{BS} serves as an agent and Wi-Fi networks are considered the environment to which the agents adapt. The proposed dynamic channel access control results in improved overall performance.
Additionally, \textcite{athukoralage2016regret} use regret-based learning \ac{DCM} ON/OFF period selection to improve network coverage in case of natural disasters. This improvement is achieved by unmanned aerial base stations coexisting with ground Wi-Fi APs. 
The area throughput improves in comparison to fixed and Q-learning-based dynamic duty cycle selection.
Furthermore, Q-learning-based multi-channel operation is proposed by \textcite{su2018lte}.
\ac{LTE-U} \acp{SBS} serve as agents to allow either independent or joint optimization of duty cycles for each channel. The mechanism ensures fairness and improves throughput for multi-channel Wi-Fi/\ac{LTE-U} coexistence.
Finally, \textcite{gao2016spectrum} have LTE-U and Wi-Fi managed by separate \ac{SDN} controllers which build decision trees. Per-technology controllers do not communicate with each other but only negotiate network sharing by playing a repeated game based on rank-order tournaments. An incentive-based approach negotiates the channel resources, i.e., there are prizes for allowing spectrum sharing and for asking the other operator for a favor. The simulation results show that it is possible to achieve harmonized coexistence of the two technologies.

\begin{figure}
\centering
\includegraphics[width=\columnwidth]{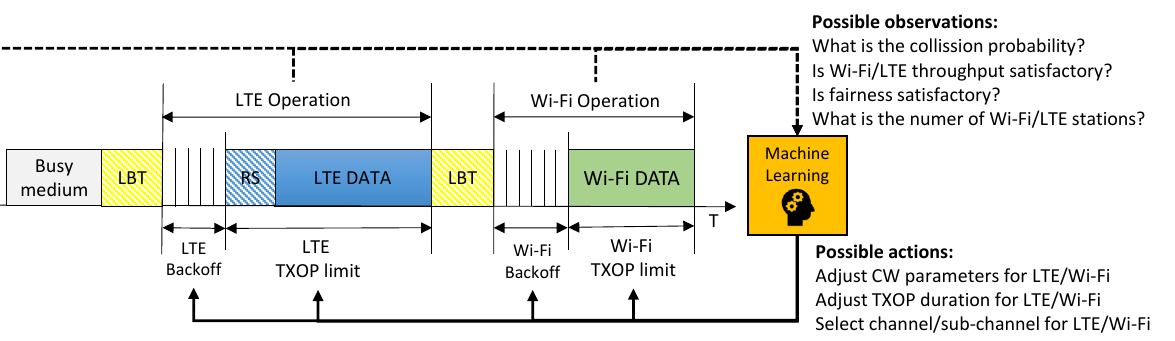}
\vspace{-1em}
\caption{LTE-LAA LBT-based coexistence mechanism. RS denotes the reservation signal, which is typically used by the \ac{LTE-LAA} devices to reserve the channel until the beginning of the next frame synchronization slot.}
\label{fig_laa}
\end{figure}

Another group of papers addresses coexistence between Wi-Fi and \ac{LTE-LAA}, which in most cases involves the adjustment of parameters of the \ac{LBT}-based channel access mechanism shown in Figure~\ref{fig_laa}. Similar to the LTE-U case, most papers are based on Q-learning.

Only a single paper proposes centralized channel access management for \ac{LTE-LAA}/Wi-Fi coexistence \cite{zhou2019lwcq}, in which Q-learning is used by \acp{MME} implemented in the LTE core to adjust the \ac{LTE-LAA} transmission duration to Wi-Fi traffic intensity. Centralized collection of data regarding \ac{LTE-LAA} and Wi-Fi systems by the LTE cloud wireless access network (C-RAN) is proposed to support \acp{MME}. 
Other papers implement distributed Q-learning to:
\begin{itemize}
    \item optimize spectral efficiency of Wi-Fi/\ac{LTE-LAA} coexistence \cite{mosleh2020dynamic}, 
    \item scale \ac{CW} parameters depending on the collision probability observed in each backoff stage by LTE \acp{UE}, as opposed to the legacy \ac{HARQ} mechanism implemented in cellular networks \cite{ali2020relbt},
    \item select optimal \ac{TXOP} and muting periods, i.e., provide opportunities for Wi-Fi transmissions, to outperform random and round-robin mechanisms \cite{maglogiannis2018q}, 
    \item adjust the \ac{TXOP} duration of coexisting Wi-Fi and \ac{LTE-LAA} systems based on buffered downlink data in APs and \acp{eNB} \cite{xu2016dynamic}, and
    \item select optimal channel and subframe numbers \cite{kishimoto2021reinforcement}. 
\end{itemize}
\textcite{xu2016dynamic} assume that both Wi-Fi and \ac{LTE-LAA} nodes have agents, which take actions (select a \ac{TXOP} of 4, 6, 8, or  \SI{10}{\milli\second}) and calculate rewards based on the target occupancy ratio.  
A different approach is considered by \textcite{han2020reinforcement}, where a \ac{MAB} improves \ac{LTE-LAA}/Wi-Fi coexistence fairness under the assumption of both cooperative and non-cooperative networks. In both cases, the \ac{CW} sizes are optimized for the two networks by using an online training technique and either throughput or the information on LTE's ON period of the other network as rewards.
Furthermore, \textcite{tan2018learning} use two-level distributed learning. At the primary level, Q-learning determines the optimal LTE transmission time in the unlicensed bands using either Wi-Fi or \ac{LTE-LAA}. At the secondary level, stochastic learning is used for LTE-LAA channel access with the protection of Wi-Fi traffic. %
Meanwhile, \textcite{challita2018proactive} improve coexistence by combining a non-cooperative game with \ac{RL} supported by the \ac{LSTM} concept to model the self-allocation of resources by \ac{LTE-LAA} \acp{SBS}. In particular, dynamic channel selection, carrier aggregation, and fractional spectrum access are considered for \acp{SBS}. Exponential backoff is used for Wi-Fi and non-exponential backoff is used for \ac{LTE-LAA} (i.e., in each epoch a static \ac{CW} is assumed, adopted from one epoch to another). This approach not only improves performance in terms of LTE's rates but also in terms of reducing disturbances in Wi-Fi's performance and achieving coexistence fairness with Wi-Fi networks and other \ac{LTE-LAA} operators. 
Finally, \textcite{kishimoto2021reinforcement}, use Q-learning for joint channel/subframe selection. Only LTE-LAA \acp{BS} perform learning and start with zero knowledge of neighboring Wi-Fi systems.

We expect that coexistence between Wi-Fi and \ac{NR-U}  will gain a growing interest of the research community in the near future \cite{bayhan2018future, gur2020expansive,kosek2021downlink, tinnirello2021no, wang2021joint}. 
In one of the first \ac{ML}-based works, \textcite{tang2020almost} use Q-learning to adjust the timing of \ac{NR-U}'s \acp{ABS} to Wi-Fi's data transmissions to achieve higher throughput and better channel utilization in comparison to static \acp{ABS} allocation. In particular, an \ac{NR-U} \ac{BS} serves as an agent which listens to Wi-Fi network parameters and learns the data transmission rules of Wi-Fi stations. Another interesting work is  by \textcite{hirzallah2021sense}, where a clustering-based
\Ac{MAB} real-time algorithm runs on NR-U/Wi-Fi nodes to adapt sensing thresholds depending on network dynamics. The sensing threshold-adaptive devices employing \ac{ML} do not harm neighboring legacy devices (with fixed sensing thresholds) and  both Wi-Fi and NR-U throughput is improved in comparison to standard and random sensing threshold settings.

For a more generic coexistence setting, \textcite{yu2020non} address a DARPA challenge on ``autonomous radios to manage the wireless spectrum.'' DQN is modified to adapt to wireless network behavior. Through centralized learning (at the gateway) and distributed execution (at the stations) it is possible to provide fairness in channel access when coexisting with other network types (like Wi-Fi). 

\subsection{Network Monitoring}
\label{sec_net_monitoring}
Efficient network monitoring is a feature that can support inter-technology coexistence by predicting the number of contending stations/technologies, which can then guide \ac{RAT} behavior adjustment. 

\textcite{yang2019machine} propose centralized monitoring. Offline \ac{DNN}-based learning from real samples predicts the number of competing Wi-Fi and IoT devices in a given area. With the inference results as input, the gateway (connected to an IEEE 802.11 \ac{AP} using an Ethernet link and to IoT IEEE 802.14.5 stations over wireless links) predicts the number of transmitting devices for each technology using a handshake-based method on the primary channel. Next, the gateway selects Wi-Fi and IoT parameters to minimize inter-technology interference, e.g., the \ac{CW} for Wi-Fi stations, the length of the contention access phase for the IoT stations, and the assignment of secondary channels for both technologies.
Meanwhile, \textcite{el-sha2021machine} install a cognitive monitoring module in each \ac{eNB} to optimize LTE operation in unlicensed bands. The monitoring module is aware of the number of coexisting eNBs and \acp{AP}. It uses an \ac{RF}-based classifier to identify the environment state and select an appropriate scheduling and resource allocation scheme which optimizes LTE throughput without deteriorating the performance of Wi-Fi networks. 
Similarly, \textcite{galanopoulos2016efficient} use centralized Q-learning and double Q-learning to improve the unlicensed spectrum utilization for carrier aggregation of \ac{LTE-A}, while providing fair coexistence with Wi-Fi stations. eNBs learn the channel occupation time by Wi-Fi users and select the least occupied channels. This procedure is further optimized with double Q-learning, in which \ac{LTE-A} transmission power is additionally adjusted to lower the impact of \ac{LTE-A} transmissions on Wi-Fi users.

\textcite{pulkkinen2020understanding} analyze deep supervised learning-based interference detection using a real testbed. The following practical recommendations to be used in future \ac{ML}-based interference detection schemes are given: (i) deep learning-based approaches require similar levels of noise in testing and training data sets or a large number of samples with different noise levels from different environments, (ii) training should include multi-label classification.

Distributed network monitoring is proposed by \textcite{yin2021learning}, where the unsupervised \ac{NN}-based estimation of the number of coexisting Wi-Fi stations is implemented in NR-U devices. The learning process builds upon the collision probability detected in the unlicensed channel. This solution outperforms Kalman filter-based solutions. 
Furthermore, \textcite{yang2018channel} use fuzzy Q-learning to either centrally (in a central unit in C-RAN) or distributively (in each eNB) learn the Wi-Fi performance to improve the scheduling decisions on the \ac{LTE-LAA} side. 

Some papers take advantage of a dedicated interface between Wi-Fi and LTE. Fakhfakh et al. \cite{fakhfakh2017incentive, fakhfakh2017optimised} have each LTE user obtain information from the 802.11k amendment on the load of the coexisting \acp{AP}. Then, supported by Q-learning, LTE offloading decisions are made. This approach is interesting from the Wi-Fi perspective, since overloaded \acp{AP} are not selected by this mechanism and therefore Wi-Fi network performance is not worsened by the offloading decisions. %

\subsection{Signal Classification}
\label{sec_sig_classification}

ML is also used for signal classification and recognition without the need for implementing a dedicated interface between technologies or knowing the per-technology operation patterns. \textcite{wu2019deep} survey wireless modulation recognition and wireless technology recognition supported by \ac{DSL}.
We review Wi-Fi-related solutions below.

\textcite{yang2019blind} use \acp{CNN} to classify \ac{LTE-U} and Wi-Fi signals while \textcite{girmay2021machine} have LTE eNBs use \acp{CNN} to classify Wi-Fi conditions (saturation, non-saturation) without the need of decoding Wi-Fi frames, based on inter-frame space histograms.
\textcite{fonseca2021radio} also use \acp{CNN}: to classify LTE and Wi-Fi signals using an SDR-based RAT classifier. The well-known object detection \ac{YOLO} model is used for transfer learning and to speed up the training process of the classifier. The only change required was the adaption of the last layer to appropriately classify LTE and Wi-Fi signals. The developed solution provides 96\% accuracy of RAT recognition. 
\textcite{gu2020deep} use 80,000 LTE-U/Wi-Fi signal samples to train a CNN and RNN to recognize LTE-U/Wi-Fi signals. Only the CNN-based approach provides satisfactory results. 
Additionally, \textcite{mosleh2020novel} use a \ac{NN} with linear regression to track \acp{KPI} and estimate the probability of \ac{LTE-LAA}/Wi-Fi coexistence, without using knowledge of the \ac{MAC}/\ac{PHY} protocols and parameters of the two technologies.
Furthermore, \textcite{sathya2020machine} use \ac{ML} to distinguish between the presence of one or two Wi-Fi \acp{AP} interfering with an \ac{LTE-U} \ac{BS}, based on detected energy levels during the OFF periods of the \ac{DCM} instead of decoding Wi-Fi frames.

Finally, WiPlus~\cite{olbrich2017wiplus} uses ML (i.e., k-means clustering) on the Wi-Fi side to detect LTE-U interference by using the spectral scan capabilities of \ac{COTS} Wi-Fi hardware. This approach allows Wi-Fi to quantify the effective available channel airtime of each Wi-Fi link (downlink/uplink) at runtime. Moreover, the obtained timing information about LTE-U's ON and OFF phases allows Wi-Fi to schedule its transmissions only during the OFF phase to avoid collisions with LTE-U.

\subsection{Cooperative Coexistence}
\label{sec_hybrid}

Inter-network coexistence can also take on a cooperative form. 
A prominent example are Wi-Fi-Li-Fi networks, where the \ac{Li-Fi} component is responsible for data transmission using light waves (the THz band). 
\Ac{VLC} has many advantages such as high bandwidth, license-free operation, and  electromagnetic safety.
However, it has a short range and is vulnerable to link outage caused by obstructions.
Therefore, it is often paired with Wi-Fi in the form of a hybrid network.

\textcite{wu2021hybrid} provide a recent survey of research on this topic. They mention one \ac{ML}-based solution related to load-balancing, where
\Ac{RL} provides centralized AP selection to avoid servicing users by overloaded APs \cite{ahmad2020load, ahmad2020reinforcement}.
In other works, \textcite{alenezi2020reinforcement, alenezi2021reinforcement} consider the optimization of a hybrid Wi-Fi-\ac{VLC} network with centralized control and Q-learning to improve network throughput. 
\textcite{wu2020novel} use an \ac{NN} to select Wi-Fi-Li-Fi APs to avoid frequent handovers. The handover decision is made based on channel quality, resource availability, and user mobility, e.g., Wi-Fi-only APs are preferred for mobile users while Wi-Fi-Li-Fi APs are selected for static users based on received signal strength and user satisfaction levels. Finally, \textcite{sanusi2020development}, combine fuzzy logic with \ac{NN} to support Wi-Fi-Li-Fi handovers.

\subsection{Open Challenges}

We have identified several open challenges in the area of network coexistence. The performance of ML-based mechanisms is mostly verified by simulations. Therefore, real testbed validation is considered an important open challenge since it would verify the ML-based operation with real radio signals. This validation will identify crucial factors which have not been implemented yet (or are impossible to be included) in the simulators and may have been overlooked by researchers. 
Additionally, only a few papers consider adjusting the behavior of both Wi-Fi and LTE nodes. In most cases, only the LTE operation was supported by \ac{ML} while the Wi-Fi operation was left unchanged. With the opening of the new 6~GHz unlicensed band, which paves the way to redefine channel access rules defined for other unlicensed bands \cite{myles2019lbt,naik2020next,sathya2020standardization}, we believe that changes in the operation of both technologies could be considered in the future. 
Furthermore, only several papers concentrate on the new features introduced by NR-U and none address the configuration possibilities introduced by the newest 802.11 amendments (like 802.11ax). We believe that, e.g., the coexistence of \ac{NR-U} with 802.11 \ac{OFDMA}/\ac{MU-MIMO} channel access gives novel options to be considered by future \ac{ML}-based mechanisms. 
Finally, following \cite{gur2020expansive,hirzallah2019matchmaker}, we strongly agree that high attention should be paid to the security of inter-technology operation, e.g., in the case of augmenting coexisting networks with federated learning.

\section{Multi-hop Wi-Fi Networks}
\label{sec_other}
\label{sec_mhop}
\begin{figure}
\centering
\includegraphics[width=0.9\columnwidth,trim=0 0 0 0, clip]{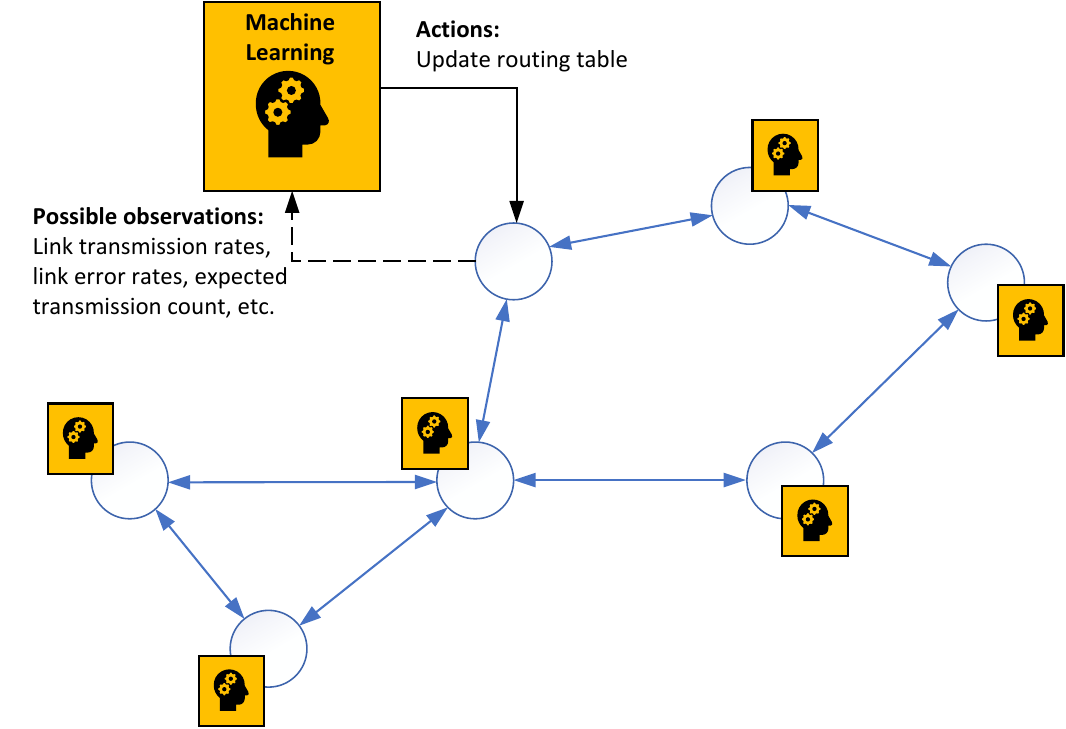}
\caption{Example of using \ac{ML} to improve routing in an ad hoc network: wireless nodes update their routing tables based on inference from observation. Each node operates a separate ML instance.}
 \label{fig_adhoc}
\end{figure}

The primary design goal for IEEE 802.11 networks is to be a single-hop access network.
However, it can also be used in a variety of multi-hop settings (e.g., ad hoc or vehicular) either using the mainline standard (802.11a/b/g/n/ac/ax) or a dedicated amendment (such as 802.11ah for \ac{IoT} and 802.11p for vehicular networks).
Research papers dealing with multi-hop settings often either do not specify the underlying technology, assume a generic \ac{CR} technology, assume heterogeneous networks (e.g., 802.11 and \ac{LTE}), or use an alternative technology (which could theoretically be replaced by Wi-Fi).
One of the reasons for this is that the key multi-hop problem, routing (Figure~\ref{fig_adhoc}), is beyond the scope of 802.11.
Therefore, in the following, we provide only a general overview of how \ac{ML} is applied in various multi-hop settings: ad hoc, mesh, sensor, vehicular, and relay networks.
We point the reader towards relevant surveys and tutorials in each area and note that a detailed overview of using \ac{ML} in multi-hop wireless settings could be the topic of a separate survey.

\subsection{Ad Hoc Networks}
The research popularity of (generic) ad hoc networks and \acp{MANET} reached its peak over a dozen years ago.
They have mostly been replaced by their more application-oriented variants (mesh, sensor, vehicular, etc.) which we will discuss further on\footnote{Ad hoc networks consisting of \acp{UAV} are a dedicated use case, for which \ac{RL}-based solutions are also being developed \cite{rezwan2021survey}.}.
An overview of applying \ac{ML} techniques to ad hoc networks is found in a 2007 paper by Forster \cite{forster2007machine}.
The state of the art reported in this paper is outdated, 
but the list of applicable \ac{ML} techniques (\ac{RL}, swarm intelligence, mobile agents, etc.) and use cases (mainly improving routing efficiency) remains current. 
Al-Rawi et al. \cite{alrawi2015application} provide an overview of applying \ac{RL} to improve routing in distributed wireless networks. 
More Wi-Fi-related examples include  
applying Q-learning to the \ac{OLSR} routing protocol \cite{mcauley2012tuning,wagen2020comparing,serhani2020routing} and 
applying \ac{RL} to 802.11-based \acp{DTN} \cite{rolla2013reinforcement} .

\Ac{ML} can also optimize ad hoc network configuration \cite{lee2021intelligent}, but this example is for a \ac{CRAHN} (i.e., without 802.11).
Another active area of research for \acp{MANET} is mobility prediction \cite{zhang2019mobility}, but again the \ac{ML}-based solutions do not explicitly consider Wi-Fi \cite{ghouti2013mobility,yayeh2018mobility}.
Similarly, research on applying Q-learning to interference cancellation in ad hoc networks also does not consider Wi-Fi \cite{mete2020qlearning}.

\subsection{Mesh Networks}
\acp{WMN} consist of static wireless nodes which are usually deployed to distribute Internet access to clients in an area.
\textcite{karunaratne2019overview} provide a recent tutorial on \ac{ML}-based approaches for \acp{WMN}.
Important problems which are solved with \ac{ML} include routing, channel assignment, and network deployment. 
\ac{ML} techniques (such as \ac{SVM}, k-means clustering, and Q-learning) are mapped to the identified \ac{WMN} problems.
Future research directions include the potential of \ac{DL}, which has been recently demonstrated in the context of network flow optimization \cite{cheng2021deep}.

\textcite{niyato2009cognitive} show how Q-learning helps clients perform  \ac{AP} selection in an IEEE 802.11 mesh network.
Decisions are based on the estimated collision probability and received signal strength.
The learning approach outperforms a best signal strength heuristic, especially under non-uniform node distribution.

Another example is training an \ac{NN} to predict link bandwidth in an 802.11 mesh network \cite{oriol2012available}. 
The inputs are the averages of important \ac{PHY} and \ac{MAC} metrics: \ac{SNR}, transmission time, \ac{MCS}, and re-transmission rate.
The approach accurately predicts link bandwidth, which can then be used as a routing metric.

Link quality prediction is also the topic of a paper by \textcite{botelorenzo2018online}.
Based on an extensive dataset from an existing 802.11-based community \ac{WMN}, four \ac{ML} algorithms for regression (online perception, online regression trees with options, fast incremental model trees with drift detection, and adaptive model rules) are evaluated.
Only the first of these outperforms a simple baseline and only under certain circumstances. 
This leads to the design of a hybrid algorithm, which supports the thesis that applying \ac{ML} is not a straightforward approach.

For heterogeneous (Wi-Fi and \ac{LTE}) mesh networks, the routing protocol is enhanced by Q-learning for \ac{RAT} selection \cite{alsaadi2016routing}.
Each node performs observations as follows: 
LTE link quality is determined by network load (measured through buffer occupancy), Wi-Fi link quality -- according to the current \ac{PHY} transmission rate.
Through appropriate \ac{RAT} selection, nodes can observe up to 200\% throughput increase compared to the single-technology case.

Finally, we comment on the dedicated 802.11s amendment for mesh networks. 
Among its features, it introduced \ac{MAC}-layer routing called \textit{path selection} in the form of the \ac{HWMP}.
However, our literature review did not identify any papers directly related to applying \ac{ML} for improving the performance of either \ac{HWMP} or other 802.11s functionalities. 
\textcite{testi2019machine} analyze network topology inference using external sensors for a simulated 802.11s network, but no specific mesh functionalities are considered.
The lack of dedicated 802.11s research is most likely the result of the limited deployment of 802.11s by the industry.

\subsection{Sensor Networks}

The application of \ac{ML} to sensor networks (i.e., the communication part of IoT) is an active research topic  \cite{alsheikh2014machine,cui2018survey,jagannath2019machine,hussain2020machine,lei2020deep, sharma2020toward, nguyen2020wireless}.
Among the most important network performance research problems for sensor networks, which are solved with \ac{ML} methods, are:
sensor grouping (clustering, data aggregation),
energy-efficient operation (scheduling, duty cycling),
resource allocation (cell/channel selection, channel access),
traffic classification,
routing, mobility prediction,
power allocation,
interference management,
and
resource discovery \cite{hussain2020machine}.
However, Wi-Fi is only one of many \ac{IoT}-enabling technologies and 802.11-related solutions are rarely mentioned in these surveys.
The only direct performance-related area mentioned in these surveys is
classifying 802.11 interference using a \ac{DCNN} \cite{kulin2018end,schmidt2017wireless}, \ac{SVM} \cite{grimaldi2017svm}, or various types of \ac{SL} classifiers: \acp{CT} and \ac{SVM} \cite{grimaldi2019realtime}.

There are two 802.11 amendments related to IoT: 802.11af and 802.11ah.
The former is a \ac{CR}-based approach to use Wi-Fi in TV white space spectrum and has not enjoyed commercial success.
Thus, there are also few research papers related to improving 802.11af performance with \ac{ML}.
A singular example is the work by \textcite{xu2019efficient,xu2020autonomous} on 802.11af rate adaptation schemes, which use \ac{DL} models, although their work is in the context of vehicular networks.

Meanwhile, the 802.11ah amendment has had more commercial success (as HaLow) and received more attention from the research community.
However, while 802.11ah permits tree-based multi-hop communication \cite{ahmed2022mac}, it is a predominantly single-hop technology.
This observation is reflected in a recent survey on 802.11ah research \cite{tian2021wifi} where, out of about 200 cited references,
only three consider multi-hop scenarios.
Moreover, surprisingly, only two papers by \textcite{tian2018restricted,tian2019optimization} deal with applying \ac{ML}: both use a form of \ac{SL} to optimize the parameters of 802.11ah's grouping functionality, \ac{RAW}.
A similar problem is also addressed by \textcite{mahesh2020ann}, where an \ac{MLP} \ac{NN} configures these parameters considering, i.a., network size and \ac{MCS} values used.
Other applications of \ac{ML} to 802.11ah include:
improving coexistence with 802.15.4g devices, a type of \ac{LR-WPAN}, by avoiding interference with their transmissions using a Q-learning-based backoff mechanism \cite{liu2018coexistence},
grouping sensors based on their traffic demands and channel conditions using a regression-based model \cite{chang2019traffic},
grouping sensors based on their data rates by classifying them with \acp{NN} \cite{mahesh2020data},
and 
improving carrier frequency offset estimation using various types of \acp{DNN} \cite{ninkovic2021deep}.

Finally, research is also being done for generic Wi-Fi (i.e., the mainline amendments). 
\textcite{zhao2020deep} propose a \ac{DQL}-based method of optimizing \ac{CW} for energy-constrained IoT networks.
\textcite{chen2020contention} also optimize \ac{CW} but using a \ac{DNN} for IoT networks using 802.11ax.
\textcite{shin2020distributed} provide a method for \ac{RAT} selection, between Wi-Fi and \ac{NB-IoT}, using \ac{RL} to optimize for per-node latency.
This has been further extended for mobile sensor networks incorporating \acp{UAV}.
\textcite{kurunathan2021deep,li2021continuous,li2021lstm} present a learning-based approach using \ac{DQN} and \ac{DDPG} for trajectory planning and integrated communication.

\subsection{Vehicular Networks}
There has been much research in the area of applying \ac{ML} to vehicular networks, with Wi-Fi being only one of the many considered wireless access technologies. 
Some recent surveys and tutorials include \cite{ye2018machine, liang2019toward, tong2019artificial, xu2019intelligent, noorarahim2020survey, posner2021federated,memedi2021vehicular}.
They point to the application of \ac{ML} in vehicular networks in the following areas of performance improvement: channel estimation, traffic flow prediction, location prediction-based scheduling and routing, network congestion control, load balancing and handovers, and resource management.
Other non-performance areas where ML is applied include vehicle trajectory prediction (for ensuring road safety), network security, and in-car infotainment \cite{hossain2020comprehensive}.

From the Wi-Fi perspective, 802.11p is the amendment dedicated to vehicular networks and is included in larger \ac{V2X} frameworks such as \ac{DSRC} and the ETSI ITS-G5 standard \cite{bazzi2017performance}. 
\textcite{noorarahim2020survey} review \ac{ML}-based resource allocation approaches in \ac{DSRC} networks.
Examples of using \ac{ML} for improving 802.11p performance include:
using \ac{DRL} for per-link band and transmission power allocation \cite{ye2018deep},
\Ac{RL} for tuning the \ac{CW} size \cite{pressas2017contention,pressas2019learning,kim2020reinforcement},
Q-learning for improving handoff decisions \cite{xu2014fuzzy},
improving \ac{TCP} performance with federated learning \cite{pokhrel2020improving},
\acp{DNN} for channel estimation \cite{pan2021channel},
and
using \ac{RL} for selecting the data transmission rate in a high-mobility scenario \cite{xu2019intelligent,xu2020augmenting}.

An emerging future research direction is applying \ac{ML} to 802.11bd, the successor to 802.11p scheduled for release in 2022 \cite{naik2019evolution}.
Beam alignment is one important problem of \ac{mmWave} bands (cf.\ \cref{sec_mmWave}). 
However, contrary to \ac{WLAN} scenarios, the knowledge of a vehicle's position supports beam sector selection \cite{va2017position}, where \ac{LtR}, also referred to as \ac{MLR}, can rank antenna pointing directions.
The extention of the input information from just the location of the receiver to the location of surrounding vehicles, called situational awareness, can improve the performance of \ac{ML}-based algorithms. 
Beam alignment is determined using classifiers \cite{wang2018training, wang2019selection} or regression models \cite{wang2018prediction,wang2018towards}.
Throughput is satisfactory even if the best beam pair is not selected, providing an accuracy-overhead trade-off.

\subsection{Relay Networks}

The typical single-hop 802.11 deployment scenario is extended to a two-hop case with \textit{cooperative communications}, where stations are allowed to relay the transmissions of others \cite{sadeghi2017survey}.
Such functionality requires appropriate coordination between the AP and stations, which is enhanced by a mechanism to support concurrent transmissions from different devices in a \ac{WLAN} setting \cite{hu2010ccmac}.
Since the \ac{AP} may not have full information of the whole network, the problem is modelled as a \ac{POMDP} and solved by an \ac{RL} algorithm. 
that can find which senders can transmit simultaneously. Results show that low-rate links, usually corresponding to distant stations, significantly improve their throughput.
Despite this singular example, \ac{WLAN}-based relay networks have received limited interest from \ac{ML} researchers.
If relay networks become an important feature of future Wi-Fi networks, solutions can be borrowed from 5G networks such as \ac{ML}-based relay selection  \cite{abdelreheem2019deep}.

\subsection{Open Challenges}
While a multitude of \ac{ML}-related open research challenges can be listed for multi-hop networks in general, much less can be named if we restrict our focus to Wi-Fi-based ones, because Wi-Fi is predominantly used in single-hop deployments.
Even the latest amendments dedicated to sensor (802.11ah) and vehicular (802.11bd) networks mainly operate over single hops.

One area where Wi-Fi is used for wireless multi-hop transmissions is \emph{providing \ac{FWA} over \ac{mmWave} links} (cf. Section~\ref{sec_mmWave}).
\ac{FWA} is an important use case for 802.11ay, where coverage is extended with a mesh-like distribution network \cite{aldubaikhy2020mmwave}. 
Research is required in developing new (or adopting existing) \ac{ML}-based solutions to this particular scenario in the areas of resource allocation and resource coordination.
An example solution is provided by \textcite{lahsencherif2021energy}: a \ac{QL}-based routing protocol optimizes energy and throughput in a backhaul \ac{WMN} scenario with directional links but Wi-Fi is not  explicitly stated as the wireless technology.

Another area with open challenges is \emph{relay selection for vehicular networks}. 
\textcite{zugno2020toward} suggest a cross-layer approach combining routing with the 802.11 stack.
\Ac{ML} could assess per-link routing cost more accurately.
Alternatively, auxiliary sources of information could support vehicular relay selection. 
A first example comes from Morocho-Cayamcela et a. \cite{morocho2020machine}, where an \ac{ML} algorithm selects relays based on satellite imagery. 
Such imagery and other types of auxiliary information, combined with the power of \ac{ML}, can potentially improve vehicular network performance.

\section{Available Tools and Datasets}

The review of research papers in the previous sections confirms that ML-based control solutions often overtake traditionally designed ones in terms of performance and efficiency.
However, to reach such high performance levels, long training is required. 
For example, an RL agent needs many interactions with an environment to learn the best policies, while in SL, the tuning of an ML model requires access to large labelled datasets.
In this section, we describe the available research tools, datasets, and testbeds that were used in the reviewed papers and are available for other researchers in the field.

\subsection{Tool Chains}

From our keyword analysis of more than 250 papers combining \ac{ML} with Wi-Fi, regarding the evaluation methodology, we found that most researchers run network simulations ($\approx$\,80\%) to validate their solutions.
Only around a quarter of them perform analytical investigations or experiments in real testbeds.
The lack of real-life experiments is understandable as they are often complex, risky, and expensive to execute.
For simulation analysis, the ns-3 network simulator\footnote{\url{https://www.nsnam.org}}, known from non-ML networking research, is the most popular with a share of 10\%.
Meanwhile, experimental studies were mostly based on \ac{SDR} platforms like Ettus USRPs\footnote{\url{https://www.ettus.com}} whereas \ac{COTS} Wi-Fi hardware, mostly with Atheros and Intel chipsets, was rarely used.
The most commonly used \ac{ML} libraries were Tensorflow (10\%) and Keras (5\%).

\begin{figure}[t!]
\centering
\includegraphics[width=0.68\linewidth]{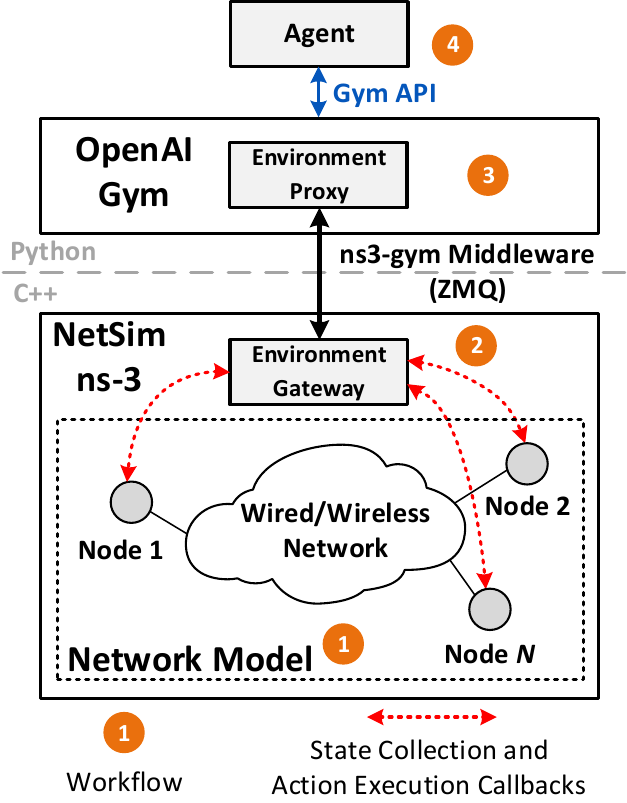}
\vspace{-2pt}
\caption{Architecture of the ns3-gym framework \cite{gawlowicz2019ns3gym}: the network model and scenario configuration is done in ns-3 (1);  ns3-gym provides an environment gateway (2) and proxy (3) for communication with the \ac{RL}-based agent; the agent is developed using standard libraries and interacts with the environment using a standard interface (4).}
\vspace{-16pt}
\label{fig:ns3gymarch}
\end{figure}

\begin{figure}[t!]
\centering
\includegraphics[width=0.68\linewidth]{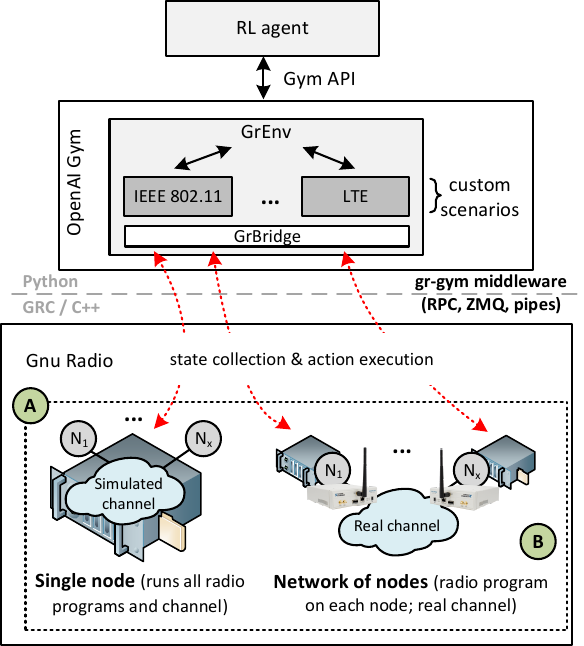}
\vspace{-2pt}
\caption{Architecture of the GrGym framework \cite{zubow2021grgym}, which provides an interface to integrate GNU Radio and OpenAI Gym.}
\vspace{-5pt}
\label{fig:grgymarch}
\end{figure}

Based on the results of our analysis, it becomes evident that the seamless support of network simulators (like ns-3) and \ac{SDR} platforms for research of \ac{ML}-based solutions for Wi-Fi is of great importance.
We have observed the first research frameworks which aim to simplify the integration of \ac{ML} and Wi-Fi.
The general role of network simulators for bridging the gap between \ac{ML} and communications systems like Wi-Fi is discussed by Wilhelmi et al.~\cite{wilhelmi2021usage}, where possible workflows for \ac{ML} in networking and the use of existing tools is presented.
Among these is ns3-gym, a software framework enabling the design of \ac{RL}-driven solutions for communication networks, proposed by Gawlowicz et al. ~\cite{gawlowicz2019ns3gym}. %
This framework is based on the OpenAI Gym toolkit\footnote{\url{https://gym.openai.com}} and provides an extension to the ns-3 network simulator (Figure~\ref{fig:ns3gymarch}).
With ns3-gym, it is possible to use any simulated communication network (e.g., mixed Wi-Fi and \ac{LTE}) as a Gym environment so that \ac{RL} agents can control the behavior of network protocols.
OpenAI Gym has also been integrated with Veins~\cite{sommer2011bidirectionally}, a popular open source vehicular networking simulator based on OMNeT++.
The resulting VeinsGym~\cite{schettler2020how} supports \ac{ML} both at the protocol as well as at the application level.
\textcite{yin2020ns3} provide ns3-ai, which offers the same functionality as ns3-gym but better performance by using shared memory for inter-process communication when running both the simulation and the Gym agent locally.
GrGym~\cite{zubow2021grgym} is a similar framework, but it builds on the GNU Radio\footnote{\url{https://www.gnuradio.org}} signal processing platform.
Any GNU Radio program can be integrated as an environment in the Gym framework (Figure~\ref{fig:grgymarch}) by exposing its state and control parameters for the agent's learning purposes. %
In contrast to ns3-gym, GrGym allows the Wi-Fi network to be a real testbed consisting of SDR nodes performing real transmissions over the air.
This enables studying the performance of an ML-based solution under real channel and interference conditions.
The downside is the higher effort required to setup a network as well as the lack of reproducibility.
Finally, Komondor\footnote{\url{https://github.com/wn-upf/Komondor}} is another network simulator which supports a subset of the 802.11ax standard.
This tool is designed for simulating complex environments in next-generation Wi-Fi networks with direct \ac{ML} support.
\textcite{barrachina2019komondor} identify several use cases and present \ac{ML}-based solutions using Komondor.

\subsection{Datasets}

The existence of open-source and standardized datasets is essential for training and comparing ML-based algorithms. Moreover, such datasets accelerate development and foster reproducible research. For example, the recent advances in image classification and recognition were enabled by the emergence of large labelled image datasets (e.g., ImageNet \cite{deng2009imagenet}). %

We have found that researchers usually rely on their own datasets. Specifically, in 49 papers, they created labelled datasets by running experiments in testbeds and/or simulators, while only in 6 articles they used publicly available datasets. 
Moreover, while being a good practice, releasing the created dataset along with the published paper is still not the case for most of the publications (i.e., only 6 datasets were released). Here, we describe datasets available online that the community can immediately use for further ML-based Wi-Fi performance optimization.

CRAWDAD\footnote{\url{http://www.crawdad.org}} is a repository with a vast set of Wi-Fi measurements. 
The datasets include traces from smartphones performing Wi-Fi scans, multipath TCP traces collected from a Wi-Fi campus network, as well as traces collected for other wireless technologies like Bluetooth and ZigBee. 
\textcite{challita2018proactive} used a subset of the  CRAWDAD dataset which included records (e.g., information about the amount of transfered data, error rates, signal strength) collected by polling Wi-Fi APs every 5 minutes in a corporate research center over several weeks. Similarly, a dataset called \textit{sigcomm2008} contains traces of wireless network measurements collected during the SIGCOMM 2008 conference.

IEEE DataPort\footnote{\url{https://ieee-dataport.org/}} is another large repository of datasets created to encourage reproducible research. 
Within this repository, \textcite{karmakar2020deep} provide the IEEE 802.11ac performance dataset\footnote{\url{https://ieee-dataport.org/documents/ieee-80211ac-performance-dataset}} that contains information regarding normalized throughput achieved under five link configuration parameters (i.e., channel bandwidth, MCS, guard interval, MIMO, and frame aggregation) and the channel quality measured as SNR.
 
Kaggle\footnote{\url{https://www.kaggle.com/}} is an online platform for data scientists and machine learning practitioners. 
The platform allows users to find and publish datasets. Moreover, it is frequently used by companies to organize competitions to solve data science challenges. At the time of writing, the Kaggle platform offers only a limited number of Wi-Fi-related datasets, e.g.,  the \textit{Wi-Fi Study}\footnote{\url{https://www.kaggle.com/mlomuscio/wifi-study}} dataset contains a study of the quality of the Wi-Fi and user perceptions of Wi-Fi conducted by students in a dormitory.

Next, we briefly describe the datasets from the reviewed papers that are  available from researchers on their individual webpages. 
Herzen \textit{et al.}~\cite{herzen2015learning} provide a dataset to predict throughput based on basic performance metrics (e.g., received power, channel width) collected in a small testbed\footnote{\url{http://www.hrzn.ch/data/lw-data.zip}}.
\textit{Cell vs. Wi-Fi}\footnote{\url{http://web.mit.edu/cell-vs-wifi/}} is a publicly available dataset based on an Android application that collects packet-level traces of TCP downlink and uplink traffic between a mobile device and a server for both Wi-Fi and cellular networks. 
The dataset is used to find hidden dependencies in low-level Wi-Fi performance data \cite{chaudhry2020finding}.
\textcite{polese2021deepbeam} provide an experimental waveform dataset\footnote{\url{http://hdl.handle.net/2047/D20409451}} generated using the NI mmWave transceiver system with 60\,GHz radio heads, as well as the source code using Keras API for training and testing \ac{ML} models\footnote{\url{https://github.com/wineslab/deepbeam}}.
Similar measurement data for indoor mmWave using 802.11ad from the papers by \textcite{aggarwal2020experimental, aggarwal2020learning} is also available\footnote{\url{http://bit.ly/60ghz-link-adaptation}}.
Rice University’s LiveLab dataset\footnote{\url{http://livelab.recg.rice.edu/traces.html}} contains long-term measurements from real-world smartphones about their usage (e.g., CPU time) as well as data collected over a Wi-Fi interface (e.g., periodic readings of available Wi-Fi access points). The dataset is used by~\textcite{chakraborty2016exbox} for admission control in wireless networks supported by light-weight machine learning.

The available datasets provide mostly raw measurements (e.g., RSSI, CSI) or traces of sniffed Wi-Fi traffic
which are used to find anomalies with ML techniques.
For example, \textcite{fulara2019machine} detect the causes of unnecessary active scanning performed by Wi-Fi stations. 
Moreover, there exist datasets meant for Wi-Fi-based applications (e.g., human detection, activity recognition, people tracing, traffic classification) which rely on \ac{ML}. 
We believe that such datasets can also be  used to improve the performance of Wi-Fi networks.  For example, if an \ac{AP} knows that a traffic flow is a long-lived flow (e.g., a video transmission), it might perform long-term optimizations to improve the flow quality that would not make sense for a short-lived flow. 
Moreover, the location tracking of Wi-Fi stations can help a Wi-Fi network prepare for a handover operation in advance, which would result in faster handover execution and a smaller number of outage events.
Example datasets containing location information and Wi-Fi signal strength are available on the Kaggle platform\footnote{\url{https://www.kaggle.com/c/indoor-location-navigation/}}.

Finally, we believe that significant efforts have to be taken to create large and high-quality datasets and encourage sharing them among the wireless research community. To this end, it would be beneficial to create standardized procedures for data collection to allow researchers to cooperatively build new and extend existing datasets. The potential use of different wireless platforms/testbeds for measurements might positively impact learning performance (e.g., avoid model overfitting). Due to diverse hardware characteristics (such as TX power), however, the created datasets have to be precisely described (i.e., provided with complete metadata) to avoid misunderstanding and unnecessary debugging of the \ac{ML} models.

\subsection{Testbeds}
\label{sec:testbeds}
To support the experimental evaluation of ML-based Wi-Fi solutions, open-access wireless testbeds are helpful \cite{bonati2022advances}.
Examples of such testbeds which support 802.11-based networking include Orbit~\cite{raychaudhuri2005overview}, COSMOS~\cite{raychaudhuri2020challenge} and POWDER~\cite{breen2021powder}. 
These testbeds provide not only hardware, but also software support (crucial for deploying \ac{ML}).
For example, in the Orbit testbed, the former requirement is addressed by having integrated \acp{GPU} in the wireless nodes which speed up the learning process. 
Moreover, preconfigured Linux images with ML tools like Keras and Tensorflow are preinstalled to accelerate the implementation of novel ML-based solutions for WiFi.
The case is similar for the other testbeds -- they support ML-related studies (even if it is not their main goal) and such research has been performed with these testbeds \cite{yang2020cosmos,reus2020trust}.

\section{Future Research Directions} \label{Sec:Future}

Through all previous sections, we have overviewed, discussed, and systematically classified many research works aiming to improve Wi-Fi through machine learning. All these works have a similar motivation: the use of ML to find what are the best decisions that a Wi-Fi network, or its different functionalities, can make to offer better performance in  changing and heterogeneous scenarios. Although we covered over 250 papers, they represent only the first step of a long path towards fully adopting ML in future Wi-Fi and wireless networks in general. In the following, we describe several general open challenges and suggest potential future research directions.

\subsection{Dealing with New and Flexible but Complex Wi-Fi Features}

In recent years, the catalog of available Wi-Fi functionalities has been rapidly expanding to include more complex features to cope with current and future user needs. For example, IEEE 802.11be will incorporate multi-link operation and, possibly, multi-AP coordination in addition to already existing features such as OFDMA, downlink and uplink MU-MIMO, spatial reuse, and channel aggregation. A common aspect of most of these functionalities is that they offer a high degree of flexibility to schedule traffic in time, space, and frequency, which, if properly used, may enable high-performance gains. 

To achieve this goal, ML techniques may play an important role, enabling self-adaptation to different situations and scenarios, as well as improving decision making by leveraging past information to predict next actions. For example, multi-band Wi-Fi devices can use \ac{ML} methods to predict link quality and select links accordingly \cite{deng2020ieee}.

\subsection{Joint Optimization of Wi-Fi Features}

Most of the discussed papers focus on the optimization of a single Wi-Fi feature like the CW of Wi-Fi's channel access function. However, it becomes clear that separate Wi-Fi features cannot be optimized in isolation. Instead, they must be jointly optimized with others to achieve the best possible performance. As an example, consider the tuning of transmit power and carrier sensing threshold to enhance spatial reuse \cite{wilhelmi2021spatial}.
Hence, the research on ML schemes suitable for joint optimization of multiple Wi-Fi features is a promising future research direction. Especially developing ML solutions with a fast learning speed is of great importance due to the high complexity involved. For example, hierarchical learning principles allow improving learning speed by decomposing complex joint optimization problems into multiple sub-problems \cite{pateria2021hierarchical}. 

\subsection{ML-enhanced Wi-Fi Features by Design}

Most of the discussed works build ML functionalities on top of current Wi-Fi features, by tuning their parameters. An open challenge and a disruptive future approach would be to redesign these functionalities by explicitly embedding ML capabilities in them. Heuristic algorithms or hard-coded rules could be replaced by ML agents able to self-configure based on gathered experience \cite{pasandi2021towards, moon2021neuro}. For example, 
providing guaranteed QoS or spatial reuse are challenges which could benefit from being designed with built-in \ac{ML} capabilities \cite{deng2020ieee}.

\subsection{ML-based Architectures and Standardized Interfaces}

Another open challenge to solve is where to perform and execute certain ML-related actions, which in the case of Wi-Fi networks may include the device, the AP, a controller in the network edge, and a controller in the cloud. In any case, the answer to this question requires knowing aspects such as the tolerable latency required to obtain the output of an ML process, the required information to perform it, and the computational resources. The design and orchestration of distributed ML solutions that adapt to the pros and cons of each case is still an open challenge, requiring the definition of new interfaces as well as how and when to exchange data and ML models between components. 

\begin{figure}
\centering
\includegraphics[width=\columnwidth,trim=0 0 0 0, clip]{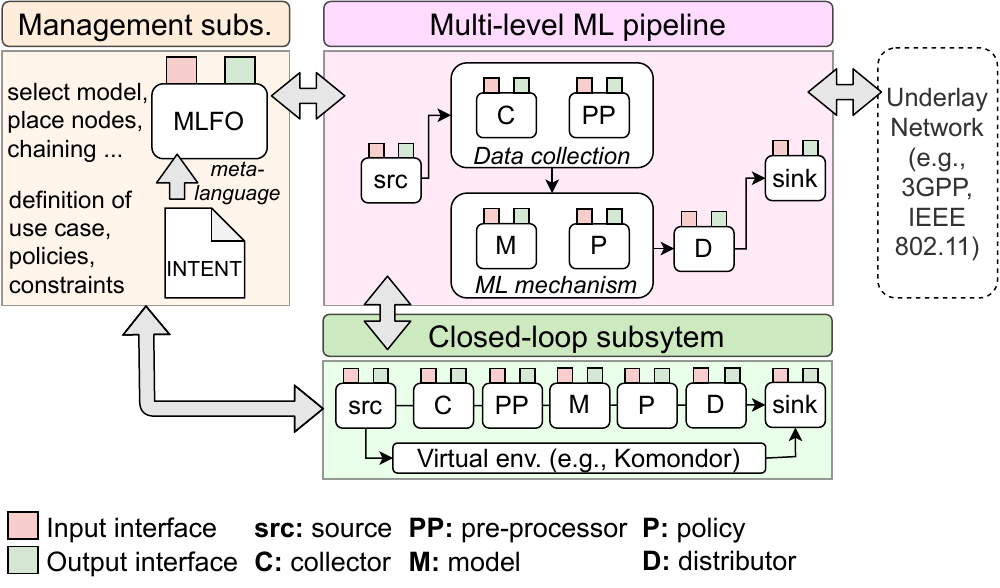}
\caption{ITU's logical architecture for future networks~\cite{wilhelmi2020flexible}.}
 \label{fig_MLWiFiArchitecture}
\end{figure}

A pioneering work dealing with these aspects for WLANs is by \textcite{wilhelmi2020flexible}, where the \ac{ITU} unified architecture for 5G and beyond is extended to support ML techniques at multiple levels, from the end device to the cloud. This work is then complemented with a  `sandbox' element of the ITU-T architecture (Figure~\ref{fig_MLWiFiArchitecture}) to execute off-line training \cite{wilhelmi2021usage}. 
Validation of ML techniques and models is further analyzed and discussed. Another framework to consider is the \ac{ETSI} \ac{GANA} \cite{chaparadza2013implementation}, which defines  decision-making entities (and their associated control-loops) where \ac{ML} can be applied.

\subsection{Reference Evaluation Scenarios and Performance Metrics}
Almost all published papers considering ML techniques conclude they can significantly improve the system performance. While we do not question these results, we point out the lack of a set of common scenarios. This situation prevents the direct comparison of the results between different papers, and therefore, makes it challenging to extract solid conclusions and track the progress in the area of using ML for enhancing Wi-Fi. Designing these scenarios in a way so they are useful to test ML solutions is challenging.
Specifically, the evaluation scenarios should cover a wide range of difficulty levels.
For example, in the beginning training phases, small stationary scenarios are helpful to illustrate and debug how ML solutions work. However, later on, the environment dynamics should be also considered, as they must be complex enough to include non-straightforward situations. Specifically, successful ML-based proposals should be tested in large, heterogeneous, and dynamic scenarios to show that they properly adapt and scale to different conditions.

Additionally, a set of common scenarios will foster another open challenge: reproducible research. This aspect is  important due to the amount of information required to reproduce exactly, step by step, the same environmental conditions and ML process responses in different places and by different actors. The use of detailed and accurate datasets may contribute to making this possible.

\subsection{ML-enhanced Network Simulation Tools}

The development and maintenance of reference scenarios is much easier with a set of simulation frameworks, standardized and commonly accepted by the research community.
However, there is still a lack of tools, which would seamlessly integrate ML solutions. Although there have been some attempts to solve this situation (e.g., the OpenAI module for ns-3 \cite{gawlowicz2019ns3gym} and Komondor \cite{barrachina2019komondor}), we are still far from a point where general networking simulators will allow including ML routines by default. Achieving a solution will be challenging, as we need to (i) define standard interfaces between Wi-Fi components and ML functions and (ii) incorporate the execution times required by ML instances as part of the virtual simulation time.

\subsection{Testbeds and Real Pilots}

The previous discussion regarding the need for scenarios and suitable simulators can be directly extended to the need for testing the correct operation of ML-enhanced functionalities in real networks, not only to validate their correct operation, but also to run experiments in conditions that simulators may not be able to reproduce accurately. 
Therefore, the development of platforms and testbeds that support the experimental research of Wi-Fi-enhanced ML networks is a crucial aspect before deploying these solutions in real networks. An important aspect to consider, and which should be included in the design of ML-aware solutions, is that they will have to coexist with non-ML-enabled solutions, and so potentially negative interactions should be considered in advance. 
An example of this is the Orbit testbed which provides access to nodes with hardware (\ac{GPU}) and software (Keras/TensorFlow) support for ML-based research (cf. Section~\ref{sec:testbeds}).

\subsection{Risks of ML Uncertainty}

Following the previous points, it is important to explicitly tackle situations in which ML techniques cause unpredictable performance and may compromise the correct operation of a certain feature or even the whole Wi-Fi network. An open challenge is to design robust ML solutions that may sacrifice performance in general to prevent unexpected behaviors in particular scenarios.

ML-based models are highly successful and provide superb performance in many complex tasks. So far, however, models are applied in a \textit{black-box} manner, i.e., no information is provided about what exactly makes them arrive at their decisions. This lack of transparency can be a major drawback and might remain a limiting factor for the broad adoption of ML-based algorithms in the area of wireless network control. Specifically, giving up human control to an \textit{intelligent black-box} brings the risk of improper behavior or unsafe decisions that might be dangerous for the operation of wireless networks, which in many cases may be considered critical infrastructure. In recent years, research on explaining and interpreting deep learning models attracted increasing attention: the work of \textcite{Samek2019} targets validation of agent behavior and establishing guarantees that they will continue to perform as expected when deployed in a real-world environment. Furthermore, by explaining the internal structures, researchers hope to learn from ML-based agents capable of learning patterns that are not tractable by humans. To conclude, the explainability of ML agents will be of significant importance for the verification and certification (i.e., checking compliance with regulations) of ML-based wireless network control systems.

\subsection{New ML Models and Distributed Learning}

\begin{figure*}[!t]
    \subfloat[]{
        \includegraphics[width=0.31\textwidth]{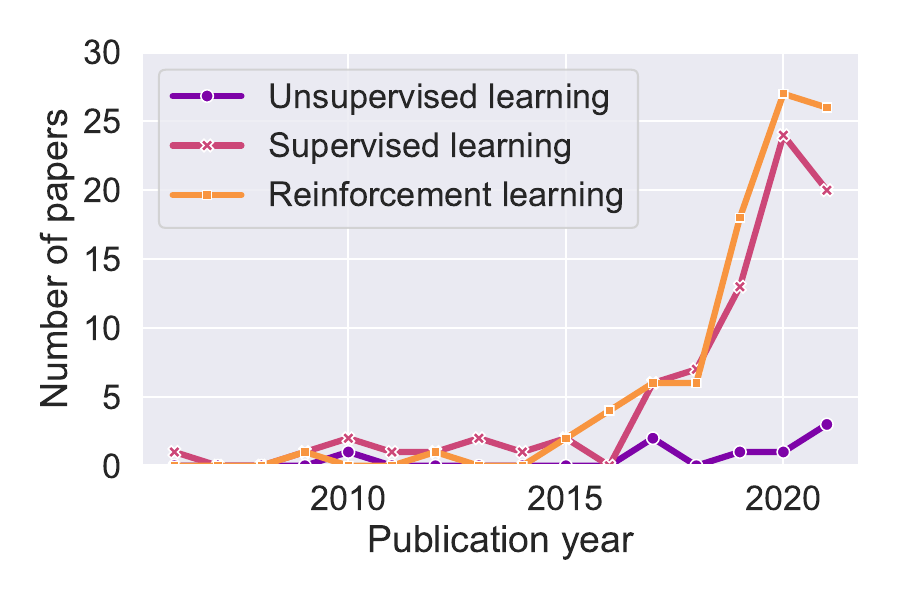}	
        \label{fig:annual_stats_ml}
    }
    \subfloat[]{
        \includegraphics[width=0.31\textwidth]{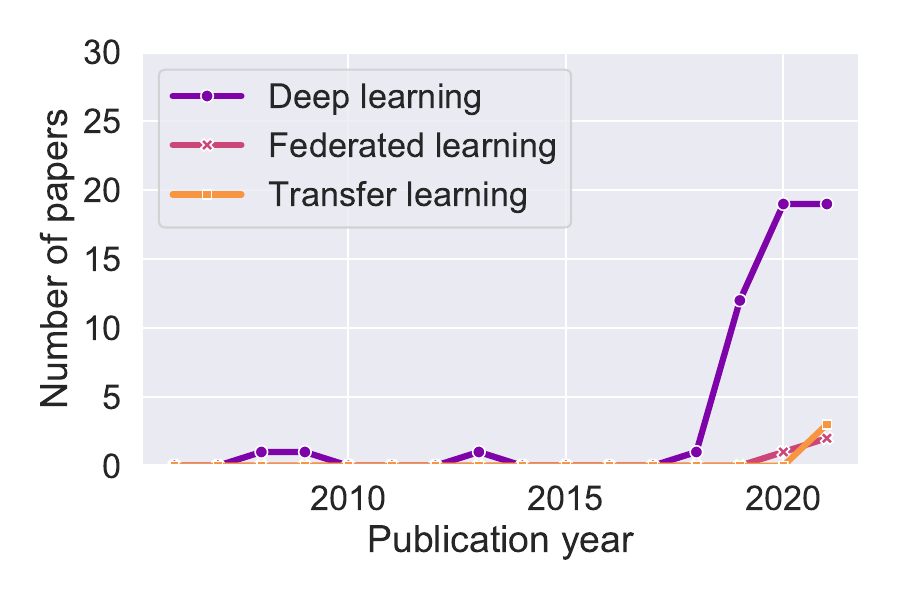}	
        \label{fig:annual_stats_new_ml}
    }  
    \subfloat[]{
        \includegraphics[width=0.31\textwidth]{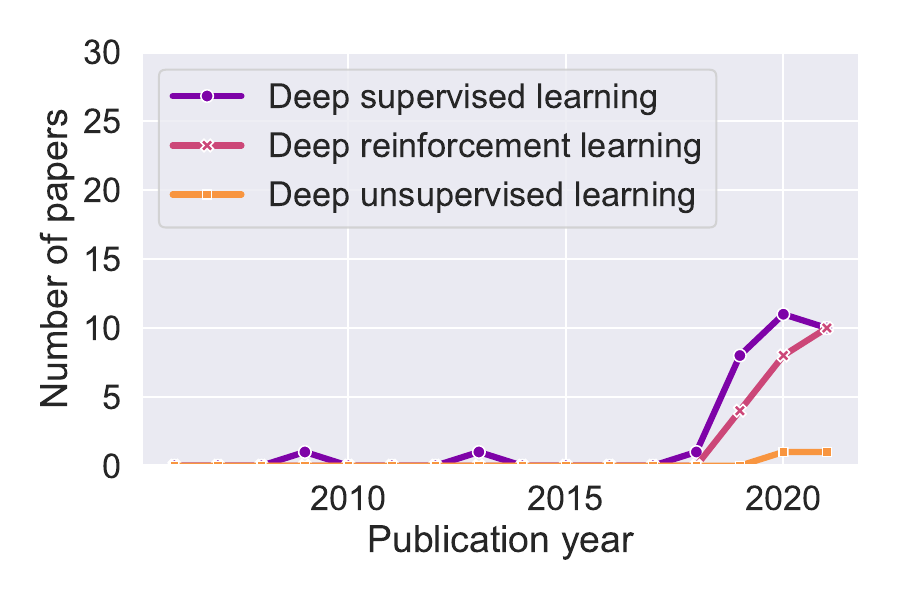}	
        \label{fig:annual_stats_dl}
    }    
    \caption{Comparison of relevant published papers by field: (a) in the three main \ac{ML} areas, (b) using new \ac{ML} paradigms, (c) using a \ac{DL} variant.}
    \label{fig:annual_stats}
\end{figure*}

Another open challenge is the need to consider recent advances in ML techniques, which will certainly go together with the definition of new ML-based architectures and Wi-Fi features. For instance, due to its recent introduction, there are still few papers considering \acf{FL} models for Wi-Fi (Figure~\ref{fig:annual_stats_new_ml}). \ac{FL} is a distributed machine learning paradigm where a set of nodes cooperatively train an \ac{ML} learning model with the help of a centralized server and without the need to share their local data~\cite{malandrino2021federated, zhang2021survey}. Specifically, nodes train their local model based on local (on-device) data, and then send the model parameters to the server, which in turn merges parameters from different nodes and sends the combined (global) parameters back to the distributed nodes. We expect that \ac{FL} is of paramount importance for the optimization of Wi-Fi networks, as it trains models with individual data (e.g., available at stations or the AP) while also preserving user privacy. However, if \ac{FL} will be implemented over wireless links, the mitigation of the adverse impact of wireless communications on \ac{FL} performance metrics becomes unavoidable \cite{pinyoanuntapong2020fedair}. 

\Acf{TL} is another concept that might be helpful for wireless networks in general. In this \ac{ML} method a model trained on one task is re-purposed on a second, related task. 
Usually, some retraining is required to fine-tune the model towards the second task. However, \ac{TL} can save time or obtain better performance in comparison to the development of a model from scratch~\cite{olivas2009handbook}. This technique works only if the model features learned from the first task are general. In the context of wireless networks, \ac{TL} might be applicable when reusing models trained in networks of a different technology (e.g., interference recognition in LTE) to boost the performance of Wi-Fi networks. A recent survey on \ac{TL} for wireless networks provides more insights regarding this important research direction \cite{nguyen2021transfer}.

\subsection{Learning from Experience}
Developing \ac{ML}-based solutions is not a straightforward process.
For example, it involves trial and error in terms of configuring satisfactory model parameters.
Just as \ac{ML} models are based on learning from previous experience, researchers deploying \ac{ML} solutions would benefit from sharing the experience gained from the development process.
Our literature review has revealed only a few explicit descriptions of such lessons learned in published papers. We share them here:
\begin{itemize}
    \item \textcite{krishnan2020optimizing} advocate a single hidden layer in agents due to their lower complexity and faster training than with multiple hidden layers. To limit data acquisition costs, \ac{DRL} agents should be trained online and provided with simple state information. Also, the duration of data collection is an important parameter to be optimized depending on the use case.
    \item \textcite{zhang2019deep}, who consider \ac{DL} for cellular networks, point out that ``deep learning solutions are not universal'' and thus not suitable in every case. They are prone to mistakes,  misinterpretation, and do not explain causality (especially in prediction models). Furthermore, the model complexity-accuracy trade-off is important for agents deployed on mobile devices and that \ac{RL} agents require training in real environments or high-fidelity simulators.
    \item \textcite{wilhelmi2021machine} warn that ``out-of-the-box DL methods may fail at capturing the relationship between interference and performance of WLANs'' and that data-driven solutions should be merged with the models used. Additionally, for prediction applications, the preprocessing of a dataset is crucial to obtain generalized solutions.
    \item \textcite{girmay2021coexistence} point out that \ac{QL} has been overused (at least in the area of network coexistence) since ``Q-learning is not an efficient solution for problems with dynamic environments''. Experience replay is proposed as an alternative. %
    \item Further recommendations regarding training \ac{ML} models include the following: the achieved model accuracy depends on the selected training features, using specific training data will lead to results which do not generalize, and, for limited training datasets, the subset selection becomes crucial \cite{fontaine2019towards, soto2021atari, pulkkinen2020understanding}.
\end{itemize}

Given the relevance of experience to avoid repeating mistakes, we  encourage researchers to always include in their works an explicit `lessons learned' section detailing new insights, or corroborating existing ones, to contribute to the development of this research area. Currently, many papers simply apply \ac{ML} methods without  clearly explaining why they are relevant for the considered problem \cite{ahmad2020machine}.
Researchers also need to discuss the challenges faced when applying ML methods.
Otherwise, the researcher contribution is unnecessarily limited.

\section{Conclusion}
\label{Sec:Conclusion}
\Ac{ML} is playing an increasing role in the field of improving Wi-Fi performance.
This survey has presented a comprehensive overview of over 250 recent \ac{ML}-based solutions for a variety of performance areas. We started with basic Wi-Fi features (such as channel access and rate adaptation), then we moved to more complex aspects (such as channel bonding, multi-band operation, and network management) and the problem of coexistence with other network technologies in shared bands. 
Next, we gave a brief overview of the application of \ac{ML} to multi-hop Wi-Fi settings. 
Finally, we summarized the tools and data sets available for researchers in this field.
To the best of our knowledge, this is the first survey to focus solely on Wi-Fi networks and to provide a detailed analysis of different Wi-Fi aspects that are supported through \ac{ML}.

A comparison of the three main \ac{ML} areas reveals
that \acl{SL} and \acl{RL} are frequently used, while \acl{USL} is less popular (Figure~\ref{fig:annual_stats}).
Meanwhile, the most often used ML mechanisms are \acl{QL}, \acl{MAB}, as well as different \acl{NN} types (mostly \ac{ANN}, \ac{DNN}, and \ac{CNN}). In most cases, these mechanisms are implemented to optimize only a constrained set of 802.11 parameters. Additionally, from reviewing the comparative Tables~\ref{tab:core} to~\ref{tab:coexistence}, we observe that with the increase in available computing power, \ac{DL} methods are gaining in popularity. About half of the  most recent papers implement \ac{DL} (cf. Figure~\ref{fig:annual_stats_ml} vs. Figure~\ref{fig:annual_stats_new_ml}).
The most commonly used \ac{DL} techniques are \acl{SL} and \acl{RL}, (Figure~\ref{fig:annual_stats_dl}). Additionally, \acl{FL} and \acl{TL} are a recent introduction in the Wi-Fi domain. We expect that they will become  more popular in the near future because they pose a chance to distribute learning tasks and improve training speed.

We believe that, as a next step, researchers will identify \ac{ML} schemes for the joint optimization of a wider range of Wi-Fi features. 
Additionally, they should investigate the coexistence of ML-controlled and legacy networks, since it poses a possible source of unfairness in channel access.
We also expect that the attractiveness of this area of research will continue to grow. To support this statement, we have identified several open research directions which could serve as a guide for researchers in their future work.

\appendices
\renewcommand{\IEEEiedlistdecl}{\IEEEsetlabelwidth{CCCCCCC}}
\section{List of Acronyms}
\label{app_acronyms}
\printacronyms[name={}]
\renewcommand{\IEEEiedlistdecl}{\relax}%

\printbibliography

\clearpage\tableofcontents

\end{document}